\documentclass{jfm}
\usepackage{graphicx}
\usepackage{newtxtext}
\usepackage{newtxmath}
\usepackage{natbib}
\usepackage{hyperref}
\usepackage{comment}
\usepackage[dvipsnames]{xcolor}
\usepackage{multirow}
\usepackage{float}
\usepackage{subcaption}
\usepackage{array}
\usepackage[normalem]{ulem}
\hypersetup{
    colorlinks = true,
    linkcolor  = blue,
    urlcolor   = blue,
    citecolor  = blue,
}

\newcommand{\RomanNumeralCaps}[1]

\newcommand\Gr{\mbox{\textit{Gr}}}
\newcommand\Sc{\mbox{\textit{Sc}}}
\definecolor{mygreen}{rgb}{0,0.5,0}
\definecolor{mymagenta}{rgb}{1 0 1}
\definecolor{myblue}{rgb}{0 0 0.75}
\definecolor{myred}{rgb}{0.7 0.11 0.11}

\newcommand{\pd}[3][]{\frac{\partial^{#1} #2}{\partial #3}}
\newcommand{\dd}[3][]{\frac{{\rm d}^{#1} #2}{{\rm d}#3}}
\DeclareMathOperator\erf{erf}

\graphicspath{{Figures/}}

\title{Self-similar scaling of variable-density Rayleigh--Taylor turbulence}

\author{Chian Yeh Goh\aff{1}
  \corresp{\email{cgoh@caltech.edu}}
 \and Daniel Brito Matehuala\aff{1}
 \and Guillaume Blanquart\aff{1}}

\affiliation{\aff{1}Department of Mechanical and Civil Engineering, California Institute of Technology, Pasadena, CA 91125, USA}

\begin{document}
\maketitle

\begin{abstract}
The dynamics of self-similar Rayleigh--Taylor (RT) mixing layers are investigated across a broad range of Atwood and Reynolds numbers using the statistically stationary Rayleigh--Taylor (SRT) flow configuration---a computational framework that enables simulation of self-similar RT flows at reduced cost compared to conventional temporally growing mixing layers. Normalizations are developed for all dominant non-transport terms in the continuity, mixed mass, and turbulent kinetic energy budgets in terms of the input parameters: the mixing layer height $h$, gravitational acceleration $g$, and fluid densities $\rho_H$ and $\rho_L$. Most normalized quantities collapse well across the parameter space. In some cases, variations in the Atwood number $A$ (or equivalently, the density ratio $R$) lead to consistent integral magnitudes but spatially shifted profiles. These shifts are primarily related to a division by density and are similarly observed in the analytical solution of the one-dimensional variable-density diffusion problem. The analysis introduces a reference density for the mixed mass, examines trends in Favre-averaged statistics, and derives a scaling law for the growth rate of the mixing layer. 
For height definitions encompassing the full extent of the layer, the conventional growth parameter, $\alpha = \dot{h}^2/4Agh$, varies with Atwood number. Our analysis leads to an alternative formulation using an effective Atwood number, $A^*= (\ln R)/2$, that is consistent with the scaling proposed by Belen'kii \& Fradkin (\emph{Trudy FIAN}, vol. 29, 1965, pp. 207--238). The corresponding growth parameter, $\alpha^*=\dot{h}^2/4A^*gh$, remains nearly constant across all Atwood numbers considered, offering a unified scaling for variable-density RT flows. 
\end{abstract}

\section{Introduction}
\label{sec:intro}

The Rayleigh--Taylor (RT) instability \citep{Rayleigh1882,Taylor1950} 
is observed in a variety of flow applications, including atmospheric science, astrophysics, and inertial confinement fusion \citep{zhou2017rayleigh1}. In its canonical configuration, two large fluid reservoirs of different densities are initially separated by an infinitely thin, planar interface at $x_2=\delta_I$. The heavier fluid of density $\rho_H$ sits atop the lighter fluid of density $\rho_L$ in the presence of a constant downward acceleration $g$. The difference between the two fluid densities is denoted by $\Delta \rho = \rho_H - \rho_L$, and their ratio is $R = \rho_H/\rho_L$.
In the presence of small interfacial perturbations, linear instabilities develop. For the case of zero viscosity and zero diffusivity, the linear growth rate $n_\lambda$ associated with a small-amplitude perturbation of wavelength $\lambda$ is given by \citet{Rayleigh1882} as
\begin{equation}
  n_\lambda^2 = \frac{2\pi A g}{\lambda},
  \label{eq:RTI}
\end{equation}
where $A$ is the Atwood number and it is related to the density ratio $R$ by 
\begin{equation}
  A = \frac{\rho_H-\rho_L}{\rho_H+\rho_L} = \frac{R-1}{R+1},
  \qquad    
  R = \frac{\rho_H}{\rho_L} = \frac{1+A}{1-A}.
  \label{eq:RandAt}
\end{equation}
Since the density ratio and the Atwood number are uniquely related, they may be used interchangeably in qualitative discussions to describe the density contrast in RT flows. 

Following the onset of linear instability, nonlinear effects become dominant, and a turbulent mixing layer develops. In an infinitely large domain, the flow eventually reaches a late-time state of self-similar growth, where viscosity and diffusivity no longer influence the large-scale evolution, and the flow dynamics become independent of its initial conditions \citep{youngs1984numerical, cook2004mixing}. \citet{youngs1984numerical} initially proposed an expression for the growth of the mixing layer height $h_i(t)$ as
\begin{equation}
  h_i = F_i(A)gt^2,
  \label{eq:Fgt2}
\end{equation}
where $F_i(A)$ is an unspecified function of the Atwood number (or density ratio). Any large-scale height definition should satisfy (\ref{eq:Fgt2}), but the expression for $F_i$ depends on this choice. Several common height definitions are listed in table~\ref{tab:hdef}. By assuming that the time scales of the turbulent mixing layer growth are proportional to the time scales of the inviscid modal growth associated with (\ref{eq:RTI}), Youngs obtained the now well-known expression
\begin{equation} 
  h_i \approx \alpha_i A g t^2,
  \label{eq:Agt2}
\end{equation} 
where $\alpha_i$ is a dimensionless growth parameter whose numerical value depends on the height definition, and $t$ is measured relative to a virtual time origin that depends on the initial conditions \citep{Snider1994,Thvenin2025}. Independently, by applying a self-similar ansatz to the governing equations in the Boussinesq limit (i.e. $A \ll 1$ or $R\approx 1$), \citet{ristorcelli2004rayleigh} derived an expression for the growth rate as $\dot{h}_i^2 \propto A g h_i$, where $\dot{h}_i = {\rm d}h_i/{\rm d}t$. By selecting the constant of proportionality as $4\alpha_i$, this relation becomes consistent with (\ref{eq:Agt2}) and can be written as
\begin{equation}
    \dot{h}_i^2 = 4 \alpha_i A g h_i.
    \label{eq:hgrowth}
\end{equation}

A different functional dependence of the mixing layer height on the density ratio was proposed much earlier by \cite{belen1965theory}. Their theoretical analysis was built on three key assumptions. First, turbulent kinetic energy $k$ scales with potential energy. Second, the flow can be described using a turbulent diffusivity of the form $D_t \propto \sqrt{k_t}h_i$. Third, for sufficiently small density ratios ($R\lesssim 4$), a self-similar solution for the mean density yields an analytical expression for the mixing layer height: $h_i \propto (\ln R) g t^2$, or equivalently,
\begin{equation}
  \dot{h}_i^2 \propto (\ln R) gh_i.
  \label{eq:hgrowth_lnR}
\end{equation}

\begin{table}
  \begin{center}
  \def~{\hphantom{0}}
    \renewcommand{\arraystretch}{1.3}
    \begin{tabular}{ll}
    Quantity & Definition \\[3pt]
    Entrainment height & \smash{$h_1 = 4\int \langle X \rangle (1-\langle X \rangle) dx_2$}
    \\
    Product height & $h_p = 2\int \min(\langle X \rangle,\langle 1-X \rangle) dx_2$ 
    \\
    Mixing height & \smash{$h_m = \frac{4}{\rho_0}\int \langle \rho Y(1-Y) \rangle dx_2 $} 
    \\
    Bubble height & $h_b = x_2(\langle X\rangle =0.99) - \delta_I$
    \\
    Spike height & $h_s = \delta_I - x_2(\langle X\rangle = 0.01)$ 
    \\
    Threshold height & $H = h_b + h_s$
    \end{tabular}
  \end{center}
  \caption{Definitions of mixing layer heights. $X$ and $Y$ are the mole and mass fractions of the heavy fluid; \smash{$\rho_0= 2\rho_H \rho_L/(\rho_H+\rho_L)$} is a normalization density; $\delta_I$ is the initial interface location. All integrals are taken from $x_2=-\infty$ to $x_2=+\infty$ and $\langle \cdot \rangle$ represents an ensemble average.}
  \label{tab:hdef}
\end{table}

\subsection{Atwood number effects}
Although \cite{belen1965theory} is commonly credited as an early proponent of $h\sim gt^2$ scaling, the associated $\ln R$ dependence has, to our knowledge, not been adopted in modern RT studies. This may be due to the original work being published in Russian and thus less accessible to the broader research community, or due to the rather specific assumptions involved that are difficult to verify without detailed data from high-Atwood-number RT flows. Instead, variable-density RT studies have been consistently presented in terms of $\alpha$ values based on (\ref{eq:Agt2}) and (\ref{eq:hgrowth}), even though neither the small-density-variation nor the small-amplitude assumptions hold.

As a result, it has been consistently observed in both experiments \citep{Dimonte2000,banerjee2010detailed} and simulations \citep{youngs2013density,zhou2019time} that $\alpha_i$ increases with Atwood number, provided $\alpha_i$ is defined using the full extent of the mixing layer. Specifically, bubble growth, represented by $\alpha_b$, is largely unaffected by the Atwood number, but spike growth $\alpha_s$ increases with the Atwood number. Overall, the growth parameter of the full mixing layer increases with Atwood number, while exhibiting Atwood-number-dependent asymmetries that can be described by $\alpha_s/\alpha_b$. Such asymmetries are not limited to the growth parameter and are also observed in one-dimensional planar-averaged density statistics. For instance, one-dimensional profiles of mixing parameters become more asymmetric as the Atwood number increases, although the nature of such asymmetries depends on the specific definition of mixedness considered \citep{livescu2013numerical,youngs2013density}. In terms of volume-averaged mixedness, \citet{zhou2016asymptotic} found that those defined in terms of the mole fraction exhibit little sensitivity to the Atwood number, while mass-fraction-based measures decrease systematically as $A$ increases. 

In the RT literature, statistics associated with the velocity field are less commonly reported. When they are, they often lack consistent normalization with respect to global flow parameters, which hinders meaningful comparisons between different studies. As an illustration, velocity fluctuations have been normalized using $\dot{h}$ \citep{zhou2019time}, $(A g h)^{1/2}$ \citep{akula2016dynamics}, $Agt$ \citep{banerjee2010detailed}, as well as scales based on the initial conditions of the simulation \citep{Livescu2010}. Perhaps due to the aforementioned issues, there have been surprisingly few comparative studies of velocity profiles across different Atwood numbers. Based on the findings of \citet{Livescu2010}, the primary effect of an increase in the Atwood number is a shift in the peaks of the Favre-averaged velocity and turbulent kinetic energy toward the light-fluid side of the mixing layer. 

\subsection{Reynolds number effects}

The late-time self-similar expressions in (\ref{eq:Agt2}) and (\ref{eq:hgrowth}) also suggest that the growth parameter is independent of time (and equivalently, the Reynolds number). In a finite domain, this self-similar state is only observed over a limited time window when the mixing layer height is sufficiently larger than the wavelengths of the initial perturbations \citep{ramaprabhu2005numerical, mueschke2009investigation,banerjee20093d,youngs2013density}, while remaining small enough such that domain confinement effects are not significant \citep{dalziel2008mixing,boffetta2012anomalous}. Given the appropriate initial conditions, regions of near-constant $\alpha_i$ values have been observed for many temporally growing RT simulations at sufficiently late times. However, in several studies \citep{Cabot2006_Nat,burton2011study,zhou2019time}, there appears to be a slight increase in $\alpha_i$ values towards the end of the simulation. Similar effects are observed for mixing parameters \citep{zhou2016asymptotic}, where a near-constant region is clearly observed but small fluctuations persist. It remains unclear whether these variations arise due to the growing Reynolds number, the diminishing influence of the initial conditions, or the effects of domain boundaries. Another global parameter that is often studied is the ratio of total kinetic energy to total potential energy loss. This ratio has been observed to exhibit a weak dependence on the Reynolds number \citep{Cabot2006_Nat,goh2025statistically}, which is commonly described as a non-equilibrium effect stemming from the increasing time lag between production and dissipation at higher Reynolds numbers.

\subsection{Statistically stationary Rayleigh--Taylor flow configuration}

A key impediment to resolving the open questions discussed above is the lack of high-quality self-similar RT flow data. Although the conditions for achieving late-time self-similar growth are now well understood, the transient behavior remains poorly characterized. Different flow quantities reach self-similarity at different times, and this issue is exacerbated at higher Atwood numbers, where convergence to self-similarity tends to be delayed \citep{burton2011study}. It remains uncertain whether any temporally growing RT simulation has fully attained a self-similar state that is truly independent of the initial conditions and domain boundaries, particularly at high Atwood numbers. This issue is discussed in detail by \citet{schilling2020progress}. For a temporally growing mixing layer, the only path toward addressing these questions about the self-similar state seems to involve longer simulation times, larger domains, and ballooning computational costs. 

Many of these challenges are substantially mitigated by the statistically stationary RT (SRT) flow configuration proposed by \citet{goh2025statistically}. A more detailed summary is provided in \S~\ref{sec:math}, but a brief introduction is presented here to highlight its computational advantages. In essence, the SRT configuration simulates a statistically stationary flow that is representative of a temporally growing RT (TRT) mixing layer. 
The flow can be evolved over arbitrarily long simulation times on a fixed grid, generating a continuous stream of flow realizations and yielding converged statistics that are independent of initial conditions. 
In addition to eliminating the cost associated with a growing Reynolds number, the SRT configuration is a minimal flow unit that requires only a fraction of the grid points for a typical self-similar TRT simulation at the same Reynolds number. For these reasons, the SRT flow configuration is uniquely suited to advancing the study of self-similar RT flows, and can be used efficiently to explore higher Atwood number regimes at considerably lower computational cost. 

\subsection{Objectives}

The primary objective of this study is to investigate the influence of Atwood and Reynolds numbers on self-similar RT flow using the SRT configuration. The analysis focuses on planar-averaged profiles of the dominant non-transport terms in the continuity, mixed mass, and turbulent kinetic energy budgets. We seek to distinguish between global (or integral) and local (or shape) effects: global effects are reflected in the integral properties of the profiles and can be accounted for by appropriate normalization, while local effects may persist post-normalization and are expressed through the shape of the normalized profiles. Normalizations will be developed through a combination of theoretical arguments and empirical observations, with the goal of identifying near-universal non-dimensional quantities that hold across a broad parameter space. Through this analysis, we aim to provide further context to some of the questions posed earlier in this section and suggest alternatives to conventional approaches.

The paper is organized as follows. In \S~\ref{sec:math}, the SRT configuration is introduced through a review of \citet{goh2025statistically}, and the methodology is extended to different Atwood numbers. Section \ref{sec:method} details the simulation cases and numerical methods, and compares SRT results with TRT results. In \S~\ref{sec:results}, normalizations for all dominant budget terms are proposed, and the normalized profiles analyzed. Section \ref{sec:discussion} builds on these results to explore three related topics: (i) the derivation of a reference density for the mixed mass, (ii) the effects of the Atwood number on Favre-averaged velocity statistics, and (iii) a proposed scaling law for the mixing layer growth rate that generalizes (\ref{eq:hgrowth}) to variable-density flows. Finally, \S~\ref{sec:conclusion} summarizes the key findings and discusses their broader implications.

\section{Mathematical description}
\label{sec:math}
The SRT flow configuration is first introduced through a review of the governing equations in \S~\ref{sec:SRT_eq}, followed by a summary of its flow properties in \S~\ref{sec:SRT_prop}. Complete details on the mathematical derivation, assumptions, and validation of the computational framework can be found in \citet{goh2025statistically}. In \S~\ref{sec:inputs}, the input parameters for the present study are discussed.

\subsection{Governing equations}
\label{sec:SRT_eq}

The governing equations for SRT flow are derived through two main steps. First, by leveraging the anticipated self-similar behavior of turbulent RT mixing layers at late times, a transformation of the vertical coordinate and velocities is applied to the low-Mach-number Navier-Stokes equations (NSE). Second, these transformed equations are then evaluated at a specified time $t=t_0$, leading to simplified equations in the transformed coordinates that resemble the original NSE, but include two sets of additional terms. The SRT governing equations corresponding to  continuity, scalar, and momentum transport are

\begin{equation} 
  \pd{\rho}{s} + \pd{\rho u_j}{x_j} = \frac{y}{\tau_0}\pd{\rho}{y},
  \label{eq:mass_SRT}
\end{equation}
\begin{equation} 
  \pd{\rho Y}{s} + \pd{\rho u_j Y}{x_j} = \pd{}{x_j}\left(\rho D \pd{Y}{x_j}\right) +\frac{y}{\tau_0}\pd{\rho Y}{y}, 
  \label{eq:scalar_SRT} 
\end{equation}
\begin{equation} 
  \pd{\rho u_i}{s} + \pd{\rho u_j u_i}{x_j} = -\pd{p}{x_i} + \pd{\tau_{ij}}{x_j} - \rho g \delta_{2i} + \frac{y}{\tau_0}\pd{\rho u_i}{y} - \frac{1}{2\tau_0}\rho u_i,
  \label{eq:mom_SRT} 
\end{equation}
where $Y$ is the mass fraction of the heavy fluid, $D$ is the kinematic diffusivity, $u_i$ are the velocity components, and $p$ is the hydrodynamic pressure. In the SRT framework, $s$ is the simulation time in the transformed coordinate system, and $\tau_0$ is an SRT-specific flow parameter. The shear stress tensor $\tau_{ij}$ is defined as
\begin{equation}
  \tau_{ij} = \rho \nu \left[\frac{\partial u_i}{\partial x_j} + \frac{\partial u_j}{\partial x_i} - \frac{2}{3} \frac{\partial u_j}{\partial x_k}\delta_{ij}\right],
\end{equation}
where $\nu$ is the kinematic viscosity and $\delta_{ij}$ is the Kronecker delta. Relating mass fraction to density, the low-Mach-number equation of state is
\begin{equation} 
  \rho(Y) = \frac{\rho_H \rho_L}{\rho_H- (\rho_H-\rho_L)Y},
  \quad
  {\rm or}
  \quad 
  Y(\rho) = \frac{\rho_H (\rho - \rho_L)}{\rho(\rho_H - \rho_L)}.
  \label{eq:EOS} 
\end{equation}

Two sets of source terms appear on the right-hand-side (RHS) of (\ref{eq:mass_SRT})--(\ref{eq:mom_SRT}). The first set of source terms, denoted generally as $S_\phi$, is derived from rescaling the vertical coordinate, and appears in all three equations (\ref{eq:mass_SRT})--(\ref{eq:mom_SRT}) in the form
\begin{equation}
    S_\phi = \frac{y}{\tau_0}\pd{\rho \phi}{y},
    \label{eq:Sphi}
\end{equation}
where $\phi = 1, u_i, {\rm or~} Y$. In (\ref{eq:Sphi}), $y=x_2-\hat{\delta}_I$ is the vertical coordinate relative to the location of the initial interface, which is specified by the input $\hat{\delta}_I$. Practically, the input $\hat{\delta}_I$ can be used to shift the mixing layer relative to the computational domain. Beyond this use (which is briefly discussed in \S~\ref{sec:simcases}), $y$ and $x_2$ can be treated equivalently. The second set of source terms arises from velocity scaling and appears only in the momentum equations (\ref{eq:mom_SRT}) as 
\begin{equation}
    T_i = -\frac{1}{2\tau_0} \rho u_i.
    \label{eq:Ti}
\end{equation}

The stationary height of the mixing layer is determined by the timescale parameter $\tau_0$, which is found in both sets of source terms defined in (\ref{eq:Sphi}) and (\ref{eq:Ti}). In \citet{goh2025statistically}, $\tau_0$ was implemented using an input mixing height $\hat{h}_m$ and the closure equation
\begin{equation} 
  \frac{1}{\tau_0 (s)} = \frac{4}{\rho_0 \hat{h}_m} \int \langle \rho \chi\rangle_{1,3}\,{\rm d}y, 
  \label{eq:closure} 
\end{equation}
where $\langle \cdot \rangle_{1,3}$ denotes an instantaneous planar average over the $x_1$ and $x_3$ directions, $\chi=2D(\partial Y/\partial x_i)^2$ is the local scalar dissipation rate, and $\rho_0 = 2\rho_H \rho_L/(\rho_H+\rho_L)$ is the harmonic mean of the reservoir densities, arbitrarily chosen as a normalization density. All integrals in this paper are evaluated over the entire domain. By solving (\ref{eq:mass_SRT})--(\ref{eq:EOS}) and (\ref{eq:closure}), a statistically stationary RT flow is achieved. 

\subsection{Properties of SRT flow}
\label{sec:SRT_prop}
To provide further context, the following section summarizes several key aspects of the SRT configuration and discusses its relationship with the TRT flow configuration.
\begin{enumerate}
    \item As mentioned earlier, the derivation of the governing SRT equations involves two main steps. The first step represents a simple shrinking of the flow in the vertical direction. The transformed equations remain valid for all $t$ and can be solved in the transformed variables to achieve a stationary height. However, the small scales would continue to shrink, and DNS of such a ``rescaled TRT'' configuration would still need to resolve the growing range of scales, offering no computational benefit to a traditional TRT simulation. The second step involves evaluating the ``rescaled TRT'' equations at a fixed time $t=t_0$, which yields the final form of the SRT equations, (\ref{eq:mass_SRT})–(\ref{eq:mom_SRT}). This approximation is crucial, as it enables stationarity not just in the large scales but also at the small scales.
    
    \item The additional terms $S_\phi$ and $T_i$ oppose the underlying TRT growth that is driven by the NSE terms. The term $S_\phi$ is zero in the core of the mixing layer ($y=0$), most active near the edges of the mixing layer, and zero in the fluid reservoirs (no gradients); $T_i$ has no explicit dependence on the spatial coordinates. On average, the source terms remove kinetic energy from the flow at a rate proportional to $1/\tau_0$. In spectral space (defined only in the homogeneous directions), this removal of energy is directly proportional to the in-plane kinetic energy spectra and preserves the wavenumber distribution of kinetic energy at each vertical location (see Appendix \ref{app:src_spectra}).
    The time average of the SRT parameter $\bar{\tau}_0$ (an overline denotes a time average) corresponds to the growth timescale of the mixing layer height in TRT flow. If the flow were self-similar, all height definitions $h_i$ would grow proportionally to each other in TRT flow and have the same timescale. This is expressed mathematically as
    \begin{equation}
        \left[ \frac{1}{\bar{\tau}_0} \right]_{\rm SRT} = \left[\frac{\dot{h}_i(t_0)}{h_i(t_0)} \right]_{\rm TRT}.
        \label{eq:tau0_TRT}
    \end{equation}
    By combining (\ref{eq:hgrowth}) and (\ref{eq:tau0_TRT}), $\bar{\tau}_0$ is related to the height of the equivalent TRT flow by
    \begin{equation}
      h_i(t_0) = 4\alpha_i Ag\bar{\tau}_0^2.
      \label{eq:alpha_SRT}
    \end{equation} 
    
    \item When solving the SRT equations, $\tau_0$ can either be prescribed as a constant input or implemented through a closure equation and a secondary input. If prescribed as a constant $\hat{\tau}_0$, the mixing layer grows until $h(s)$ reaches a stationary value that is not known \emph{a priori}. Alternatively, using the closure (\ref{eq:closure}) with an input height $\hat{h}_m$ ensures that the instantaneous mixing layer height 
    \begin{equation}
        h_m(s) = \frac{4}{\rho_0}\int \langle \rho Y (1-Y) \rangle_{1,3} \,{\rm d}y
        \label{eq:hm_def}
    \end{equation}
    convergences to the constant input $\hat{h}_m$. As shown by \citet{goh2025statistically}, the choice of method does not impact the converged flow statistics. That being said, an input height is preferable to an input time scale. First, the mixing layer height is an intuitive and common reference in the RT literature. Second, a known stationary height allows the user to define the computational grid efficiently without needing to iterate or allocate excess grid cells.
    
    \item Similar to observations from other stationary configurations \citep{rosales2005linear,Chung2010_JFM,Carroll2015_TCFD,dhandapani2020isotropic}, the SRT configuration exhibits box-filling tendencies in the homogeneous directions. In a domain of lateral size $\mathcal{L}_{\rm SRT}$, the largest flow structures are expected to expand laterally over time and eventually interact with the periodic boundaries, leading to the formation of a single large-scale structure or minimal flow unit (MFU). In this statistically stationary state, the integral lengthscales of the flow are bounded vertically by $\ell_2 \sim \hat{h}$ and horizontally by $\ell_1 \sim \mathcal{L}_{\rm SRT}$, The resulting stationarity of $\ell_2/\ell_1$ in SRT mirrors that observed in unconfined, self-similar TRT flows. Specifically, \citet{zhou2019time} found $\ell_2/\ell_1$ to be approximately constant across Atwood number and Reynolds numbers, despite $h(t)/\mathcal{L}_{\rm TRT}$ growing in time. In an unconfined TRT, a large domain allows the growing mixing layer sufficient time to approach its intrinsic large-scale aspect ratio through mode-coupling processes. In a confined SRT configuration, $\mathcal{L}_{\rm SRT}$ (along with input height $\hat{h}$) constrains the flow and selects $\ell_2/\ell_1$ directly. \citet{goh2025statistically} reported that the mixing layer aspect ratio $\hat{h}_m/\mathcal{L}_{\rm SRT}$ affects not only the integral lengthscales, but also other global quantities, e.g. $\alpha$ and mixedness. For $A=0.5$, all SRT flow statistics from $\bar{h}_m/\mathcal{L} \approx 1.5$ agree well with self-similar TRT flow.

    \item Since the resulting flow is statistically stationary in all quantities, it can be simulated indefinitely. SRT flow has a finite time memory, which can be quantified by its integral timescale $\tau_{\rm int}$. After $\Delta s \sim \tau_{\rm int}$, the flow becomes de-correlated, and a subsequent, independent realization of the flow begins, each of which are representative of TRT flow in the vicinity of $t=t_0$. Over simulation time $s$, SRT flow generates continuous, independent realizations of a single large-scale structure which are equivalent to realizations (generated over $x_1$, $x_3$, and initial conditions) in a TRT flow ensemble at a specific time.  Mathematically, any flow quantity $\Omega$ averaged over $x_1$, $x_3$, and $s$ in SRT flow is (approximately) equal to its ensemble average in self-similar TRT flow at time $t_0$,  
    \begin{equation}
        \langle \Omega_{\rm SRT} \rangle_{1,3,s}(y) \approx \langle \Omega_{\rm TRT} \rangle (t_0,y).
    \end{equation}
    Henceforth, unless the domain of averaging is specified as a subscript, $\langle \cdot \rangle = \langle \cdot \rangle_{1,3,s}$ is assumed for SRT flow. Finally, due to its finite time memory, converged SRT flow is independent of its initial conditions. In practice, this is typically achieved within $s<5\tau_0$.

\end{enumerate}

\subsection{Input parameters}
\label{sec:inputs}

To extend the framework from \citet{goh2025statistically} to a range of Atwood numbers, a different input height definition from (\ref{eq:hm_def}) is chosen. The reasons for this choice are first presented, followed by a summary of all non-dimensional inputs.

\subsubsection{Input height}
\label{sec:input_h1}

The mixing height $h_m$, defined in (\ref{eq:hm_def}), includes the density within the integral and thus requires a reference density for normalization; this reference density generally depends on the reservoir densities $\rho_H$ and $\rho_L$. In \citet{goh2025statistically}, this reference density was chosen arbitrarily as the harmonic mean of the reservoir densities $\rho_0$. Because all simulations were conducted at the same reservoir densities (i.e. same Atwood number), the choice of normalization density did not influence the results. In contrast, in the present study, the Atwood number (and consequently, at least one of $\rho_H$ and $\rho_L$) is varied; the physically appropriate normalization density is not an obvious one.  To ensure consistency across
SRT configurations of different Atwood numbers, a different height definition that does not require an arbitrary density normalization is used. 

Heights can be defined in terms of the mole fraction, $X = (\rho-\rho_L)/\Delta \rho$, or the mass fraction $Y$. In the analytical self-similar solution for one-dimensional variable-density diffusion (shown in Appendix~\ref{app:1dvd}), the mole fraction profile $X(y/h_1)$ does not change with Atwood number, but the mass fraction profile does. For three-dimensional, turbulent SRT flow, it will be shown in \S~\ref{sec:results_cont} that this observation remains true. More specifically, the mean mole fraction $\langle X \rangle$ is approximately constant with respect to the Atwood number, but the Favre average of the mass fraction is not. Therefore, we seek height definitions that involve only the mean mole fraction. 

There are several common height definitions based on the mean mole fraction. These include the entrainment height, $h_1 = 4\int \langle X \rangle (1-\langle X \rangle) \,{\rm d}y$, the product height, $h_p = 2\int \min(\langle X \rangle,\langle 1-X \rangle) \,{\rm d}y$, or threshold-based heights that are derived from the $y$-locations of the mean mole fraction field at specified thresholds, e.g. $H = y(\langle X\rangle=0.99) - y(\langle X\rangle=0.01)$. Threshold-based heights are determined from averages on a single plane and are more prone to statistical noise. Conversely, integral height definitions are computed over the entire domain. Between the two integral heights mentioned, we choose to use $h_1$ because its integrand varies smoothly with $\langle X \rangle$ and an analytical equation for its time evolution can be derived. This leads to a closure equation for $\tau_0$ that functions similarly to (\ref{eq:closure}), but for a prescribed input $\hat{h}_1$ instead of $\hat{h}_m$, 
\begin{equation}
  \frac{1}{\tau_0(s)} = \frac{8}{\Delta \rho \hat{h}_1} \int \langle X \rangle_{1,3} \pd{\langle \rho v \rangle_{1,3}}{y} dy.
  \label{eq:closureh1}
\end{equation}
The derivation of this closure equation is presented in Appendix \ref{app:closure_h1}.
Solving (\ref{eq:closureh1}) alongside (\ref{eq:mass_SRT})--(\ref{eq:EOS}), $h_1(s)$ converges to a stationary value, and $\bar{h}_1 \approx \hat{h}_1$.

The minimal flow unit aspect ratio of $\bar{h}_m/\mathcal{L}=1.5$ for $A = 0.5$ quoted by \citet{goh2025statistically} corresponds to $\bar{h}_1/\mathcal{L} = 1.7$. In the present study, we seek to keep $\bar{h}_1/\mathcal{L}=1.7$ approximately constant while studying the effect of other parameters.

\subsubsection{Non-dimensional inputs}
\label{sec:nondim}

The SRT equations (\ref{eq:mass_SRT})--(\ref{eq:mom_SRT}) require the specification of gravity ($g$) and fluid properties ($\rho_H$, $\rho_L$, $\nu$, $D$), while the closure equation (\ref{eq:closureh1}) requires input $\hat{h}_1$. Additionally, as discussed in \S~\ref{sec:SRT_prop}, the lateral domain size $\mathcal{L}$ is physically significant in SRT flow, while the initial conditions are not. In accordance with the Buckingham Pi theorem, the flow can be fully defined by four non-dimensional numbers,
\begin{equation} 
  A = \frac{\rho_H-\rho_L}{\rho_H+\rho_L}, 
  \quad 
  \Sc = \frac{\nu}{D}, 
  \quad
  \Gr = \frac{A g\hat{h}_1^3}{\nu^2},
  \quad 
  \lambda = \frac{\hat{h}_1}{\mathcal{L}},
  \label{eq:parameters}
\end{equation}
where $A$ is the Atwood number, $\Sc$ the Schmidt number, $\Gr$ the Grashof number, and $\lambda$ the aspect ratio of the mixing layer. In the present study, the Atwood and Grashof numbers are varied, while $\Sc$ and $\lambda$ are held constant. 

The choice of these non-dimensional quantities is not unique. First, the density ratio $R$ could have been used instead of $A$. They are uniquely related by (\ref{eq:RandAt}) and either choice is equivalent. Second, the inclusion of $A$ in the definition of the Grashof number is based on convention. While $\Gr$ is listed as the input parameter in (\ref{eq:parameters}), the results are primarily presented in terms of the output Reynolds number. Both the Grashof number and the Reynolds number characterize the ratio of the buoyancy-driven large scales to the viscous scales and their values correlate with turbulence intensity. However, the Reynolds number is preferred because it is more commonly used in the turbulence literature and widely understood beyond the field of buoyancy-driven flows. It is defined in terms of the time-averaged output parameters as
\begin{equation}
  \Rey = \frac{\bar{h}_1^2}{\bar{\tau}_0\nu}.
  \label{eq:Re_def}
\end{equation}
To maintain consistency throughout this paper, $h_1$ is used in (\ref{eq:Re_def}), but any large-scale height definition would be reasonable for a self-similar RT flow. For instance, a Reynolds number based on $H$ was used in \citet{Cabot2006_Nat} and \citet{goh2025statistically}.

\section{Simulation cases and numerical method}
\label{sec:method}

This section is divided into three parts. First, the setup of the computational domain and the list of simulation cases are presented. Second, the numerical method is described. Finally, several global parameters are compared between SRT and self-similar TRT results, extending the comparisons beyond what was done by \citet{goh2025statistically} for $A=0.5$.

\subsection{Simulation cases}
\label{sec:simcases}

Simulations are performed on a three-dimensional rectangular domain. The top and bottom boundaries are implemented as Dirichlet boundary conditions and the lateral boundaries as periodic. The computational grid is centered at $x_i=0$ with domain lengths $\mathcal{L}_i$ and grid spacing $\Delta_i$. In the horizontal directions, $\mathcal{L}_1= \mathcal{L}_3=\mathcal{L}$, and $\Delta_1 = \Delta_3 = \mathcal{L}/N_1$, where $N_1=N_3$ is the number of grid points in each direction. In all simulations, $\mathcal{L} = 1$ is fixed. In the vertical direction, $\Delta_2$ varies with $x_2$. A uniformly spaced ($\Delta_2 = \Delta_1$) core grid of length $\mathcal{L}_{2c} = N_{2c}\Delta_1$ is used to resolve the bulk of the mixing layer ($|x_2|\le \mathcal{L}_{2c}/2$). This core grid region is defined to be at least 10\% wider than the region that contains the 1\% and 99\% mean mole fraction locations. Because of known asymmetries associated with high Atwood number flows \citep{cabot2013statistical,zhou2019time}, the optional parameter $\hat{\delta}_I$ is used to shift the mixing layer so that it lies within the refined core grid. Outside the core, the vertical spacing $\Delta_2(x_2)$ is stretched with a factor of 1.1. The total domain height satisfies the condition $\mathcal{L}_2>4\mathcal{L}_{2c}$. In the vertical direction, the results were verified to be insensitive to a larger domain or core grid region. 

A list of the simulation cases considered in this study is summarized in tables~\ref{tab:simcases1} and \ref{tab:simcases0}. All simulations are performed at $\Sc = 1$ with constant $\nu$ and $D$. Table~\ref{tab:simcases1} summarizes the set of new simulations in which the Atwood number is varied at an approximately constant $\Rey \approx 1100$ while table~\ref{tab:simcases0} lists the simulation cases from \citet{goh2025statistically} that represent a sweep of the Reynolds number at constant $A = 0.5$. As discussed in \S~\ref{sec:SRT_eq}, the input height used by \citet{goh2025statistically} was $\hat{h}_m$, and the associated non-dimensional inputs were $A$, $\Sc$, $\Gr_m = Ag\hat{h}_m^3/\nu^2$, and $\lambda_m = \hat{h}_m/\mathcal{L}$. To account for this difference in definitions, each set of simulations is listed in terms of their respective input parameters. The output values of $\bar{h}_1/\mathcal{L}$ and $\Rey$ are included in the final two columns of both tables for a consistent comparison across both sets of simulations. 

\begin{table}

\begin{center}
\begin{tabular}{lcccccccccc}
  \textrm{Case} & \multicolumn{1}{c}{$A$} & \multicolumn{1}{c}{$\Gr (\times 10^6)$} & $\lambda$ & $\hat{\delta}_I/\mathcal{L}$ & $N_1$ & $N_{2c}$ & $N_2$  & $\kappa_{\max}\eta$ & $\bar{h}_1/\mathcal{L}$ & $\Rey$\\ [3pt]
  A0 & 0.01 & 11.4 & 1.7 & 0 & 128 & 456  & 546 & 3.05 & 1.72 & 1021\\
  A2 & 0.2  & 11.6 & 1.7 & 0.080 & 128 & 456  & 546 & 3.00 & 1.71 & 1051 \\
  A5 & 0.5  & 10.8 & 1.7 & 0.178 & 128 & 456  & 546 & 2.99 & 1.69 & 1084 \\
  A8 & 0.8  & 8.61 & 1.7 & 0.370 & 128 & 456  & 546 & 2.98 & 1.71 & 1127 \\
\end{tabular}
\end{center}
\caption{Simulation cases performed in the current study using $\hat{h}_1$ as input \label{tab:simcases1}}
\end{table}

\begin{table}
\begin{center}
\begin{tabular}{lcccccccccccc}
  \textrm{Case} & \multicolumn{1}{c}{$A$} & \multicolumn{1}{c}{$\Gr_m (\times 10^6)$} & $\lambda_m$ & $\hat{\delta}_I/\mathcal{L}$ & $N_1$ & $N_{2c}$ & $N_2$ & $\kappa_{\max}\eta$ & $\bar{h}_1/\mathcal{L}$ & $\Rey$ \\ [3pt]
  G1 & 0.5 & 0.183 & 1.5 & 0.178 & 32  & 114  & 176 & 2.83 & 1.82 & 245 \\
  G2 & 0.5 & 1.16  & 1.5 & 0.178 & 64  & 228  & 304 & 2.91 & 1.77 & 496\\
  G3 & 0.5 & 3.43  & 1.5 & 0.178 & 96  & 342  & 426 & 2.95 & 1.76 & 791\\
  G4 & 0.5 & 7.39  & 1.5 & 0.178 & 128 & 456  & 546 & 2.99 & 1.74 & 1061 \\
  G5 & 0.5 & 46.9  & 1.5 & 0.178 & 256 & 910  & 1014 & 3.04 & 1.73 & 2543 \\
  G6 & 0.5 & 138   & 1.5 & 0.178 & 384 & 1364 & 1476 & 3.06 & 1.74 & 4390 \\
\end{tabular}
\end{center}
\caption{Simulation cases from \citet{goh2025statistically} performed using $\hat{h}_m$ as input \label{tab:simcases0}}
\end{table}

In table~\ref{tab:simcases1}, A0--A8 have increasing Atwood numbers with A0 representing the Boussinesq limit. The Grashof number is reduced with increasing Atwood number to ensure that the ratio $\bar{h}_1/\eta$ is constant, where $\eta = (\nu^3/\tilde{\epsilon})^{1/4}$ is the Kolmogorov scale and $\tilde{\epsilon}$ is the Favre-averaged viscous dissipation (defined later in (\ref{eq:tkebudget})). Doing so results in a near-constant Reynolds number across Atwood numbers. Although the variation in the Grashof number is not large, $\Gr$ does not scale directly with $\bar{h}_1/\eta$ or $\Rey$ due to the somewhat arbitrary inclusion of $A$ in its definition. This slight difference in trends between $\Rey$ and $\Gr$ is explained in \S~\ref{sec:Astar}.

In table~\ref{tab:simcases0}, the Reynolds number increases from G1 to G6. There is a gradual decrease of the aspect ratio $\bar{h}_1/\mathcal{L}$ with Reynolds number; this is because $\hat{h}_m$ was used as the input parameter and $h_1$ was not prescribed directly. It will be shown in \S~\ref{sec:scaling_budget} that the small differences in $\bar{h}_1/\mathcal{L}$ do not affect the results, once quantities are suitably normalized. 

To ensure that simulations are well-resolved, all simulations are performed at $\kappa_{\max}\eta \gtrsim 3.0$, where $\kappa_{\max} = \pi/\Delta_2$ is the maximum resolved wavenumber. The minimum values of $\kappa_{\max}\eta$ across the mixing layer are listed in tables~\ref{tab:simcases1} and \ref{tab:simcases0}.

\subsection{Numerical method}
\label{sec:numerics}

Simulations are performed using the computational solver, NGA \citep{Desjardins2008_JCP}. The numerical code solves the conservative-variable formulation of the low Mach number NSE with staggered finite difference operators and uses a fractional step method to enforce continuity.  The NSE are solved using a second order semi-implicit iterative midpoint scheme and uses staggering in time between the momentum field and the scalar and density fields. The scalar is advanced first, the density field is updated, and the momentum equations are then advanced. The resulting computational framework conserves kinetic energy discretely (i.e. there is no numerical viscosity).

The bounded cubic Hermite polynomial (BCH) \citep{verma2014improved} scheme is used for scalar transport. This scheme was chosen because it ensures scalar boundedness and has less numerical diffusion than other schemes, including weighted essentially non-oscillatory (WENO) schemes. In particular, \citet{verma2014improved} found that other bounded schemes require at least twice the spatial resolution to resolve small-scale scalar features as well as BCH. The Courant-Friedrichs-Lewy condition of ${\rm CFL}\le 0.8$ is imposed for all simulations in the current study. Further details on the implementation of the original NSE and scalar transport can be found in \citet{Desjardins2008_JCP} and \citet{verma2014improved}, while the implementation of the SRT terms $S_\phi$ and $T_i$ is addressed by \citet{goh2025statistically}.

\subsection{Comparison with self-similar TRT}

The SRT results for $A=0.5$, $\lambda=1.7$ have been validated comprehensively against TRT studies by \citet{goh2025statistically}. We extend the analysis here to other Atwood numbers. In particular, it is crucial to verify that $\lambda=1.7$ remains the minimal flow unit for self-similar RT turbulence for the range of Atwood numbers considered. To this end, global parameters obtained from the SRT results for $A=$ 0.2, 0.5, and 0.8 are compared against similar quantities from DNS studies of self-similar TRT flows \citep{zhou2016asymptotic, zhou2019time}. The growth parameter $\alpha_p$ is computed using (\ref{eq:alpha_SRT}) and the definition of $h_p$ (see table \ref{tab:hdef}), while the mixing parameters are defined as
\begin{gather}
  \Xi = \frac{\int \langle \min(X,1-X)\rangle \,{\rm d}y}{\int \min(\langle X \rangle, \langle 1- X \rangle) \,{\rm d}y}, 
  \label{eq:Xi}
  \\
  \Theta = \frac{\int \langle X(1-X) \rangle \,{\rm d}y}{\int \langle X \rangle (1- \langle X \rangle) \,{\rm d}y},
  \label{eq:Theta}
  \\ 
  \Psi = \frac{\int \langle \rho Y (1-Y) \rangle \,{\rm d}y}{\int \langle \rho \rangle \langle Y \rangle (1-\langle Y \rangle) \,{\rm d}y}.
  \label{eq:Psi}
\end{gather}
The reference TRT values in table~\ref{tab:TRTvalidate} are obtained from figure~2 of \citet{zhou2016asymptotic} and figure~5 of \citet{zhou2019time}. 

Comparisons involving TRT values should be considered in the context of two main issues. First, there is no established criteria for determining the onset of self-similarity in a TRT flow. The reference TRT values in table \ref{tab:TRTvalidate} correspond to the range of values observed for $t>10$, which is visually estimated to be the time from which global quantities start to vary more slowly in time. Second, there is variability between different studies owing to differences in the initial conditions, lateral confinement effects, numerical methods, and data processing. Therefore, comparison of numerical values are typically only meaningful up to one or two significant figures. The observed spread in these global values and the associated issues have been discussed in several review articles \citep{dimonte2004,zhou2017rayleigh1,schilling2020progress}. The TRT values reported in table~\ref{tab:TRTvalidate} are from a single set of simulations and are provided for purposes of comparison rather than validation. Nonetheless, the good agreement between SRT and TRT results indicates that SRT flow at $\lambda=1.7$ is representative of self-similar TRT flow up to ${\rm A = 0.8}$.

\begin{table}
  \begin{center}
  \begin{tabular}{lccc}
    \textrm{Parameter} & \multicolumn{1}{c}A & Reference TRT values & SRT \\ [3pt]
    \multirow{3}{*}{Growth parameter, $\alpha_p$}  & 0.2 & 0.016 -- 0.023 & 0.0168 \\
    & 0.5 & 0.017 -- 0.023 & 0.0199 \\
    & 0.8 & 0.020 -- 0.026 & 0.0255 \\\\
  
    \multirow{3}{*}{Mixedness, $\Xi$}  & 0.2 & 0.80 -- 0.82 & 0.828 \\
    & 0.5 & 0.80 -- 0.82 & 0.807\\
    & 0.8 & 0.78 -- 0.80 & 0.770\\\\
  
    \multirow{3}{*}{Mixedness, $\Theta$}  & 0.2 & 0.78 -- 0.80 &  0.809 \\
    & 0.5 & 0.78 -- 0.80 & 0.793 \\ 
    & 0.8 & 0.77 -- 0.79 & 0.767 \\\\
  
    \multirow{3}{*}{Normalized mixed mass, $\Psi$}  & 0.2 & 0.78 -- 0.80  & 0.806 \\
    & 0.5 & 0.76 -- 0.78 & 0.768\\
    & 0.8 & 0.67 -- 0.70 & 0.678\\

  \end{tabular}
  \end{center}
  \caption{Comparison of global parameters between SRT and self-similar TRT simulations \citep{zhou2016asymptotic, zhou2019time}}
\label{tab:TRTvalidate}
\end{table}

\section{Results}
\label{sec:results}

The ultimate objective of this work is to develop scalings in terms of the dimensional inputs introduced in \S~\ref{sec:inputs}, which are $\rho_H$, $\rho_L$, $\nu$, $D$, $g$, and $h_1$. In \S~\ref{sec:scaling_budget}, scalings are initially developed with the introduction of two intermediate global variables: the SRT timescale $\tau_0$ and the integrated mixed mass $M$. In \S~\ref{sec:scaling_global}, expressions for relating each of the intermediate global variables to the input variables are proposed.
All scaling relationships are intended to be applicable to both statistically converged SRT flows and late-time TRT flows. To reflect this equivalence, time-averaging in SRT flows is assumed and the overline notation is omitted throughout the remainder of this paper.

\subsection{Scaling of SRT budget terms}
\label{sec:scaling_budget}

The ensemble (or Reynolds) average of a general flow variable $\phi$ is denoted by $\langle \phi \rangle$ and evaluated in SRT flow as averages over the homogeneous directions $x_1$, $x_3$, and time $s$. The density-weighted (or Favre) average of $\phi$ is denoted by $\tilde{\phi} = \langle \rho \phi \rangle/ \langle \rho \rangle$; the Reynolds and Favre fluctuations are defined as $\phi' = \phi - \langle \phi \rangle$ and $\phi'' = \phi - \tilde{\phi}$, respectively. 

Scalings (notated with the $\sim$ symbol) are proposed by examining the dominant balances in each of the ensemble-averaged budget equations. Formally, the $\sim$ symbol or the phrase ``scales with'' is interpreted as  
\begin{equation}
  \langle \phi \rangle (y) \sim f(h_1, \rho_H, \rho_L, g, \ldots) 
  \quad
  \Leftrightarrow
  \quad 
  \int \langle \phi \rangle \,{\rm d}\left(\frac{y}{h_1}\right)\approx K_{\langle \phi \rangle} \,  f(h_1, \rho_H, \rho_L, g, \ldots),
  \label{eq:scaling_def}
\end{equation}
where $f(h_1, \rho_H, \rho_L, \ldots)$ is a dimensional parameter computed from global parameters and $K_{\langle \phi \rangle}$ is a dimensionless constant specific to the field of interest. The spatial coordinate $y$ is normalized based on the assumptions of self-similarity (i.e. all lengthscales scale proportionally to each other) and $h_1$ is the only input length scale.

To quantify the collapse of profiles, their spread is quantified using a spatially integrated root-mean-square deviation $E_{\rm rms}$. For a quantity $\Phi(y/h_1)$, this is defined as
\begin{equation}
  E_{\rm rms} = \sqrt{\frac{1}{n}\sum_{i=1}^n \int \left[\Phi_i(y/h_1) - \bar{\Phi}(y/h_1) \right]^2 {\rm d}\left(\frac{y}{h_1}\right)},
\end{equation}
where $n$ is the number of cases considered and $\bar{\Phi}(y/h_1)$ is the mean profile. A normalized spread $\mathcal{E}_{\rm rms}$ is defined as one of two possibilities. If $\Phi(\pm\infty)=0$, $E_{\rm rms}$ is normalized by the area under the curve of the mean profile, $\mathcal{E}_{\rm rms} = E_{\rm rms}/\left|\int\bar{\Phi}  {\rm d}(y/h_1)\right|$. If $\Phi\ne0$ in the fluid reservoirs (e.g. the mole fraction has a value of 1 in the heavy reservoir), $E_{\rm rms}$ is normalized by the range of $\Phi$ as $\mathcal{E}_{\rm rms} = E_{\rm rms}/(\Phi_{\rm max}-\Phi_{\rm min})$.

\subsubsection{Continuity}
\label{sec:results_cont}

\begin{figure}
  \begin{minipage}{0.498\textwidth}
    (\emph{a})\\
    \includegraphics[width=\textwidth]{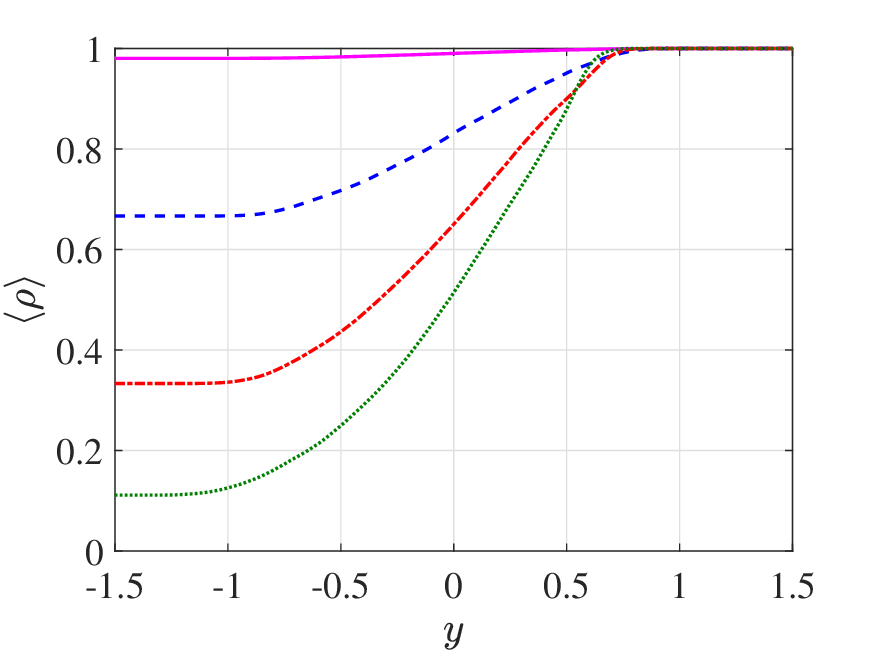}
  \end{minipage}
  \hfill
  \begin{minipage}{0.498\textwidth}
    (\emph{b})\\
    \includegraphics[width=\textwidth]{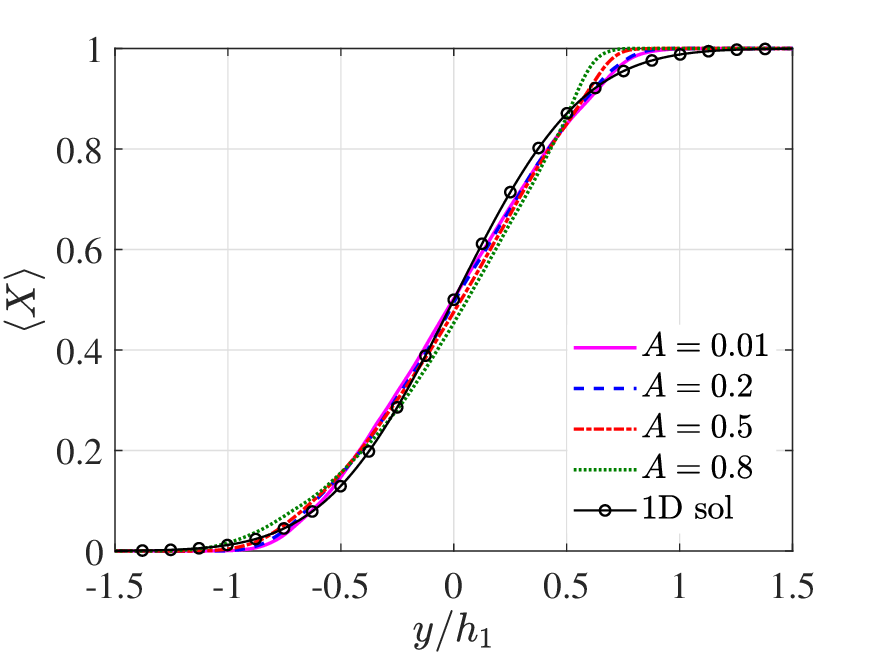}
  \end{minipage}
  \begin{minipage}{0.498\textwidth}
    (\emph{c})\\
    \includegraphics[width=\textwidth]{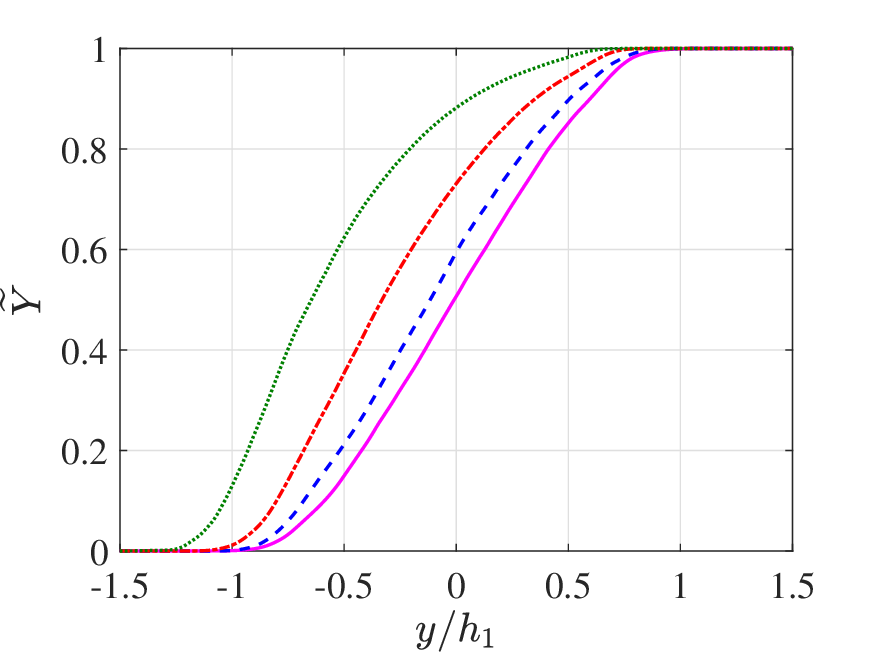}
  \end{minipage}
  \hfill
  \begin{minipage}{0.498\textwidth}
    (\emph{d})\\
    \includegraphics[width=\textwidth]{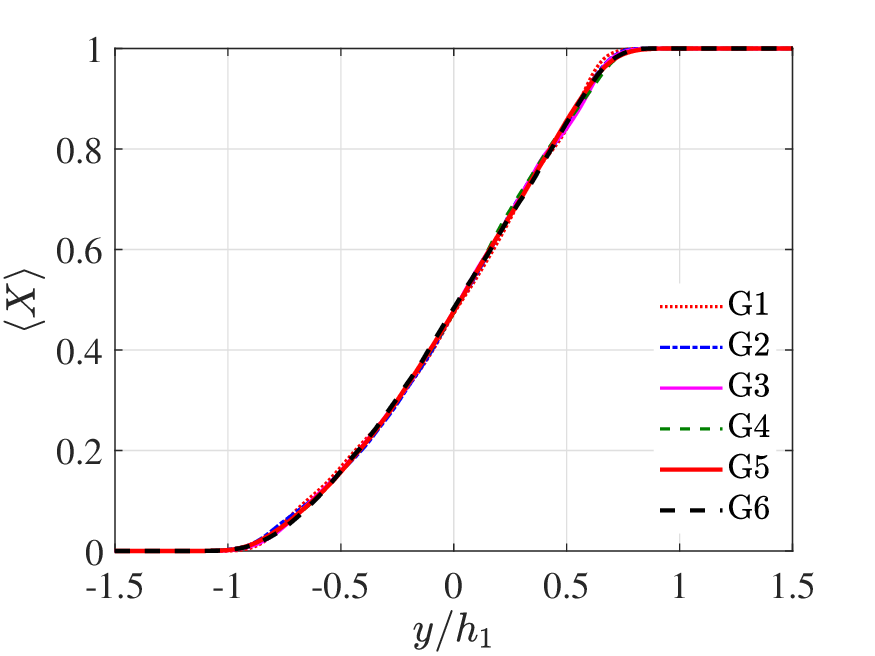}
  \end{minipage}
  \caption{Effect of the Atwood number on the (\emph{a}) mean density, (\emph{b}) mean mole fraction, and (\emph{c}) Favre-averaged mass fraction. Legend: $A=$ 0.01 (magenta solid), 0.2 (blue dashed), 0.5 (red dash-dotted), 0.8 (green dotted), analytical 1D solution (black circles); (\emph{d}) Effect of the Reynolds number on the mean mole fraction. Legend: $\Rey\approx$ 245 (red dotted), 496 (blue dash-dotted), 791 (magenta solid), 1060 (green dashed), 2540 (thick red solid), and 4390 (thick black dashed).
  \label{fig:density}}
\end{figure}

First, the scaling of the mean density $\langle \rho \rangle$ is considered. In all simulation cases, $\rho_H = 1$ is chosen to be constant and $\rho_L$ varies according to the Atwood number. As a result, cases A0--A8 have different densities in the light fluid reservoir, evident from the mean density profiles shown in figure~\ref{fig:density}(a). 

A straightforward normalization of the mean density is to stretch its profiles linearly along the vertical axis to take values between 0 and 1; this normalization results precisely in the definition of the heavy fluid mole fraction, $\langle X \rangle = (\langle \rho \rangle - \rho_L)/\Delta\rho$. The mean mole fraction for cases A0--A8 are shown in figure~\ref{fig:density}(\emph{b}) and they are approximately independent of Atwood number with $\mathcal{E}_{\rm rms}\approx 0.02$. This apparent collapse of the profiles was an important justification for our selection of $h_1$ as the lengthscale for defining the mixing layer aspect ratio (see \S~\ref{sec:input_h1}). Some Atwood number effects remain noticeable, particularly at the edges of the mixing layer. As the Atwood number increases, gradients are larger on the heavy side of the mixing layer and smaller on the light side. This results in $h_s/h_b$ ratios that grow with the Atwood number, which are consistent with observations from other TRT studies \citep{banerjee2010detailed,burton2011study,youngs2013density,akula2016dynamics,zhou2019time}.

Another possible normalization of the mean density is the Favre-average of the mass fraction, which is shown in figure~\ref{fig:density}(\emph{c}). The Favre-averaged mass fraction exhibits significant variations with the Atwood number, which is expected because $\tilde{Y}$ is algebraically related to $\langle X \rangle$ via the equation of state (\ref{eq:EOS}) such that 
\begin{equation}
  \tilde{Y} = \frac{\rho_H \langle X \rangle}{\langle \rho \rangle} = \frac{R\langle X \rangle}{1 + (R-1)\langle X \rangle}.
  \label{eq:Yfav}
\end{equation}
Based on (\ref{eq:Yfav}), if $\langle X \rangle$ is independent of the density ratio $R$, then $\tilde{Y}(\langle X \rangle,R)$ is not. By extension, any function of $\tilde{Y}$ is not, in general, independent of $R$. The observed Atwood-invariance of the mean mole fraction (and not the mass fraction) is also consistent with the one-dimensional (1D) variable-density temporal diffusion problem, whose solution is derived in Appendix~\ref{app:1dvd} and included in figure~\ref{fig:density}(\emph{b}). 

The effect of the Reynolds number on the mean mole fraction is illustrated in figure~\ref{fig:density}(\emph{d}) using cases G1--G6. The profiles appear nearly identical ($\mathcal{E}_{\rm rms}\approx 0.007$), indicating that the Reynolds number has a negligible influence on the mean density field. 

Based on (\ref{eq:scaling_def}), the mean density gradient scales as
\begin{equation}
  \dd{\langle \rho \rangle}{y} \sim \frac{\Delta \rho}{h_1}.
  \label{eq:drhodx}
\end{equation}
The scaling of the mean mass flux can be determined using the observed scaling of the mean density gradient and the continuity equation (\ref{eq:mass_SRT}). The ensemble average of (\ref{eq:mass_SRT}) is
\begin{equation}
  0 = - \dd{\langle \rho v \rangle}{y} + \frac{y}{\tau_0}\dd{\langle\rho\rangle}{y},
  \label{eq:contMean}
\end{equation}
where partial derivatives with respect to $x_1$, $x_3$, and $s$ are eliminated due to statistical homogeneity and stationarity. Combining (\ref{eq:drhodx}) and (\ref{eq:contMean}), the expected scaling for the mean mass flux is
\begin{equation}
  \langle \rho v \rangle \sim \frac{\Delta \rho h_1}{\tau_0}.
  \label{eq:rhouscaling}
\end{equation}
In (\ref{eq:rhouscaling}), and throughout \S~\ref{sec:scaling_budget}, the SRT parameter $\tau_0$ is left as a characteristic timescale for scaling of the budget equations. As described by (\ref{eq:tau0_TRT}), $\tau_0$ is equal to $[h_1/\dot{h}_1]_{t=t_0}$ in a TRT flow, and all normalizations involving $\tau_0$ can be applied analogously to a self-similar TRT flow with this substitution. Nonetheless, the final objective remains to express all scalings in terms of $h_1$ (without $\tau_0$ or $\dot{h}_1$), and the scaling of $\tau_0$ (or $\dot{h}_1$ for the case of TRT flow) with respect to $h_1$ will be addressed in \S~\ref{sec:scaling_global}.

\begin{figure}
  \begin{minipage}{0.498\textwidth}
    (\emph{a})\\
    \includegraphics[width=\textwidth]{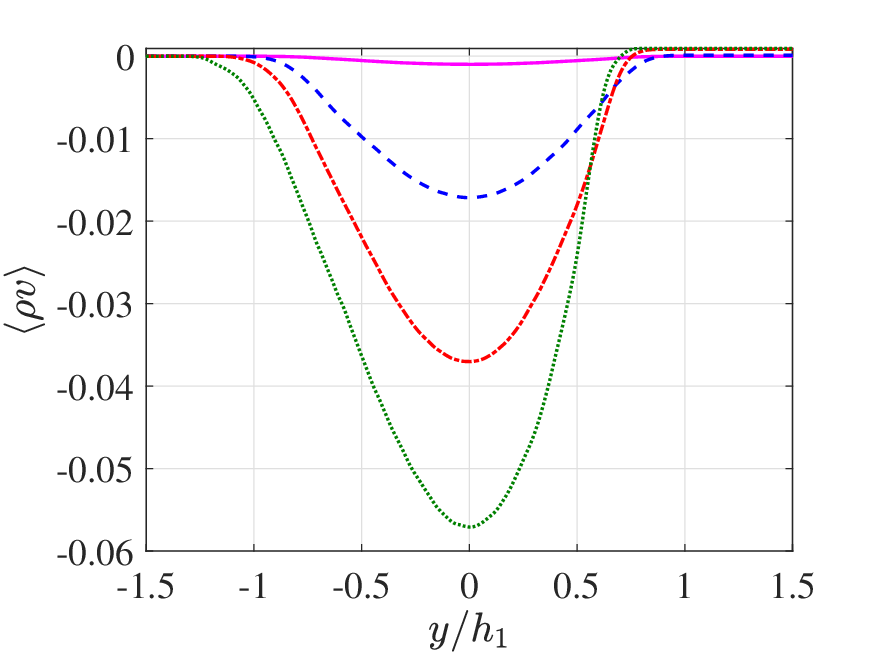}
  \end{minipage}
  \hfill
  \begin{minipage}{0.498\textwidth}
    (\emph{b})\\
    \includegraphics[width=\textwidth]{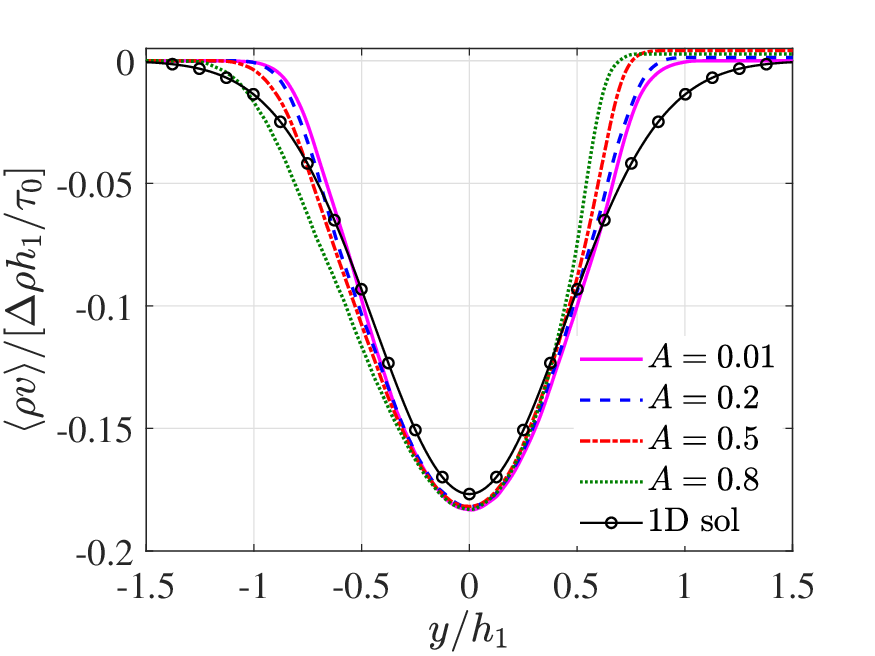}
  \end{minipage}
\caption{Effect of the Atwood number on the (\emph{a}) mass flux and (\emph{b}) normalized mass flux. Legend: $A=$ 0.01 (magenta solid), 0.2 (blue dashed), 0.5 (red dash-dotted), 0.8 (green dotted), analytical 1D solution (black circles)}
\label{fig:velocity}
\end{figure}

The variation of the mean mass flux $\langle \rho v\rangle$ with Atwood number is shown in figures~\ref{fig:velocity}(\emph{a}) and \ref{fig:velocity}(\emph{b}), corresponding to profiles before and after normalization, respectively. The location of the peak magnitude ($y\approx 0$) seems to be independent of Atwood number, and the normalized mean mass flux has a universal minimum value of approximately $-0.18$, much similar to the value of $-1/\sqrt{32}\approx-0.177$ for the analytical 1D solution, which is included in figure~\ref{fig:velocity}(\emph{b}) for reference. While the collapse is generally successful with a normalized spread of $\mathcal{E}_{\rm rms}\approx0.07$, differences due to the Atwood number are noticeable in the normalized profiles near the edges of the mixing layer, with a slight shift of the tails towards the side of the lighter fluid. Because the mean mass flux is determined entirely by the mean density gradient according to (\ref{eq:contMean}), the excellent collapse in the core of the mixing layer and the minor deviations towards the edges are consistent with observations from figure~\ref{fig:density}(\emph{b}). 

\begin{figure}
  \begin{minipage}{0.498\textwidth}
    (\emph{a})\\
    \includegraphics[width=\textwidth]{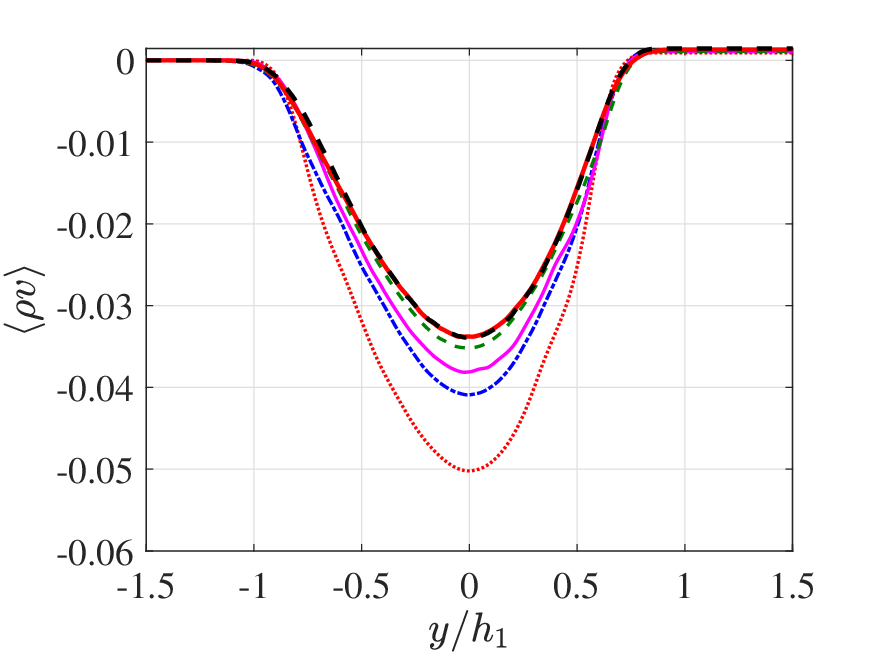}
  \end{minipage}
  \hfill
  \begin{minipage}{0.498\textwidth}
    (\emph{b})\\
    \includegraphics[width=\textwidth]{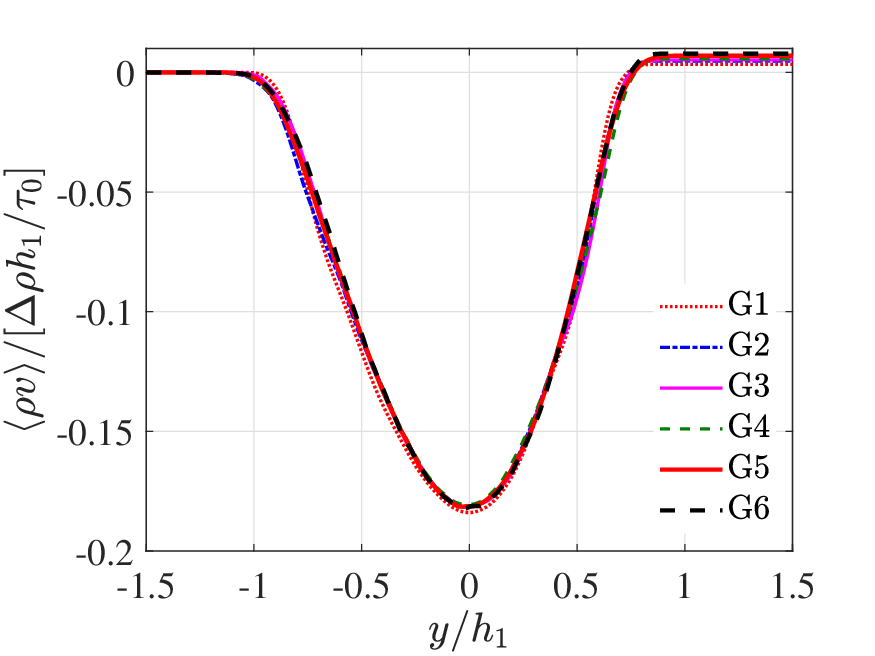}
  \end{minipage}
\caption{Effect of the Reynolds number on the (\emph{a}) mass flux and (\emph{b}) normalized mass flux. Legend: $\Rey\approx$ 245 (red dotted), 496 (blue dash-dotted), 791 (magenta solid), 1060 (green dashed), 2540 (thick red solid), and 4390 (thick black dashed).}
\label{fig:velocity_Re}
\end{figure}

The effect of the Reynolds number on the mean mass flux is shown in figure~\ref{fig:velocity_Re}. Similar to previous observations about the mean mole fraction, the normalized mean mass flux appears to be insensitive to changes in the Reynolds number. Profiles are nearly identical ($\mathcal{E}_{\rm rms} \approx 0.02$) and the same universal minimum value of $-0.18$ is observed.

\subsubsection{Turbulent kinetic energy}

Starting with (\ref{eq:mom_SRT}), the pressure gradient is decomposed into a hydrostatic background component $-\langle\rho\rangle g \delta_{2i}$ and a fluctuating component $\partial p^*/\partial x_i$ such that $\partial p/\partial x_i = \partial p^*/\partial x_i - \langle \rho \rangle g \delta_{2i}$. This results in a modified momentum equation
\begin{equation}
  \pd{\rho u_i}{s} + \pd{\rho u_i u_j}{x_j} = - \pd{p^*}{x_i} + \pd{\tau_{ij}}{x_j} -\rho'g\delta_{2i} + \frac{y}{\tau_0}\pd{\rho u_i }{y} - \frac{\rho u_i}{2\tau_0},
  \label{eq:mom2}
\end{equation}
in which all terms are only active within the mixing layer and zero in the fluid reservoirs. Using this pressure decomposition, the ensemble-averaged equation for the turbulent kinetic energy 
$k$ is
\begin{figure}
  \begin{center}
  \begin{minipage}{0.6\textwidth}
    \includegraphics[width=\textwidth]{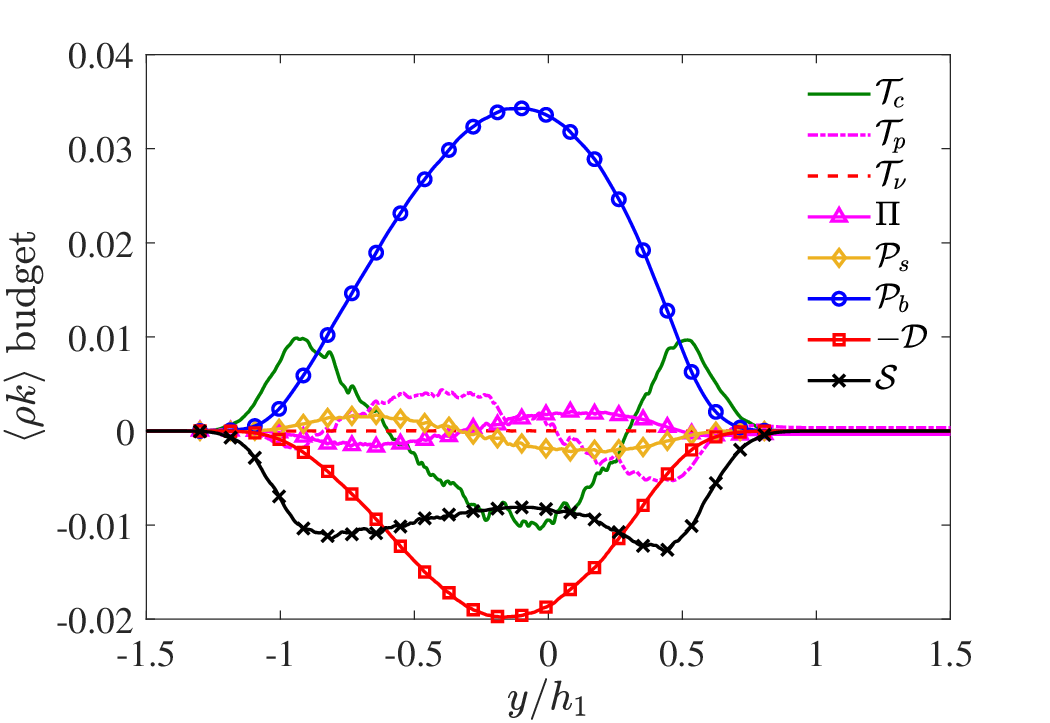}
  \end{minipage}
  \end{center}
  \caption{Turbulent kinetic energy budget for case G5. Legend: convective transport (green solid), pressure transport (magenta dash-dotted), viscous transport (red dashed), pressure dilatation (magenta triangles), production by mean shear (yellow diamonds), production by buoyancy (blue circles), viscous dissipation (red squares), and SRT source terms (black crosses).
  \label{fig:budgets}}
\end{figure}

\begin{multline}
  0 = \underbrace{- \dd{\langle \rho v k \rangle}{y}}_{\mathcal{T}_c}
  \underbrace{- \dd{\left\langle p^*v''\right\rangle}{y}}_{\mathcal{T}_p}
  \underbrace{+ \dd{\left\langle \tau_{i2} u''_i\right\rangle}{y}}_{\mathcal{T}_\nu} 
  \underbrace{+ \left\langle p^* \pd{u''_i}{x_i} \right\rangle}_{\Pi}
  \underbrace{- \left\langle\rho v''^2 \right\rangle\dd{\tilde{v}}{y}}_{\mathcal{P}_s}
  \\
  \underbrace{- \langle\rho'v\rangle g}_{\mathcal{P}_b}
  \underbrace{- \langle \rho \epsilon \rangle}_{-\mathcal{D}}
  \underbrace{+\frac{y}{\tau_0}\dd{\langle \rho k\rangle}{y} - \frac{\langle \rho k\rangle}{\tau_0}}_{\mathcal{S}},
  \label{eq:tkebudget}
\end{multline}
where $ \epsilon  = \tau_{ij}(\partial u''_i/\partial x_j)/\rho $ is the viscous dissipation rate. The budget terms correspond to convective transport $\mathcal{T}_c$, pressure transport $\mathcal{T}_p$, viscous transport $\mathcal{T}_\nu$, pressure dilatation $\Pi$, TKE production by mean shear $\mathcal{P}_s$, TKE production by buoyancy $\mathcal{P}_b$, viscous dissipation $\mathcal{D}$, and the SRT source term contribution $\mathcal{S}$, respectively. The budget terms for case G5 are shown in figure~\ref{fig:budgets}(\emph{b}). Excluding the transport terms, the magnitudes of pressure dilatation ($\Pi$) and production by shear ($\mathcal{P}_s$) are much smaller than the production by buoyancy ($\mathcal{P}_b$), dissipation ($\mathcal{D}$), and SRT source terms ($\mathcal{S}$). Integrating (\ref{eq:tkebudget}) vertically and considering only the dominant terms, 
\begin{equation}
  \frac{2\int \langle \rho k \rangle \,{\rm d}y}{\tau_0} 
  \approx 
  \int -\langle \rho' v \rangle g \,{\rm d}y
  - \int \langle \rho \epsilon \rangle \,{\rm d}y,
  \label{eq:tkeInt}
\end{equation}
which describes the integrated TKE balance as one between TKE growth, production by buoyancy, and viscous dissipation. 

\begin{figure}
  \begin{minipage}{0.498\textwidth}
    (\emph{a})\\
    \includegraphics[width=\textwidth]{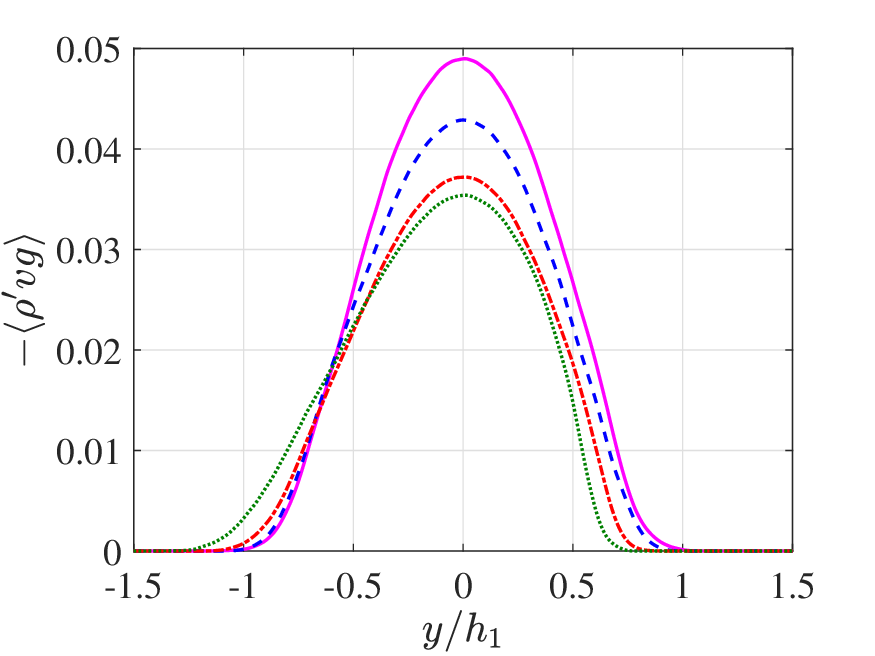}
  \end{minipage} 
  \hfill
  \begin{minipage}{0.498\textwidth}
    (\emph{b})\\
    \includegraphics[width=\textwidth]{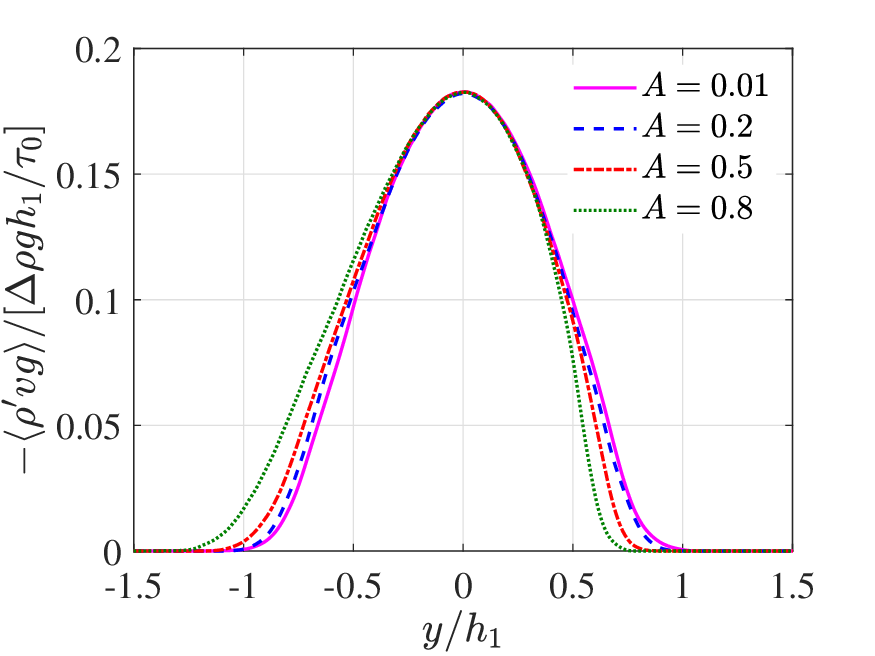}
  \end{minipage} 
  \begin{minipage}{0.498\textwidth}
    (\emph{c})\\
    \includegraphics[width=\textwidth]{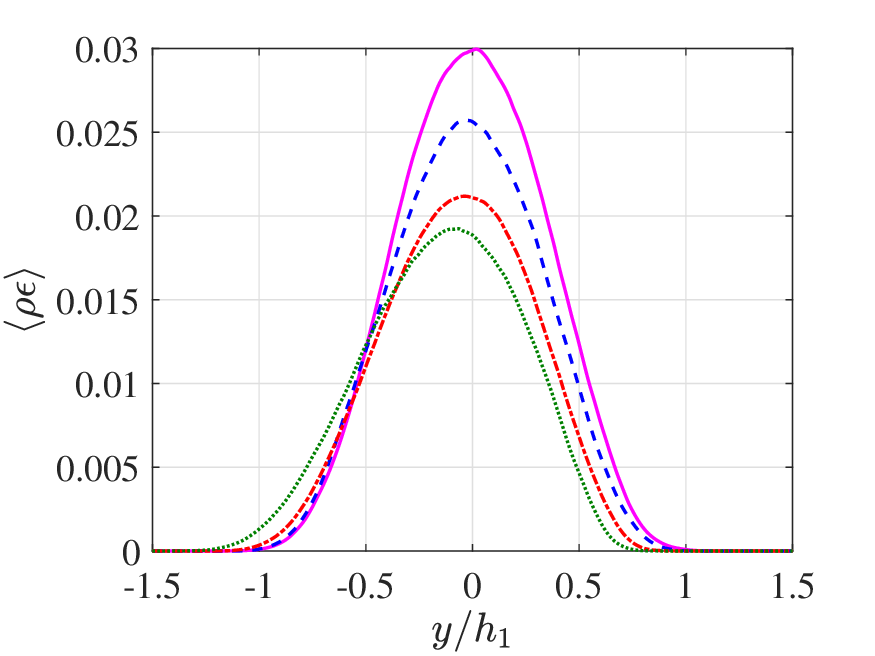}
  \end{minipage}
  \hfill
  \begin{minipage}{0.498\textwidth}
    (\emph{d})\\
    \includegraphics[width=\textwidth]{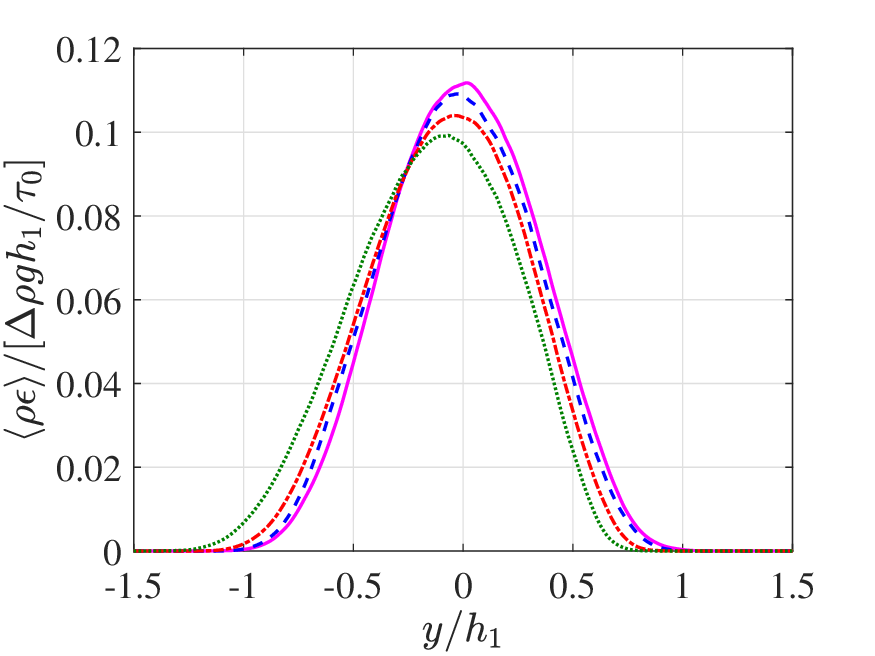}
  \end{minipage} 
  \begin{minipage}{0.498\textwidth}
    (\emph{e})\\
    \includegraphics[width=\textwidth]{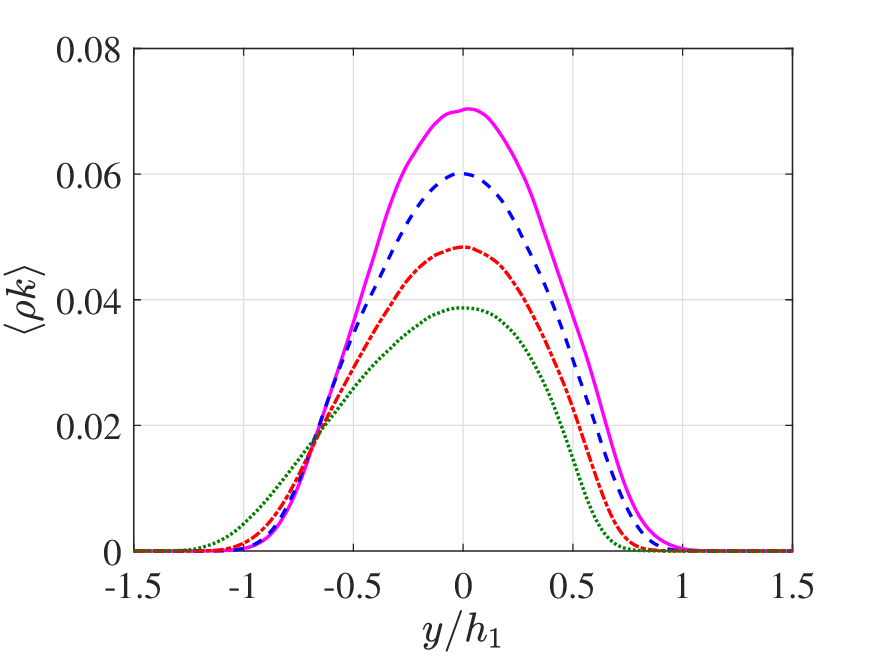}
  \end{minipage}
  \hfill
  \begin{minipage}{0.498\textwidth}
    (\emph{f})\\
    \includegraphics[width=\textwidth]{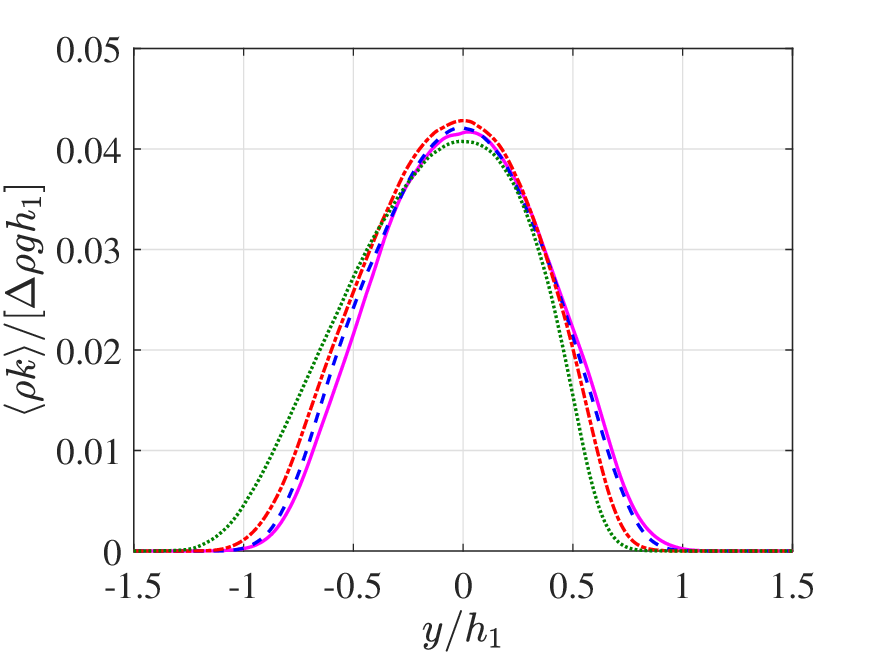}
  \end{minipage}
\caption{Effect of the Atwood number on the (\emph{a}) TKE production by buoyancy, (\emph{b}) normalized production, (\emph{c}) viscous dissipation, (\emph{d}) normalized viscous dissipation, (\emph{e}) TKE, and (\emph{f}) normalized TKE. Legend: $A=$ 0.01 (magenta solid), 0.2 (blue dashed), 0.5 (red dash-dotted), 0.8 (green dotted) }
\label{fig:tke}
\end{figure}

\begin{figure}
  \begin{minipage}{0.498\textwidth}
    (\emph{a})\\
    \includegraphics[width=\textwidth]{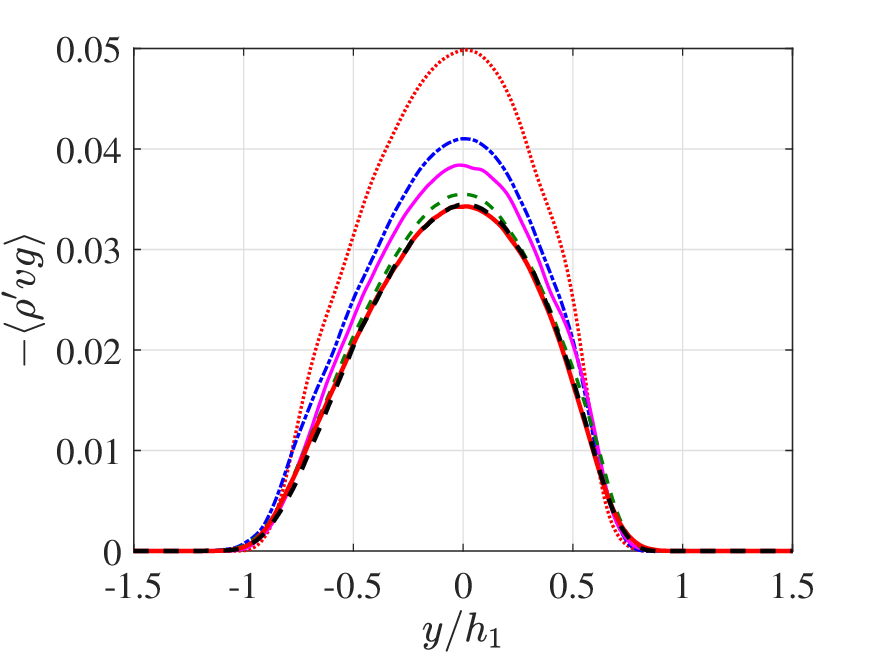}
  \end{minipage}
  \hfill
  \begin{minipage}{0.498\textwidth}
    (\emph{b})\\
    \includegraphics[width=\textwidth]{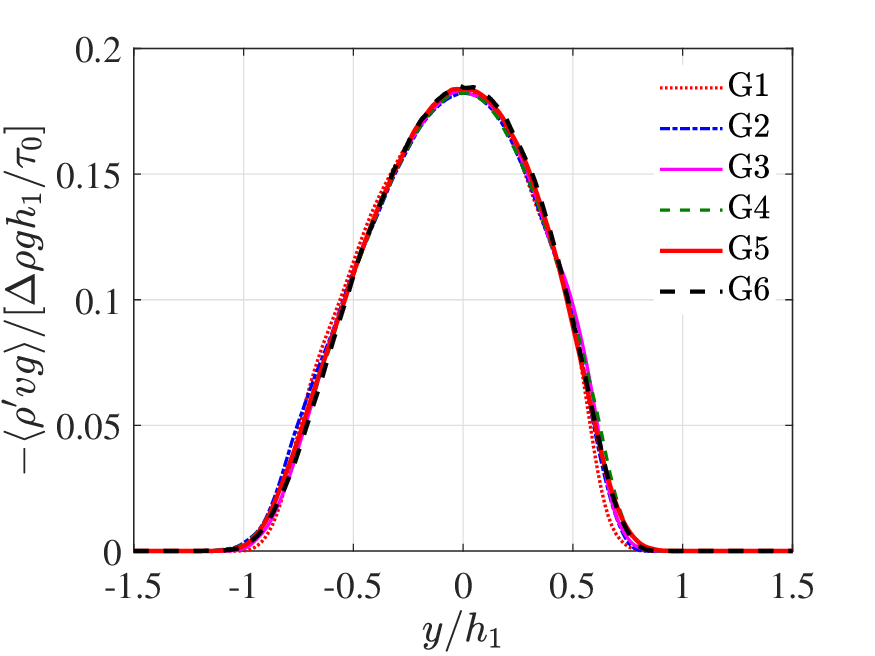}
  \end{minipage} 
  \begin{minipage}{0.498\textwidth}
    (\emph{c})\\
    \includegraphics[width=\textwidth]{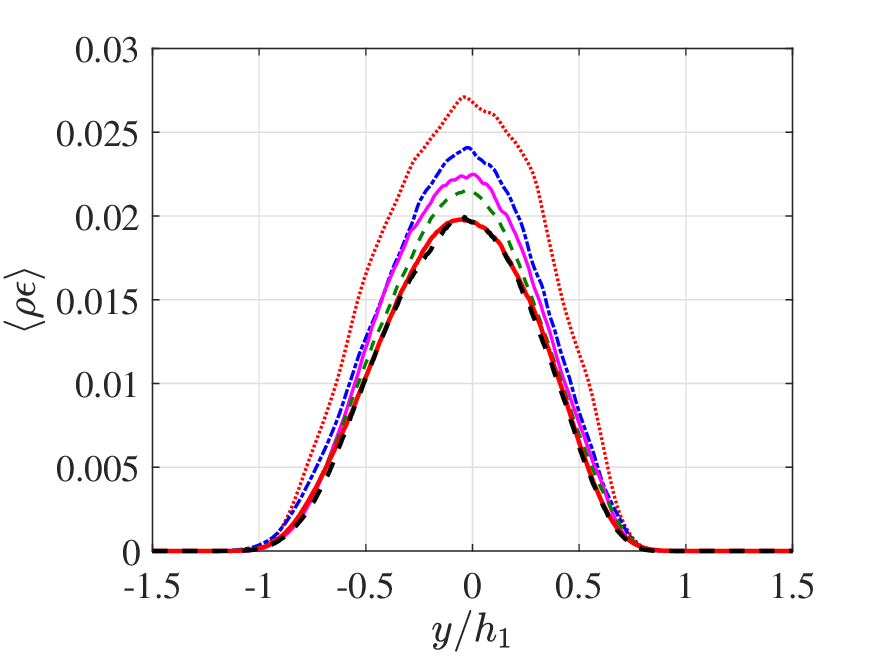}
  \end{minipage}
  \hfill
  \begin{minipage}{0.498\textwidth}
    (\emph{d})\\
    \includegraphics[width=\textwidth]{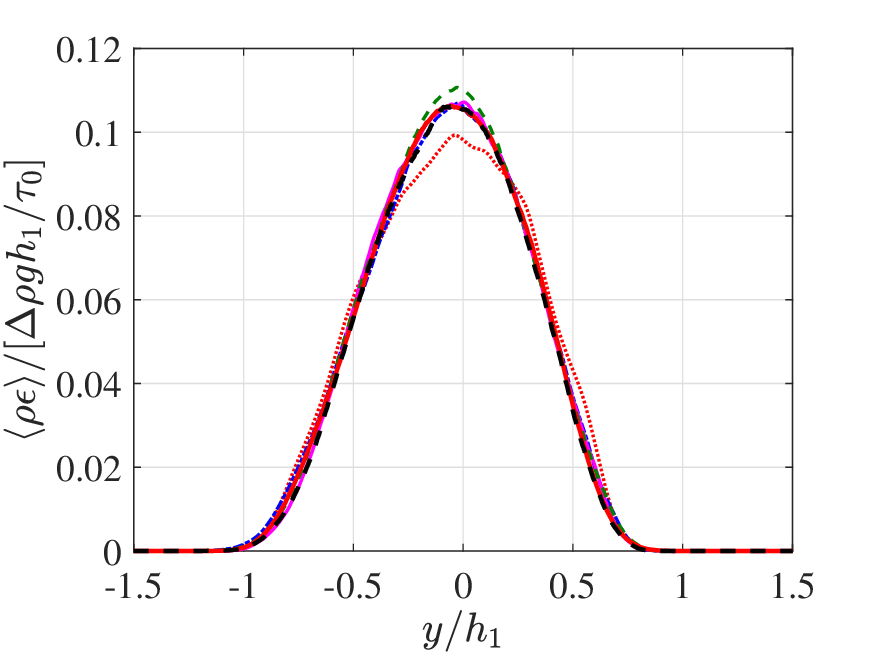}
  \end{minipage} 
  \begin{minipage}{0.498\textwidth}
    (\emph{e})\\
    \includegraphics[width=\textwidth]{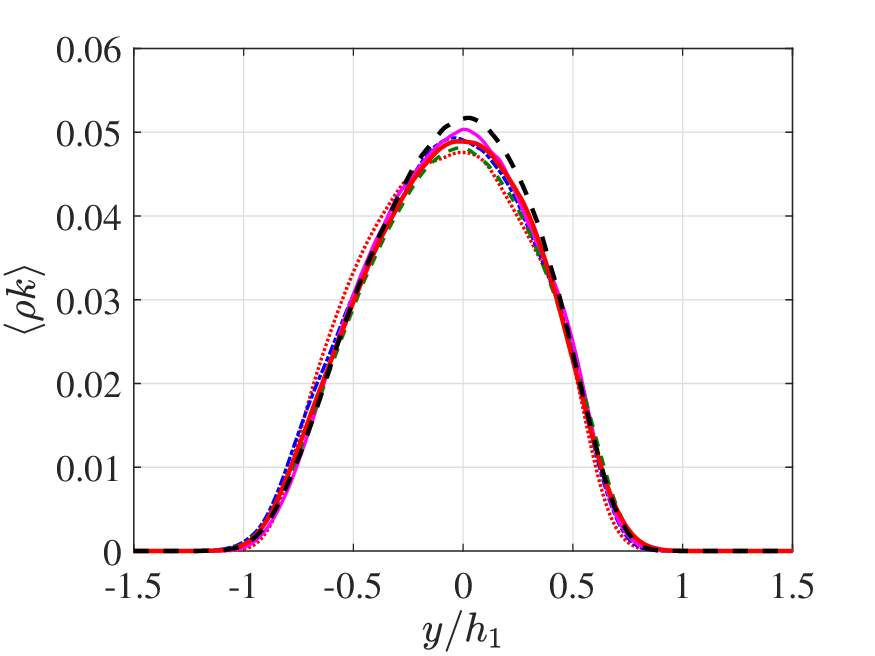}
  \end{minipage}
  \hfill
  \begin{minipage}{0.498\textwidth}
    (\emph{f})\\
    \includegraphics[width=\textwidth]{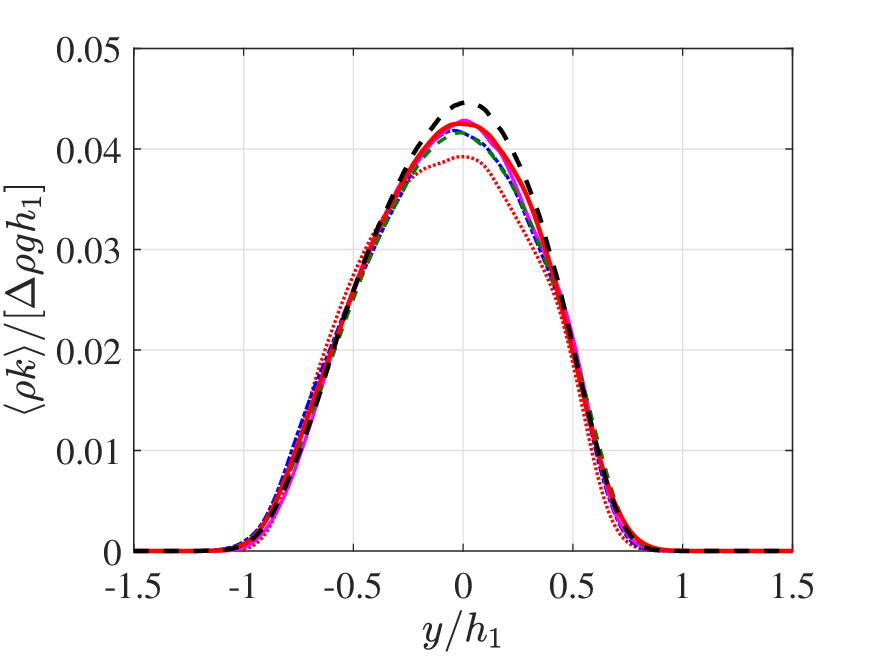}
  \end{minipage}
\caption{Effect of the Reynolds number on the (\emph{a}) TKE production by buoyancy, (\emph{b}) normalized production, (\emph{c}) viscous dissipation, (\emph{d}) normalized viscous dissipation, (\emph{e}) TKE, and (\emph{f}) normalized TKE. Legend: $\Rey\approx$ 245 (red dotted), 496 (blue dash-dotted), 791 (magenta solid), 1060 (green dashed), 2540 (thick red solid), and 4390 (thick black dashed).}
\label{fig:tke_Re}
\end{figure}

A scaling for the production term can be derived from the scaling for $\langle \rho v\rangle$. It is observed that $\langle \rho \rangle \langle v \rangle \ll \langle \rho v\rangle$ such that $\langle \rho' v \rangle \approx \langle \rho v \rangle$. Hence, applying the result from (\ref{eq:rhouscaling}), the scaling for the production term is 
\begin{equation}
  \langle \rho' v \rangle g \sim \frac{\Delta \rho gh_1}{\tau_0},
  \label{eq:prodscaling}
\end{equation}
where $\Delta \rho g h_1$ can intuitively be interpreted as the total loss of potential energy of the flow. The effectiveness of this normalization is demonstrated in figure~\ref{fig:tke}(\emph{a},\emph{b}) for cases A0--A8 and figure~\ref{fig:tke_Re}(\emph{a},\emph{b}) for cases G1--G6. The normalized production term collapses well across both Atwood numbers ($\mathcal{E}_{\rm rms} \approx 0.06$) and Reynolds numbers ($\mathcal{E}_{\rm rms} \approx 0.02$), which justifies our neglect of the $\langle \rho \rangle \langle v\rangle$ term. 

The same scaling from buoyancy production is applied to the viscous dissipation rate. There are two possible justifications for this. Mathematically, all terms in (\ref{eq:tkeInt}) are positive, and the integrated dissipation is bounded by the integrated production. Second, production-to-dissipation ratios are observed to be approximately constant (or vary slowly) in TRT flows at high Reynolds numbers \citep{Livescu2009_JoT,schilling2010analysis}. Hence,
\begin{equation}
  \langle \rho \epsilon \rangle \sim \frac{\Delta \rho gh_1}{\tau_0}.
  \label{eq:epsscaling}
\end{equation}
The effect of the Atwood number on the viscous dissipation is shown in figure~\ref{fig:tke}(\emph{c},\emph{d}) for cases A0--A8. Relative to the unnormalized profiles in figure~\ref{fig:tke}(\emph{c}), the normalized profiles in figure~\ref{fig:tke}(\emph{d}) show an improved collapse with a reduction of the normalized spread from $\mathcal{E}_{\rm rms}\approx$ 0.16 to 0.08, although some Atwood number effects remain visible. Specifically, as Atwood number increases, the normalized profile broadens, its peak shifts to the left, and the maximum value decreases by approximately $10\%$ across the range of Atwood numbers considered. Figure \ref{fig:tke_Re}(\emph{c},\emph{d}) show the equivalent curves for cases G1--G6, in which the Reynolds number is varied. All Reynolds number effects present in the unnormalized profiles seem to be accounted for by the proposed normalization, except for minor differences in case G1. As the scalings are derived under assumptions of high Reynolds numbers, observing some discrepancies in the low-$\Rey$ cases is not surprising.

Finally, combining (\ref{eq:tkeInt})--(\ref{eq:epsscaling}), the proposed scaling for the TKE is 
\begin{equation}
  \langle \rho k \rangle \sim \Delta \rho g h_1,
  \label{eq:kscaling}
\end{equation}
and the influence of the Atwood number on the TKE is shown in figure~\ref{fig:tke}(\emph{e},\emph{f}). In figure~\ref{fig:tke}(\emph{e}), the unnormalized TKE magnitudes decrease systematically with Atwood number. After normalization, the peak values in figure~\ref{fig:tke}(\emph{f}) are similar across Atwood numbers and are consistently located at $y\approx 0$. The observed asymmetry in the tails are similar to observations made about the mean mole fraction and the mean mass flux in \S~\ref{sec:results_cont}. The effect of Reynolds number on the normalized TKE is shown in figure~\ref{fig:tke_Re}(\emph{f}), the normalized TKE shows a minor $12\%$ increase in the peak values as the Reynolds number is increased by more than an order of magnitude. This gradual increase is consistent with the slow growth of kinetic-to-potential-energy ratios commonly observed in other studies \citep{Cabot2006_Nat,goh2025statistically}---a finite-Reynolds-number non-stationary effect that has been attributed to the increasing time lag between production and dissipation \citep{Cabot2006_Nat,Livescu2009_JoT}. 

\subsubsection{Mixed mass}
\label{sec:results_mix}

The specific mixed mass $m = Y(1-Y)$ is dimensionless, suggesting that the product $\langle \rho m \rangle$ should be normalized by a density scale $\rho_m(\rho_H,\rho_L)$ that we do not define explicitly at this point. Without further assumptions, we introduce the integrated mixed mass of the flow $M = \int \langle \rho m \rangle \,{\rm d}y$ as an additional scaling parameter so that 
\begin{equation}
  \langle \rho m \rangle \sim \frac{M}{h_1} \sim \rho_m.
  \label{eq:mmscaling}
\end{equation}
The normalization scale $M/h_1$ can be interpreted as a vertically-averaged mixed mass, or a turbulent mixing density $\rho_m$. Results will be presented first using $M/h_1$ scaling in this section; then, an \emph{a posteriori} definition for $\rho_m$ will be proposed in \S~\ref{sec:rhom}.

\begin{figure}
  \begin{minipage}{0.498\textwidth}
    (\emph{a})\\
    \includegraphics[width=\textwidth]{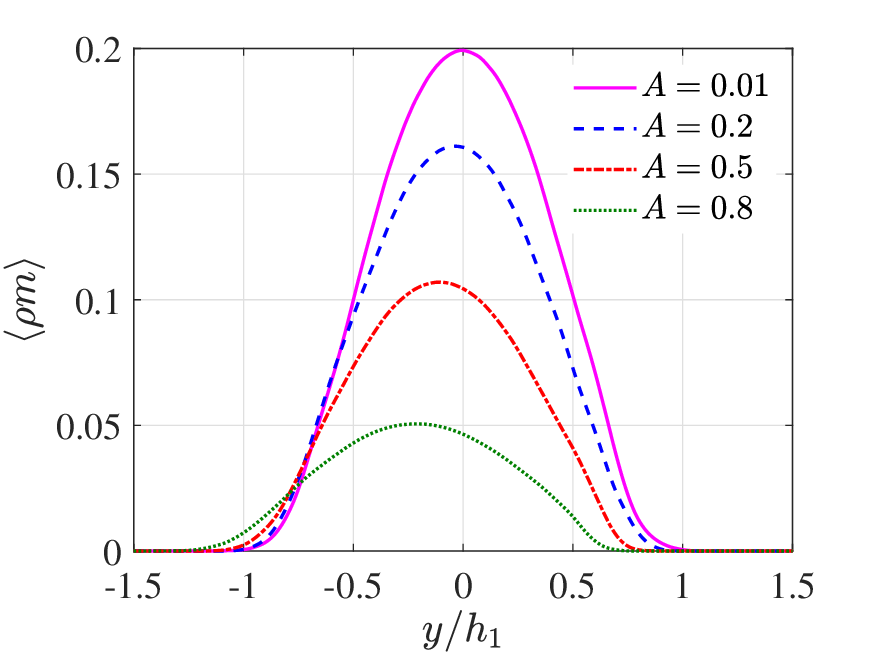}
  \end{minipage}
  \hfill
  \begin{minipage}{0.498\textwidth}
    (\emph{b})\\
    \includegraphics[width=\textwidth]{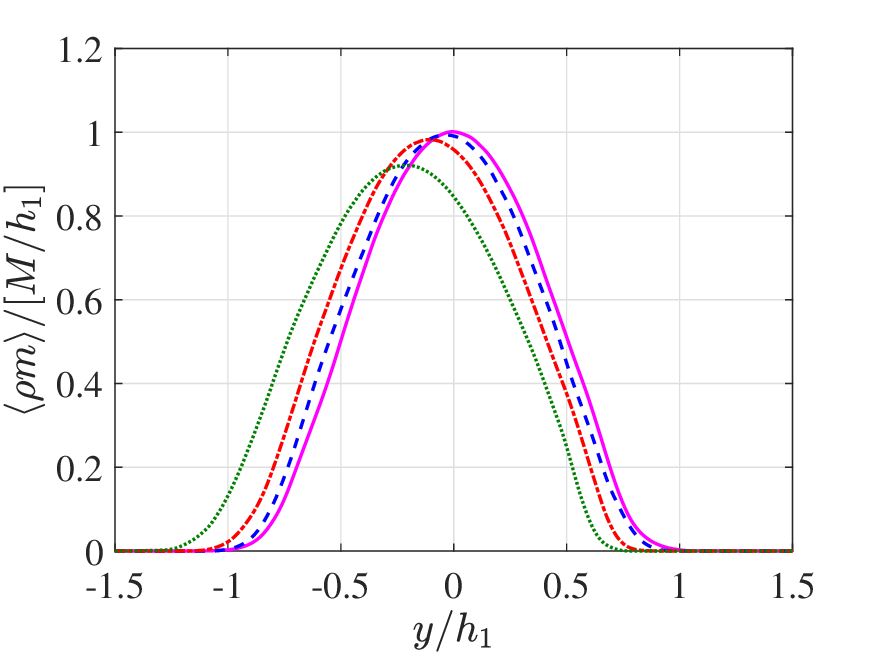}
  \end{minipage}
  \begin{minipage}{0.498\textwidth}
    (\emph{c})\\
    \includegraphics[width=\textwidth]{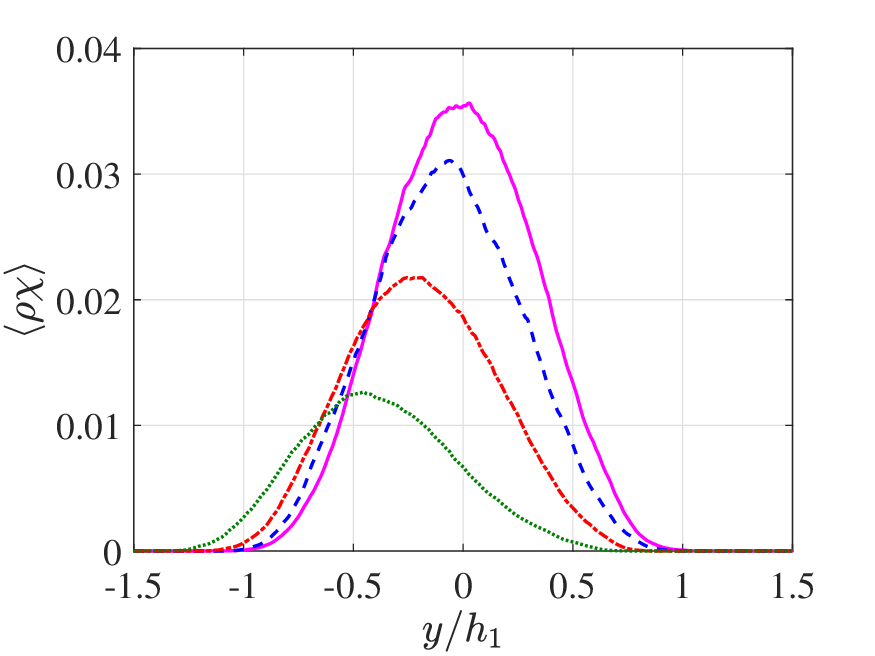}
  \end{minipage}
  \hfill
  \begin{minipage}{0.498\textwidth}
    (\emph{d})\\
    \includegraphics[width=\textwidth]{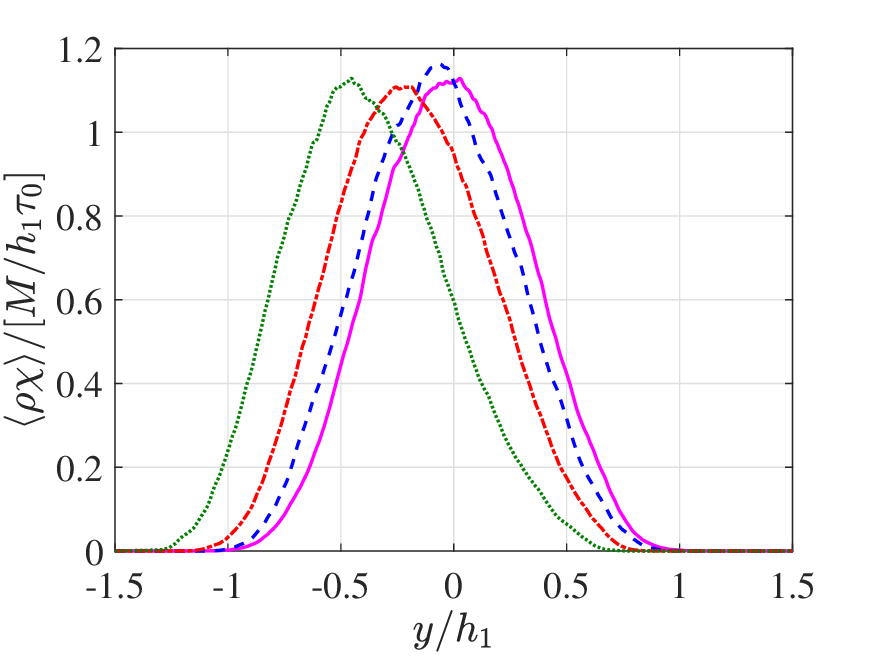}
  \end{minipage}
  \caption{Effect of Atwood number on (\emph{a}) mixed mass, (\emph{b}) normalized mixed mass, (\emph{c}) scalar dissipation, and (\emph{d}) normalized scalar dissipation, based on $M/h_1$ scaling. Legend: $A=$ 0.01 (magenta solid), 0.2 (blue dashed), 0.5 (red dash-dotted), 0.8 (green dotted)}
\label{fig:mm}
\end{figure}

The effect of the Atwood number on the mixed mass is shown in figure~\ref{fig:mm}(\emph{a},\emph{b}), for the raw and normalized profiles, respectively. Since the mixed mass is normalized by its integral $M$, the areas under the curves in figure~\ref{fig:mm}(\emph{b}) are identical by definition. After normalization, some Atwood number effects remain evident, which is reflected in a normalized spread of $\mathcal{E}_{\rm rms}\approx 0.12$. Specifically, an increase in the Atwood number reduces the peak value of the mixed mass, shifts the peak to the left, and broadens the profiles. These effects are similar to the analytical 1D solution discussed in Appendix \ref{app:1dvd}, where the equivalent 1D mixed mass profile is shown in figure~\ref{fig:sol1D}(\emph{d}). In \S~\ref{sec:results_cont}, we observed that the mean density field (represented by $\langle \rho \rangle$, $\langle X \rangle$, or $\tilde{Y}$) behaves in a similar manner to the 1D solution. By extension, the mixed mass of the mean field, $\langle \rho \rangle\tilde{Y}(1-\tilde{Y})$, would also resemble the mixed mass of the 1D solution. The observations in figure~\ref{fig:mm}(\emph{b}) not only contain this expected result but further indicate that such quasi-1D variable density effects on the mean field are also inherited by the second order statistic, $\langle \rho m \rangle = \langle \rho Y(1-Y)\rangle$.

\begin{figure}
  \begin{minipage}{0.498\textwidth}
    (\emph{a})\\
    \includegraphics[width=\textwidth]{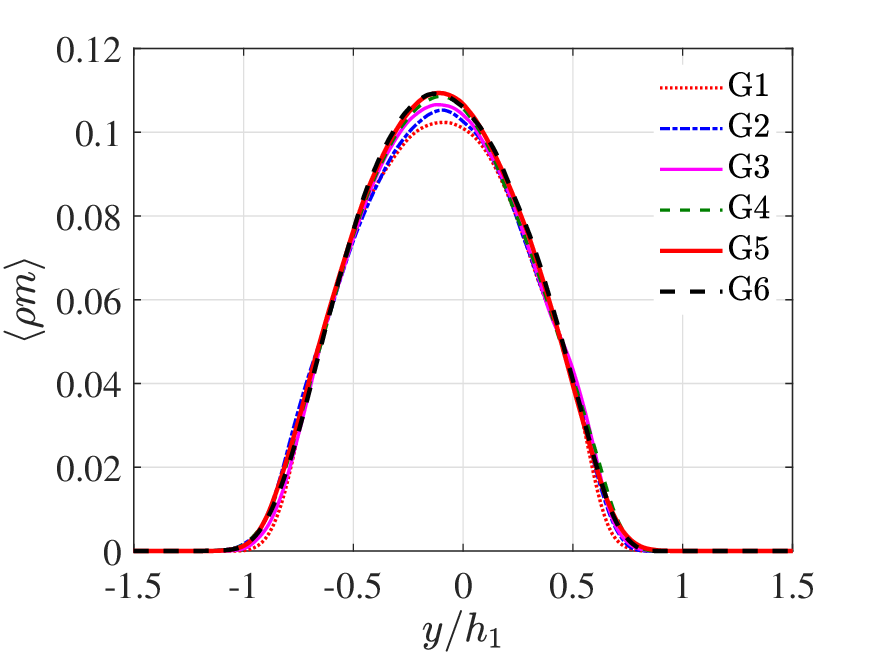}
  \end{minipage}
  \hfill
  \begin{minipage}{0.498\textwidth}
    (\emph{b})\\
    \includegraphics[width=\textwidth]{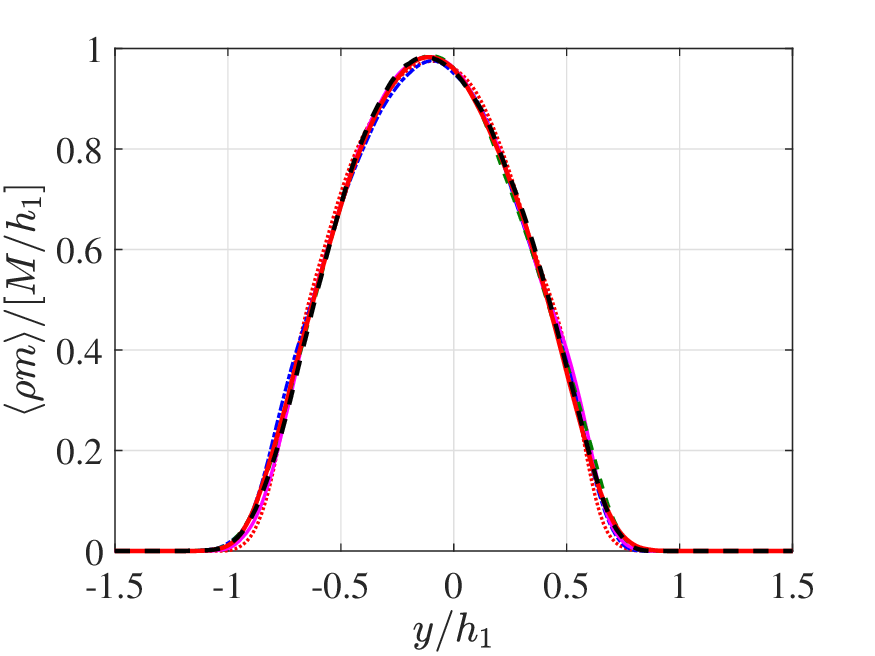}
  \end{minipage}
  \begin{minipage}{0.498\textwidth}
    (\emph{c})\\
    \includegraphics[width=\textwidth]{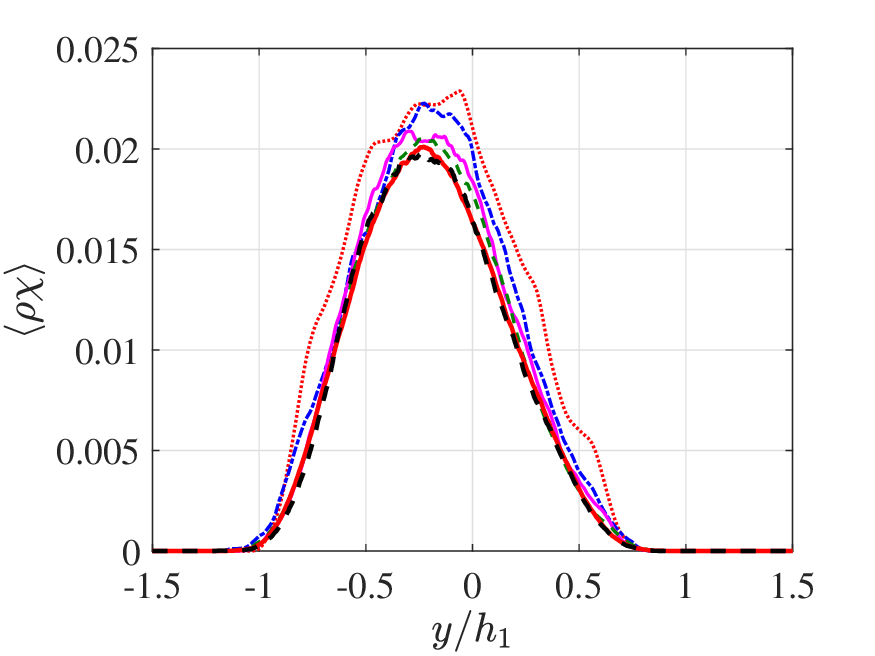}
  \end{minipage}
  \hfill
  \begin{minipage}{0.498\textwidth}
    (\emph{d})\\
    \includegraphics[width=\textwidth]{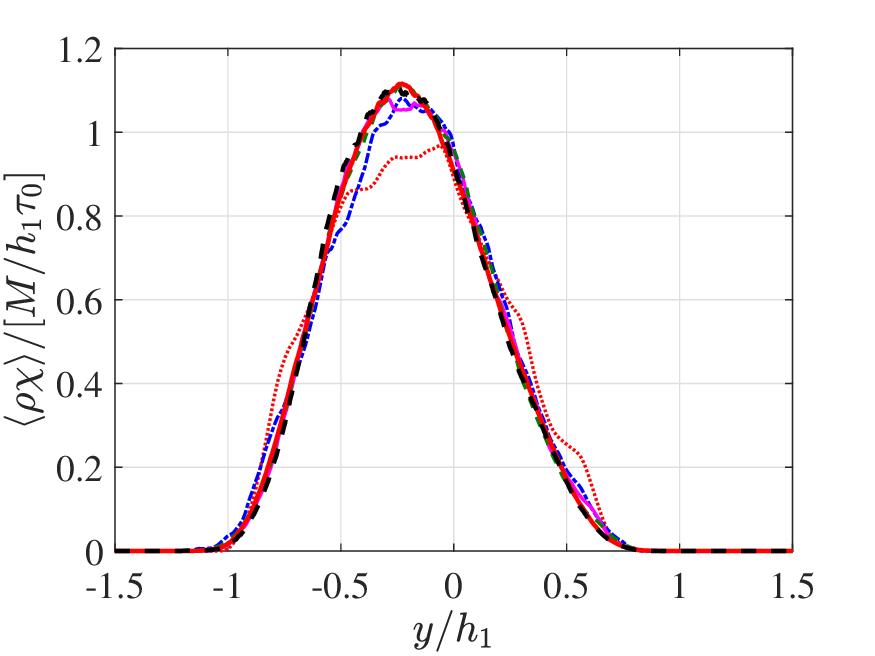}
  \end{minipage}
  \caption{Effect of Reynolds number on (\emph{a}) mixed mass (\emph{b}) scalar dissipation, (\emph{c}) normalized mixed mass, and (\emph{d}) normalized scalar dissipation. Legend: $\Rey\approx$ 245 (red dotted), 496 (blue dash-dotted), 791 (magenta solid), 1060 (green dashed), 2540 (thick red solid), and 4390 (thick black dashed).}
\label{fig:mm_Re}
\end{figure}

The effect of the Reynolds number on the mixed mass are shown in figure~\ref{fig:mm_Re}(\emph{a},\emph{b}). There are minor differences in the unnormalized mixed mass profiles in figure~\ref{fig:mm_Re}(\emph{a}), possibly due to the small differences in $h_1/\mathcal{L}$ (see table~\ref{tab:simcases0}). However, the normalized profiles in figure~\ref{fig:mm_Re}(\emph{b}) are virtually identical ($\mathcal{E}_{\rm rms} \approx 0.02$).

The governing equation for the local mixed mass is derived by multiplying the scalar transport equation (\ref{eq:scalar_SRT}) with $2Y$ and making simplifications using the continuity equation (\ref{eq:mass_SRT}). The ensemble-averaged mixed mass equation is 
\begin{equation}
  0 = 
  - \dd{ \langle \rho v m\rangle}{y}
  + \dd{}{y}\left\langle \rho D \pd{m}{y}\right\rangle
  + \langle \rho \chi \rangle
  +\frac{y}{\tau_0}\dd{ \langle \rho m \rangle}{y},
  \label{eq:mmbudget}
\end{equation}
where the terms correspond to convective transport, diffusive transport, scalar dissipation, and the SRT source term, respectively. 
Neglecting the transport terms, a scaling for the dissipation rate can be derived from (\ref{eq:mmscaling}) and (\ref{eq:mmbudget}), yielding 
\begin{equation}
  \langle \rho \chi \rangle \sim \frac{M}{h_1\tau_0}.
  \label{eq:chiscaling}
\end{equation}
The scalar dissipation rates for A0--A8 are presented in figure~\ref{fig:mm}(\emph{c},\emph{d}) for the raw and normalized profiles, respectively. Unlike the normalized mixed mass profiles of figure~\ref{fig:mm}(\emph{b}), the normalized peak values for scalar dissipation in figure~\ref{fig:mm}(\emph{d}) are similar across Atwood numbers, showing no discernible trend and agreeing to within $\pm 3\%$. The leftward shift of the peak is more pronounced in the normalized scalar dissipation than in the normalized mixed mass. Compared to the analytical 1D solution in figure~\ref{fig:sol1D}(\emph{e}), this shift of the peak in SRT turbulence is quantitatively similar. However, unlike the 1D solution, the maximum value of the scalar dissipation rate does not increase with Atwood number for SRT turbulence. 

Finally, the effect of the Reynolds number on the scalar dissipation rate are presented in figure~\ref{fig:mm_Re}(\emph{c},\emph{d}) for the raw and normalized profiles, respectively. Similar to observations on the viscous dissipation rate, Reynolds number effects are minimal after normalization, except in the less turbulent (i.e. lower $\Rey$) cases. The overall spread of values is $\mathcal{E}_{\rm rms}\approx 0.05$.

\subsection{Scaling of global parameters}
\label{sec:scaling_global}
In \S~\ref{sec:scaling_budget}, scalings were proposed for the budget terms in terms of six global parameters, namely, the mixing layer height $h_1$, gravity $g$, timescale $\tau_0$, the integrated mixed mass $M$, and the reservoir densities $\rho_H$ and $\rho_L$. While $h_1$, $g$, $\rho_H$, and $\rho_L$ are input parameters, $\tau_0$ and $M$ are outputs of the SRT simulation. In this section, we relate $\tau_0$ and $M$ to the other parameters by using several simplifying assumptions, further reducing the total number of parameters required to the minimum set of $h_1$, $g$, $\rho_H$, and $\rho_L$.

\subsubsection{Mixing layer timescale}
\label{sec:tau0_scaling}

Consistent with the findings of TRT studies \citep{Livescu2009_JoT,zhou2019time}, we postulate that the large-scale growth rate of the mixing layer scales with the magnitude of the turbulent velocity fluctuations such that  
\begin{equation}
  \frac{h_1}{\tau_0} \sim \tilde{k}^{1/2},
  \label{eq:tau0andkfav}
\end{equation}
where the scaling for $\tilde{k}$ can be derived from previously established results. The Favre average is first written as
\begin{equation}
  \tilde{k} = \frac{\langle \rho k \rangle}{\langle \rho \rangle} = \dd{\ln \langle \rho \rangle}{\langle \rho \rangle} \langle \rho k \rangle = \frac{\langle \rho k \rangle}{{\rm d}\langle \rho \rangle/{\rm d} y}\dd{\ln \langle \rho \rangle}{y}.
  \label{eq:kfavscaling0}
\end{equation}
To arrive at a scaling relationship that is consistent with (\ref{eq:scaling_def}), two specific assumptions are invoked. First, the evolution of mean density is assumed to be diffusion-like, i.e. $\langle \rho \rangle$ and $\langle \rho v\rangle$ can be approximated by an error function (see figure \ref{fig:density}(\emph{b})) and gaussian distribution (see figure \ref{fig:velocity}(\emph{b})), respectively. Second, the TKE profile $\langle \rho k \rangle$ inherits the shape of the mean mass flux through local buoyancy production scaling and can also be approximated as Gaussian (see figures \ref{fig:tke}, \ref{fig:tke_Re}, and Appendix \ref{app:tke_scaling}). If $\langle \rho \rangle$ and $\langle \rho k \rangle$ were exact error and Gaussian functions, respectively, the ratio $\langle \rho k\rangle/[{\rm d}\langle \rho \rangle/ {\rm d}y] $ would be independent of $y$ and equal to the ratio of their integrals. Utilizing this property in an approximate sense, and substituting (\ref{eq:drhodx}) and (\ref{eq:kscaling}), the scaling for $\tilde{k}$ becomes
\begin{equation}
  \int \tilde{k} {\rm d}\left(\frac{y}{h_1}\right) = \int \frac{\langle \rho k \rangle}{{\rm d}\langle \rho \rangle/{\rm d} y} \dd{\ln \langle \rho \rangle}{y} {\rm d}\left(\frac{y}{h_1}\right) \approx \frac{1}{h_1}\frac{\int \langle \rho k \rangle {\rm d}y}{\int {\rm d} \langle \rho \rangle} \int {\rm d}\ln \langle \rho \rangle \propto gh_1 \ln R
  \label{eq:kfavscaling1}
\end{equation}
or, equivalently,
\begin{equation}
  \tilde{k} \sim gh_1 \ln R.
  \label{eq:kfavscaling}
\end{equation}
Combining (\ref{eq:tau0andkfav}) and (\ref{eq:kfavscaling}), the SRT timescale $\tau_0$ can be related to $h_1$, $g$, $\rho_H$, and $\rho_L$ via
\begin{equation}
  {\tau_0} \sim \left( \frac{h_1} {g \ln R}\right)^{1/2}.
    \label{eq:tau0scaling}
\end{equation}

\begin{figure}
  \begin{minipage}{0.498\textwidth}
    (\emph{a})\\
    \includegraphics[width=\textwidth]{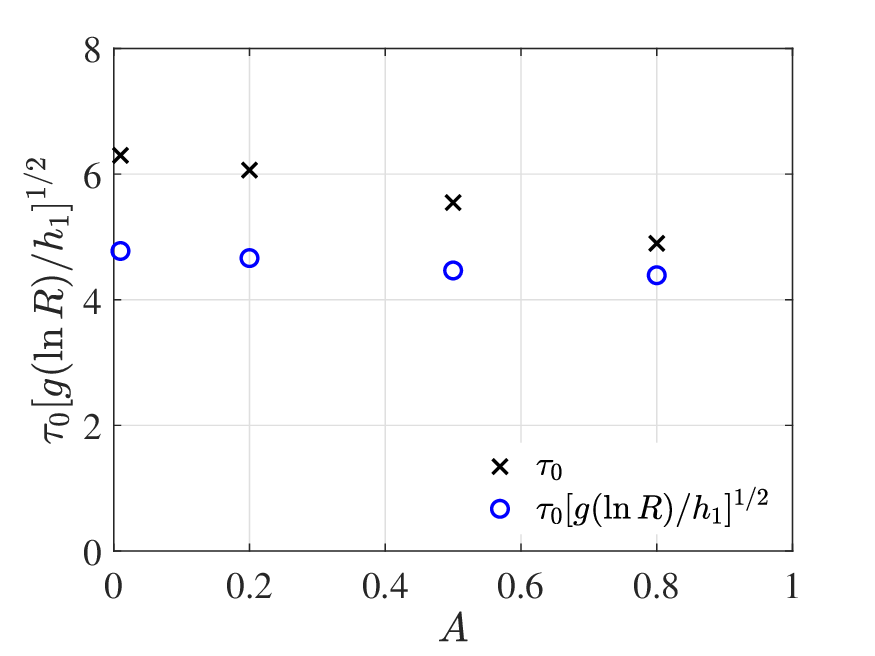}
  \end{minipage}
  \hfill
  \begin{minipage}{0.498\textwidth}
    (\emph{b})\\
    \includegraphics[width=\textwidth]{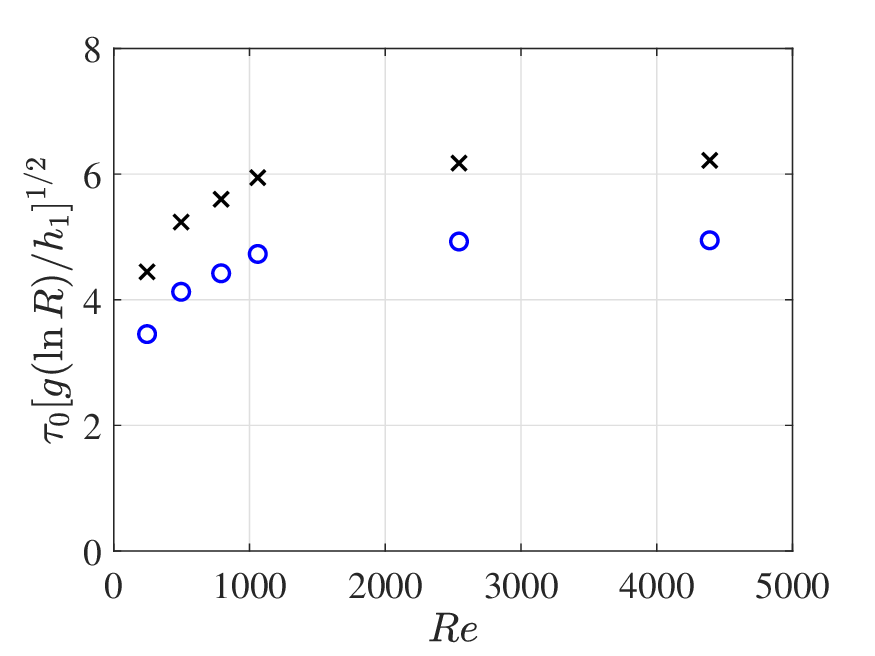}
  \end{minipage}
\caption{Verification of $\tau_0$ scaling as a function of the (\emph{a}) Atwood number (A0--A8) and (\emph{b}) Reynolds number (G1--G6). Legend: unnormalized $\tau_0$ (black crosses), normalized $\tau_0[(g \ln R)/h_1]^{1/2}$ (blue circles)}
\label{fig:tau0scaling}
\end{figure}
This scaling is evaluated using all simulation cases and presented in figures~\ref{fig:tau0scaling}(\emph{a}) and (\emph{b}) for cases A0--A8 and G1--G6, respectively. In figure~\ref{fig:tau0scaling}(\emph{a}), the unnormalized values of $\tau_0$ decrease with increasing Atwood number. After normalization, they become approximately constant. In figure~\ref{fig:tau0scaling}(\emph{b}), the normalization has little effect on the Reynolds number trend. This is because $h_1$, $g$, and $R$ are nearly constant across G1--G6. Nevertheless, the observed values of $\tau_0$ approach a high Reynolds number limit. The proposed scaling does not ``correct'' for Reynolds number effects, because (\ref{eq:tau0andkfav}) assumes a sufficiently high Reynolds number regime where inertial effects dominate over viscous and diffusive processes. In the limiting case of pure 1D diffusion (see Appendix~\ref{app:1dvd}), $\tau_0 = \pi h_1^2/16D$. Thus, it is reasonable to expect that $\tau_0$ may exhibit some dependence on $D$ (and $\nu$) for low Reynolds number SRT turbulence. Despite this $\Rey$-dependence of $\tau_0$ at low Reynolds numbers, we note that all profiles for cases G1--G6 shown in \S~\ref{sec:scaling_budget} are almost identical after normalization, suggesting that the influence of the Reynolds number is almost entirely captured by the global parameter $\tau_0$.

\subsubsection{Integrated mixed mass}
\label{sec:M}

A scaling for the integrated mixed mass is developed in two main steps. First, the integrated mixed mass of the full SRT field, $M = \int \langle \rho Y(1-Y) \rangle\,{\rm d}y$, is related to the integrated mixed mass of the mean density field, $M_{\langle \rho \rangle} = \int \langle \rho \rangle \tilde{Y}(1-\tilde{Y})\,{\rm d}y$. Next, an estimate for $M_{\langle \rho \rangle}$ is proposed using a modeled mean mole fraction profile.

\begin{figure}
  \begin{center}
  \begin{minipage}{0.498\textwidth}
    \includegraphics[width=\textwidth]{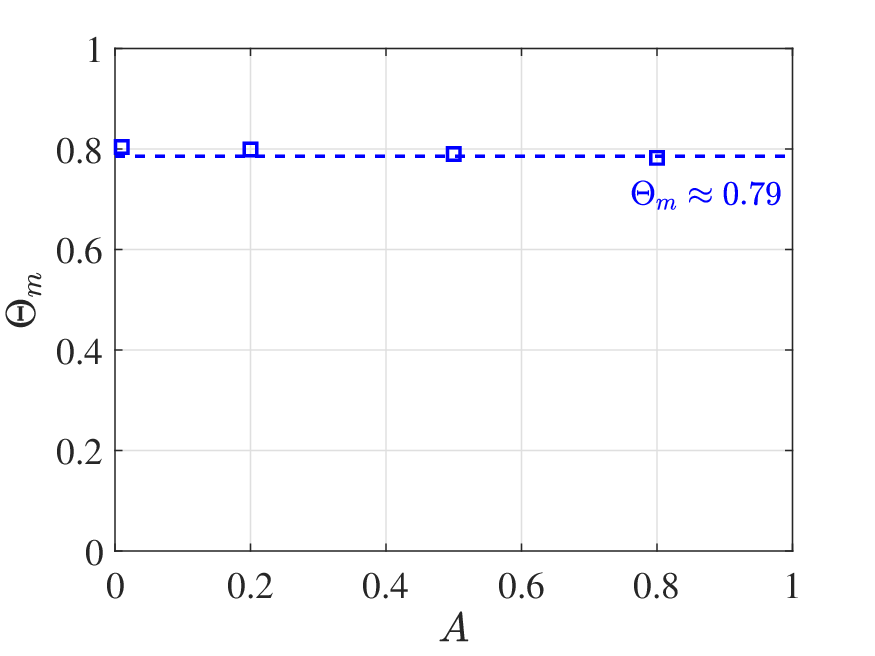}
  \end{minipage}
  \end{center}
  \caption{Mixedness parameter $\Theta_m= M/M_{\langle \rho \rangle}$
  \label{fig:Mscaling}}
\end{figure}

In \S~\ref{sec:results_mix}, it was observed that the average mixed mass, $\langle \rho Y(1-Y) \rangle$, behaves similarly to the mixed mass of the mean density field, $\langle \rho \rangle \tilde{Y}(1-\tilde{Y})$. To quantify the relationship between their integrals, a new measure of mixedness is introduced and defined as $\Theta_m = M/M_{\langle\rho\rangle}$. The values of $\Theta_m$ for cases A0--A8 are shown in figure~\ref{fig:Mscaling}; they remain approximately constant at $\Theta_m \approx 0.8$. Hence, this ratio is modeled as a constant $\Theta^*_m$, and any proposed scaling for $M_{\langle \rho \rangle}$ can be applied directly to $M$.

The next step is to establish a scaling relationship between $M_{\langle \rho \rangle}$ and the input parameters. Since both $\langle\rho\rangle$ and $\tilde{Y}$ are fully determined by the mean mole fraction $\langle X \rangle$ and the reservoir densities, so is the derived integral 
\begin{equation}
  M_{\langle \rho \rangle}
  = \int \langle\rho\rangle \tilde{Y} (1-\tilde{Y}) \,{\rm d}y
  = \frac{h_1}{4} \frac{\int \langle\rho\rangle\tilde{Y} (1-\tilde{Y}) \,{\rm d}y}{\int \langle X\rangle (1-\langle X\rangle) \,{\rm d}y}  
  = \frac{\rho_H h_1}{4} \frac{\int \frac{\langle X\rangle(1-\langle X\rangle)}{1+(R-1)\langle X\rangle}\,{\rm d}y}{\int \langle X\rangle (1-\langle X\rangle) \,{\rm d}y}.
  \label{eq:Mrhoav}
\end{equation}
To estimate $M_{\langle \rho \rangle}(\langle X \rangle, \rho_H, \rho_L)$ \emph{a priori}, a model expression for the mean mole fraction $\langle X \rangle^*$ is necessary, where the $^*$ superscript denotes a modeled quantity. In \S~\ref{sec:results_cont}, it was observed that the mean mole fraction does not change significantly with Atwood number and is well-approximated by the 1D solution. Thus, the 1D solution (derived in Appendix \ref{app:1dvd}) can be adopted as a model for the mean mole fraction:
\begin{equation}
  \langle X \rangle^* = X_{1D} = \frac{1}{2} \left[1 + \erf\left( \sqrt{\frac{8}{\pi}}\frac{y}{h_1} \right)\right],
  \label{eq:Xfit}
\end{equation}
where the mixing layer height is matched, i.e. $h_1 = 4\int \langle X\rangle^*(1-\langle X\rangle^*)\,{\rm d}y$. Once $\langle X \rangle^*$ is specified, the integrated mixed mass of the modeled mean density, $M^*_{\langle \rho \rangle}(\langle X\rangle^*,\rho_H,\rho_L)$, is obtained by substitution of $\langle X\rangle = \langle X^*\rangle$ in (\ref{eq:Mrhoav}). 

Combining the two steps, the integrated mixed mass $M$ can be expressed as 
\begin{equation}
  M 
  \approx \Theta^*_m M_{\langle \rho \rangle}(\langle X\rangle,\rho_H,\rho_L) 
  \approx \Theta_m^* M^*_{\langle \rho \rangle}(\langle X\rangle^*,\rho_H,\rho_L),
  \label{eq:Mmodel}
\end{equation}
where the RHS of (\ref{eq:Mmodel}) is solely a function of $\rho_H$ and $\rho_L$.

\section{Discussion}
\label{sec:discussion}

This section presents three extensions to the analysis of \S~\ref{sec:results}. First, a reference density scale is inferred from the scaling of the integrated mixed mass. Second, normalized Favre-averaged velocity statistics are examined. Finally, a variable-density scaling law for the self-similar growth rate is derived and compared to the commonly used scaling expression (\ref{eq:hgrowth}) in the literature. 

\subsection{Density scale for the mixed mass}
\label{sec:rhom}
\begin{figure}
  \begin{center}
  \begin{minipage}{0.498\textwidth}
    \includegraphics[width=\textwidth]{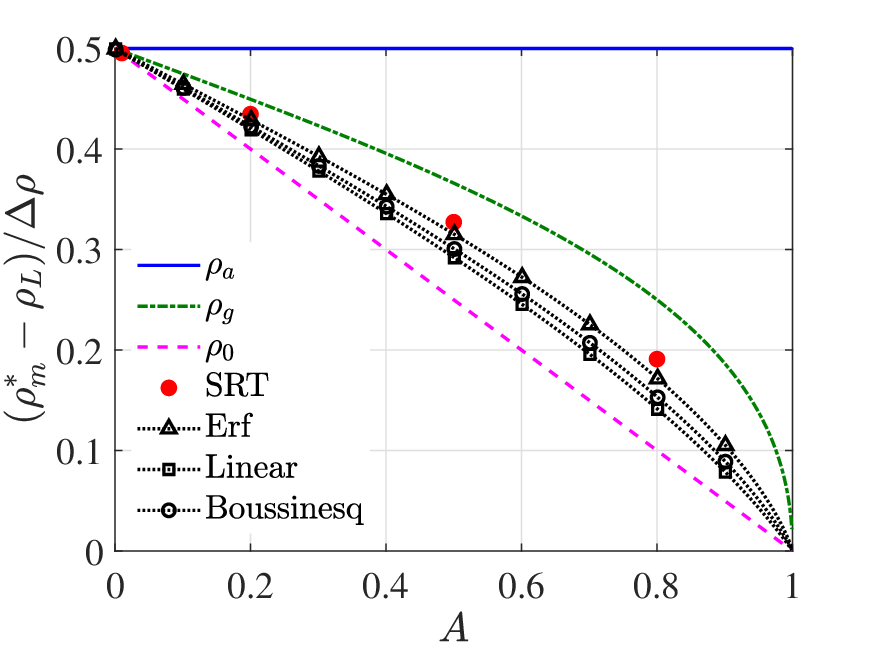}
  \end{minipage}
  \end{center}
  \caption{Density scale for the mixed mass. Mean definitions of the reservoir densities are shown without markers: arithmetic mean $\rho_a$ (blue solid), geometric mean $\rho_g$ (green dash-dotted), harmonic mean $\rho_0$ (magenta dashed); SRT data  $\rho_m = 4M_{\langle \rho \rangle}/h_1$ is represented by red filled markers; density scales derived from modeled expressions $\langle X \rangle^*$ are shown as black dotted lines with markers: triangles (error function/1D solution), squares (piecewise linear), circles (SRT mean mole fraction for case A0)   
  \label{fig:rhomscaling}}
\end{figure}

As discussed in \S~\ref{sec:results_mix}, the integrated mixed mass $M$ is expected to scale with the product of a mixing density $\rho_m$ and the height $h_1$. Leveraging the observation that $\Theta_m$ is approximately constant, we define
\begin{equation}
  \rho_m(\langle X\rangle,\rho_H,\rho_L) 
  = \rho_H \frac{\int \frac{\langle X\rangle(1-\langle X\rangle)}{1+(R-1)\langle X\rangle}\,{\rm d}y}{\int \langle X\rangle (1-\langle X\rangle) \,{\rm d}y},
  \label{eq:rhom}
\end{equation}
such that (\ref{eq:Mrhoav}) is expressed as $M_{\langle\rho\rangle} = \rho_m h_1/4$. Values of $\rho_m$ are extracted from cases A0--A8 and shown in figure~\ref{fig:rhomscaling}, alongside several common averages of the reservoir densities: the arithmetic mean $\rho_a = (\rho_H+\rho_L)/2$, geometric mean $\rho_g = \sqrt{\rho_H \rho_L}$, and harmonic mean $\rho_0=2/(\rho_H^{-1}+\rho_L^{-1})$. The extracted SRT values of $\rho_m$ are significantly different from any of these simple averages considered.

Extending the analysis of \S~\ref{sec:M}, modeled mole fraction profiles can be applied to (\ref{eq:rhom}) to obtain possible models for the mixing density $\rho_m^*(\langle X\rangle^*,\rho_H,\rho_L)$. The error function solution proposed in (\ref{eq:Xfit}) is an obvious candidate, but it is also possible to use alternative expressions for $\langle X \rangle^*$. To assess the sensitivity of $\rho_m^*$ to the assumed profile, three models are considered: the analytical 1D solution (error function), a piecewise linear approximation, and the mean mole fraction from the Boussinesq SRT case (A0). All three models perform noticeably better than any of the simple averages, which highlights the importance of accounting for the shape of the mole fraction profile. Among them, the error function yields the best agreement with the SRT data. 

Finally, the definition of the mixing height $h_m$ used in \citet{goh2025statistically} is revisited. Starting with its definition in (\ref{eq:hm_def}), the mixing height $h_m$ can be related to $h_1$ using (\ref{eq:Mrhoav}) and (\ref{eq:rhom}) such that 
\begin{equation}
  h_m 
  = \frac{4M}{\rho_0}
  = \Theta_m \frac{4M_{\langle\rho\rangle}}{\rho_0} 
  = \Theta_m \frac{\rho_m}{\rho_0} \frac{4M_{\langle\rho\rangle}}{\rho_m}
  = \Theta_m \frac{\rho_m}{\rho_0} h_1.
  \label{eq:hmvsh1}
\end{equation}
 The concerns highlighted in \S~\ref{sec:inputs} about the use of the harmonic mean as an arbitrary density normalization are illustrated by (\ref{eq:hmvsh1}). While $\Theta_m$ is approximately constant across Atwood numbers (see figure \ref{fig:Mscaling}), $\rho_m/\rho_0$ is not (see figure \ref{fig:rhomscaling}). Proportionality of the large lengthscales is a feature of self-similar flows, but the existing definition of $h_m$ does not preserve similarity across Atwood numbers. A more physically appropriate definition of $h_m$ is
 \begin{equation}
   h_{m,{\rm new}} = \frac{4M}{\rho_m^*} = \frac{4}{\rho_m^*}\int \langle \rho Y (1-Y)\rangle\,{\rm d}y ,   
 \end{equation}
where $\rho_m^*(\langle X \rangle^*, \rho_H,\rho_L)$ is computed based on the model mole fraction profile using the error function of the 1D solution shown in (\ref{eq:Xfit}).

\subsection{Favre-averaged velocity statistics}

In \S~\ref{sec:intro}, the scaling of velocity statistics was mentioned as an open question in the study of variable-density RT flows. In \S~\ref{sec:results}, the scaling of the conserved variables $\langle \rho \phi \rangle$ were addressed, where $\phi$ represents $v$, $k$, or $\epsilon$. Here, we build on the scalings developed in \S~\ref{sec:results} to consider the effect of the Atwood number on Favre-averaged velocity statistics. 

\begin{figure}
  \begin{minipage}{0.498\textwidth}
    (\emph{a})\\
    \includegraphics[width=\textwidth]{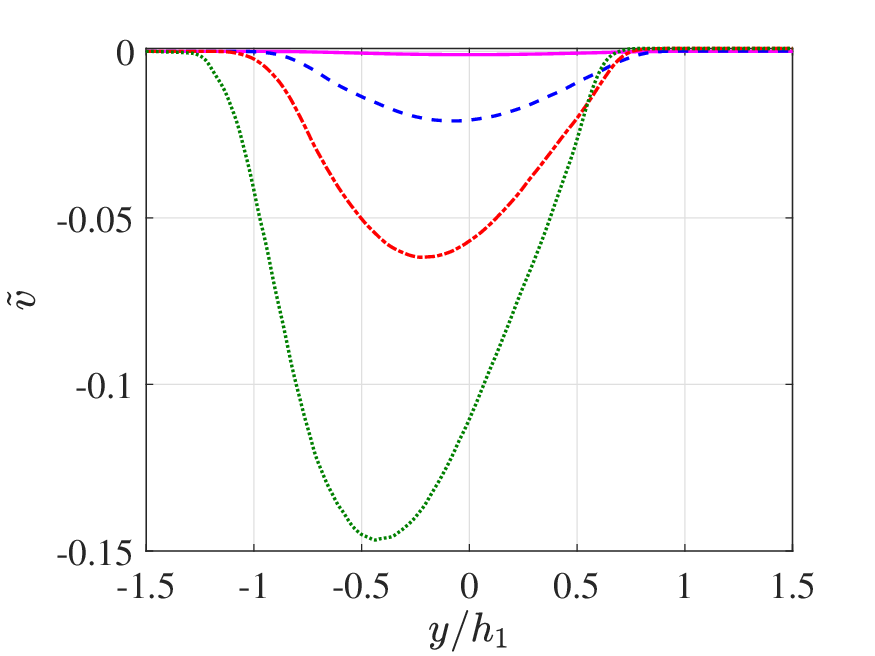}
  \end{minipage}
  \hfill
  \begin{minipage}{0.498\textwidth}
    (\emph{b})\\
    \includegraphics[width=\textwidth]{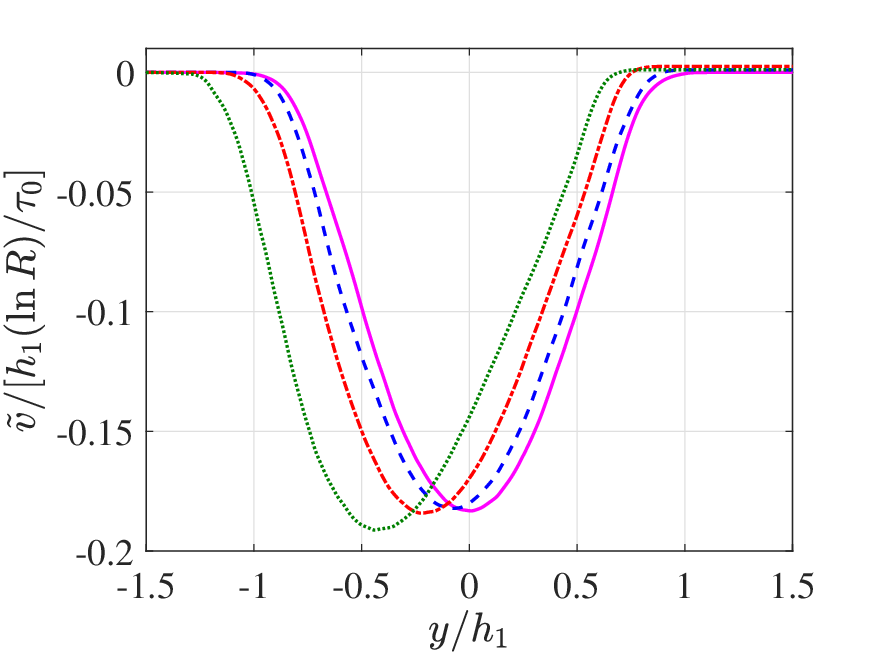}
  \end{minipage}
  
  \begin{minipage}{0.498\textwidth}
    (\emph{c})\\
    \includegraphics[width=\textwidth]{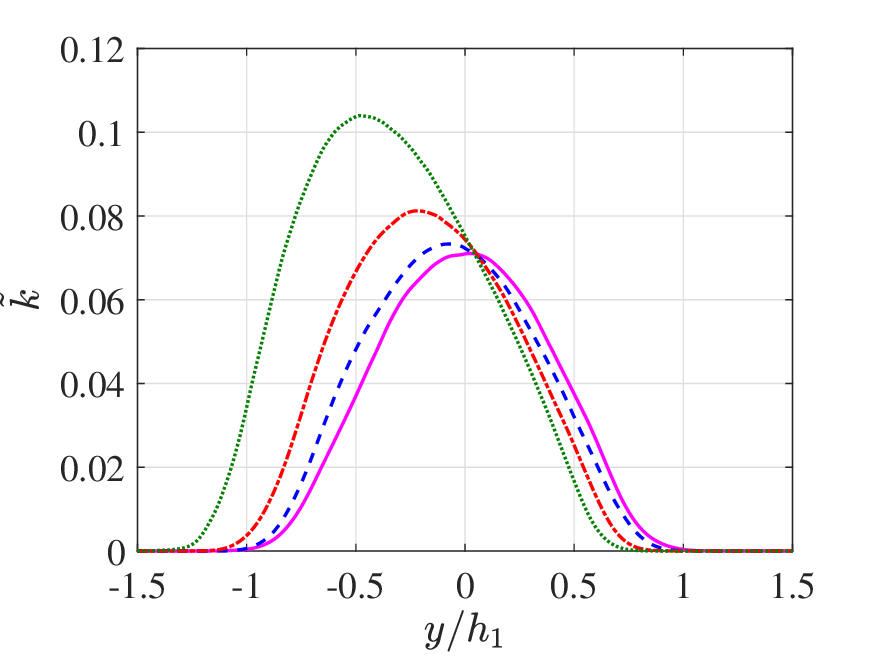}
  \end{minipage}
  \hfill
  \begin{minipage}{0.498\textwidth}
    (\emph{d})\\
    \includegraphics[width=\textwidth]{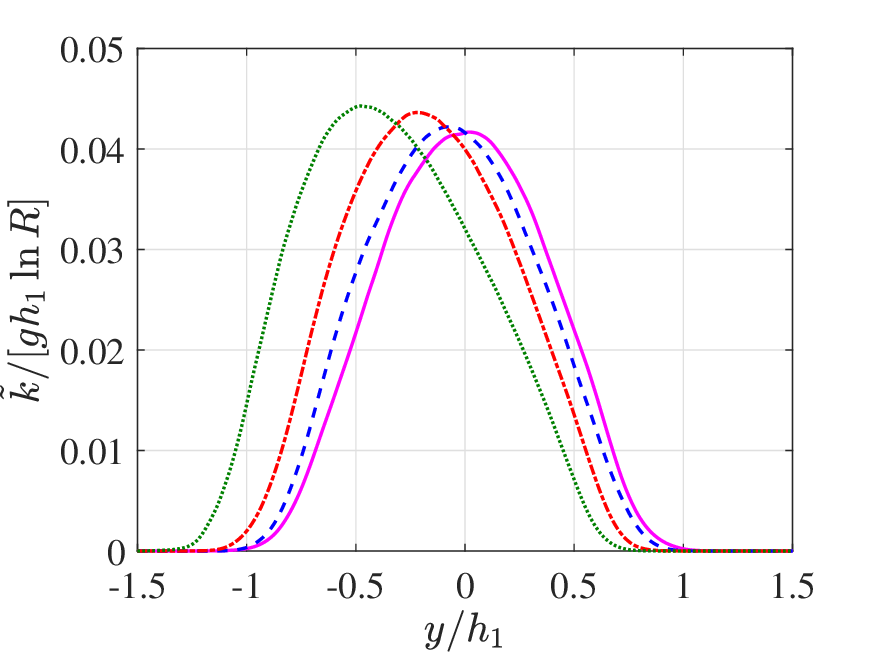}
  \end{minipage} 

  \begin{minipage}{0.498\textwidth}
    (\emph{e})\\
    \includegraphics[width=\textwidth]{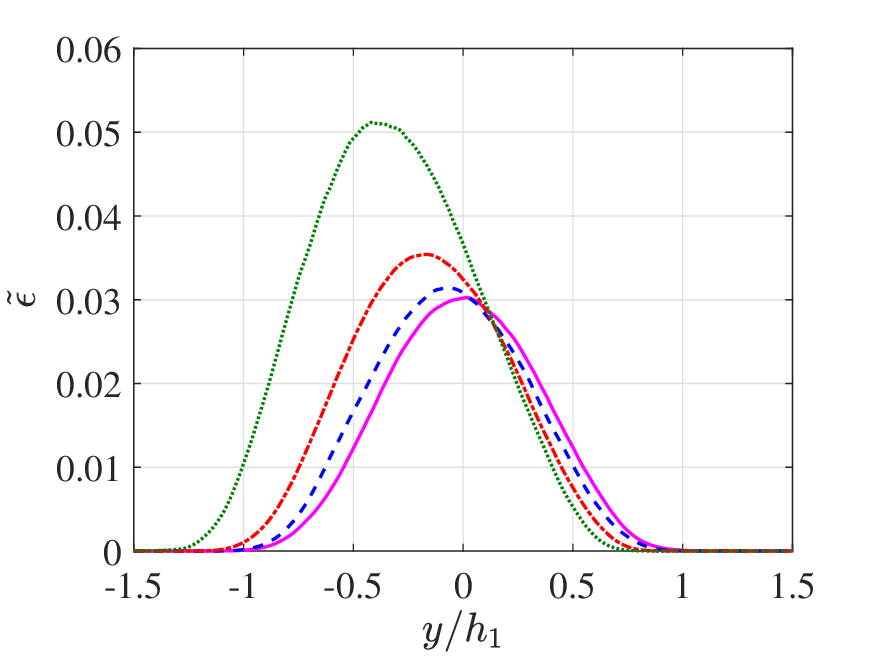}
  \end{minipage}
  \hfill
  \begin{minipage}{0.498\textwidth}
    (\emph{f})\\
    \includegraphics[width=\textwidth]{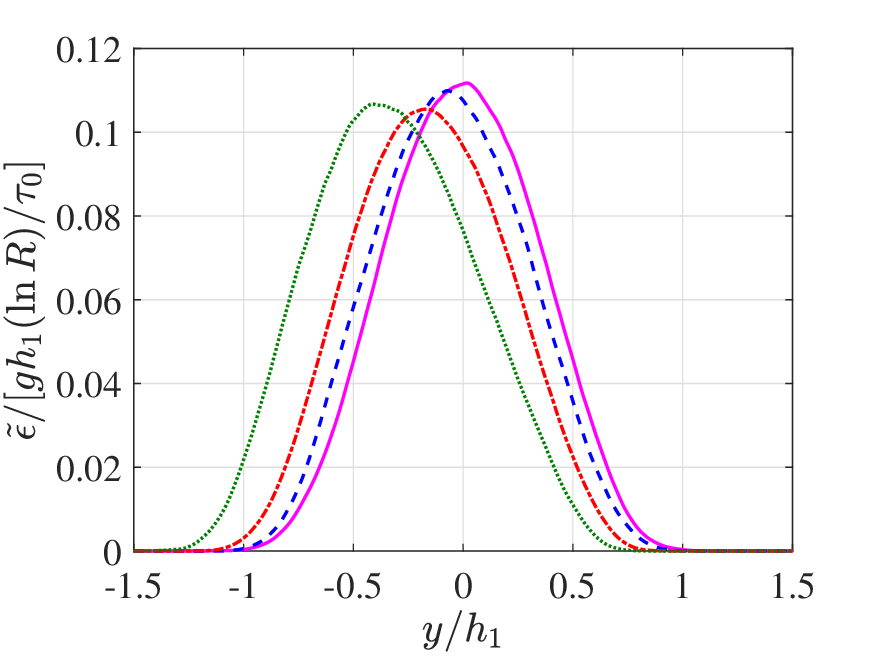}
  \end{minipage} 

\caption{Effect of the Atwood number on the (\emph{a}) Favre-averaged velocity, (\emph{b}) normalized Favre-averaged velocity, (\emph{c}) Favre-averaged TKE, (\emph{d}) normalized Favre-averaged TKE, (\emph{e}) Favre-averaged viscous dissipation, and (\emph{f}) normalized Favre-averaged dissipation. Legend: $A=$ 0.01 (magenta solid), 0.2 (blue dashed), 0.5 (red dash-dotted), 0.8 (green dotted)}
\label{fig:favre}
\end{figure}

In (\ref{eq:kfavscaling}) of \S~\ref{sec:tau0_scaling}, the scaling for $\tilde{k}$ was derived from $\langle \rho k\rangle$ by using the result $d \ln \langle \rho \rangle = d\langle\rho\rangle/\langle\rho\rangle$, and approximating $\langle \rho k \rangle$ as a Gaussian function. Because the mean mass flux $\langle \rho v \rangle$ and viscous dissipation rate $\langle \rho \epsilon \rangle$ profiles have similar shapes to the TKE, the assumptions for the derivation of (\ref{eq:kfavscaling}) are expected to be satisfied approximately. Hence, the same procedure is used to obtain scalings for $\tilde{v}$ and $\tilde{\epsilon}$ leveraging the scalings for $\langle \rho v \rangle$ and $\langle \rho \epsilon \rangle$. The unnormalized Favre-averaged profiles are shown in figure~\ref{fig:favre}(\emph{a},\emph{c},\emph{e}), and the normalized profiles are shown in figure~\ref{fig:favre}(\emph{b},\emph{d},\emph{f}). There are three key observations.

First, the peak magnitude of each of the unnormalized quantities increases significantly with Atwood number. The peak magnitudes of the normalized profiles agree to within 5\%, which demonstrates the effectiveness of the $\ln R$ scaling proposed for all velocity statistics considered.

Second, while the conserved quantities, $\langle \rho v \rangle$, $\langle \rho k\rangle$, and $\langle \rho \epsilon \rangle$, all have peak values at $y\approx0$, there is a leftward shift of the peaks for all Favre-averaged profiles as the Atwood number increases, similar to trends observed by \citet{Livescu2010}. This is merely a consequence of dividing near-symmetric $\langle \rho\phi \rangle$ profiles by an asymmetric, monotonically increasing $\langle \rho \rangle$ profile. Dividing by smaller mean density values on the left of $y=0$ than on the right causes the shift. This effect is not dissimilar to the leftward shift of the Favre-averaged mass fraction profiles shown in figure~\ref{fig:density} and is also evident in the normalized velocity of the 1D diffusion solution shown in figure \ref{fig:sol1D}(d). 

Third, the apparent shift of the peaks has implications on the interpretation of turbulence statistics. For instance, the Kolmogorov scale is commonly defined in variable density flows using the Favre-averaged dissipation, $\eta = (\nu^3/\tilde{\epsilon})^{1/4}$. In figure~\ref{fig:favre}(\emph{f}), the location of the smallest turbulence scales (i.e. maximum dissipation) shifts towards the light-fluid side as the Atwood number increases. Based on this definition of $\eta$, the smallest scales are not found at $y=0$, even though this is commonly assumed to be the ``core'' of the mixing layer. Any comparison of small-scale statistics across different Atwood numbers should take this shift into consideration. 

\subsection{Effective Atwood number for non-Boussinesq flows}
\label{sec:Astar}

\begin{figure}
  \begin{minipage}{0.498\textwidth}
    (\emph{a})\\
    \includegraphics[width=\textwidth]{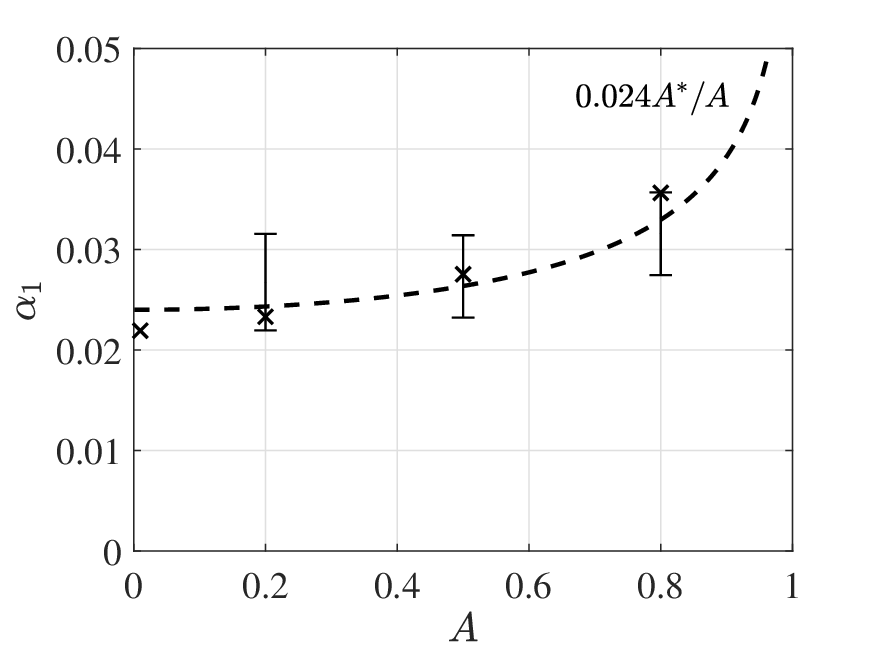}
  \end{minipage}  
  \begin{minipage}{0.498\textwidth}
    (\emph{b})\\
    \includegraphics[width=\textwidth]{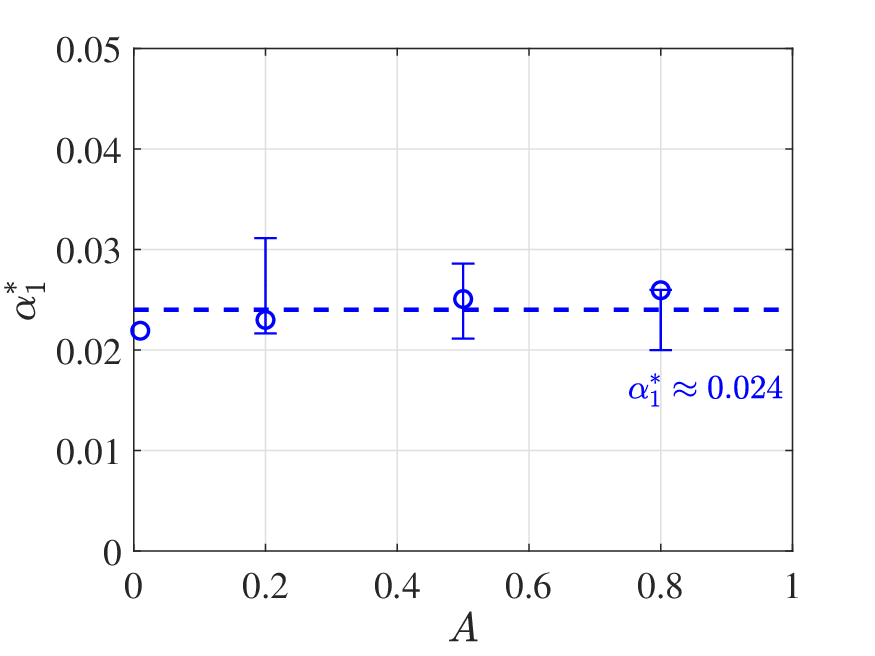}
  \end{minipage}
  \caption{(\emph{a}) Standard growth parameter $\alpha_1$ and (\emph{b}) effective growth parameter $\alpha_1^*$. Legend: SRT results (markers), range of TRT values from \citet{zhou2019time} (bars), constant-$\alpha_1^*$ model (dotted line).
\label{fig:alpha}}
\end{figure}

The final discussion addresses the scaling of the growth rate for self-similar variable-density RT flows, conventionally represented by (\ref{eq:hgrowth}). Figure~\ref{fig:alpha}(\emph{a}) shows the growth rate parameter $\alpha_1$ extracted from two separate sources. First, it is evaluated from SRT cases A0--A8 using $\alpha_1 = h_1/(4Ag\tau_0^2)$. Second, the range of $\alpha_p$ values from the TRT simulations of \citet{zhou2019time} (summarized in table \ref{tab:TRTvalidate}) are converted to $\alpha_1$ using $h_1/h_p$ ratios derived from the SRT data. Both sets of data indicate that $\alpha_1$ increases with the Atwood number, consistent with observations from experimental studies \citep{Dimonte2000,banerjee2010detailed}. The observed $\alpha_1(A)$ trend is a direct consequence of the scaling of $\tau_0$ presented in (\ref{eq:tau0scaling}).  

Equation (\ref{eq:hgrowth}) was derived using the Boussinesq approximation \citep{ristorcelli2004rayleigh}. For variable-density flows, (\ref{eq:hgrowth}) can be generalized as  
\begin{equation}
  \dot{h}_i^2 = 4F_i(A)gh_i,
  \label{eq:4Fgh}
\end{equation}
where $F_i(A) = \alpha_i(A)A$. Building on the demonstrated scaling of (\ref{eq:tau0scaling}), we seek an alternative expression for $F_i$ that accounts for variable-density effects using a modified or effective Atwood number. Specifically, we propose the form $F_i(A) = \alpha_i^*A^*(A)$, where all density effects are captured by the effective Atwood number $A^*$ and $\alpha_i^*$ is a universal constant that is truly independent of $A$. The functional form of $A^*(A)$ is derived below.  

Multiplying $h_1$ to both sides of (\ref{eq:tau0scaling}), squaring them, and including a constant of proportionality $4\alpha_1^*C$ yields
\begin{equation}
  \left(\frac{h_1}{\tau_0}\right)^2 \approx 4\alpha_1^* C g h_1 \ln R,
  \label{eq:growthscaling}
\end{equation}
which has a $\ln R$ dependence that is consistent with (\ref{eq:hgrowth_lnR}) proposed by \citet{belen1965theory}---a similarity that will be discussed in \S~\ref{sec:summary_disc}. To maintain consistency of the $\ln R$ scaling with the large number of studies done in the Boussinesq limit, a constant of proportionality $4\alpha_1^*C$ is included and $C$ is chosen such that $\alpha_i^*(\Delta\rho \to 0) = \alpha_i(\Delta\rho \to 0)$. The RHS of (\ref{eq:hgrowth}) is evaluated for $h_i=h_1$ and expanded in a Taylor series as 
\begin{equation}
  4\alpha_1 A g h_1 = 
  4\alpha_1 g h_1 \left[ \frac{1}{2}\left(\frac{\Delta \rho}{\rho_L}\right) - \frac{1}{4}\left(\frac{\Delta \rho}{\rho_L}\right)^2 +  \frac{1}{8}\left(\frac{\Delta \rho}{\rho_L} \right)^3 + \ldots \right],
\end{equation}
while the corresponding expansion for the RHS of (\ref{eq:growthscaling}) is
\begin{equation}
  4\alpha_1^*Cgh_1 \ln R = 
  4\alpha_1^* g h_1 C\left[ \left(\frac{\Delta \rho}{\rho_L}\right) - \frac{1}{2} \left(\frac{\Delta \rho}{\rho_L}\right)^2 + \frac{1}{3} \left(\frac{\Delta \rho}{\rho_L} \right)^3 - \ldots\ \right].
\end{equation}
Applying the constraint $\alpha_1^*(\Delta\rho \to 0) = \alpha_1(\Delta\rho \to 0)$ and matching the leading order terms, we find $C=1/2$. Hence, (\ref{eq:growthscaling}) can be used to define an effective Atwood number,
\begin{equation}
  A^* = \frac{1}{2}\ln R,
\end{equation}
such that 
\begin{equation}
  \left( \frac{h_1}{\tau_0}\right)^2 = 2\alpha_1^*gh_1 \ln R = 4\alpha_1^* A^* g h_1.
  \label{eq:newalphadef}
\end{equation}

\begin{figure}
  \begin{minipage}{0.498\textwidth}
    (\emph{a})\\
    \includegraphics[width=\textwidth]{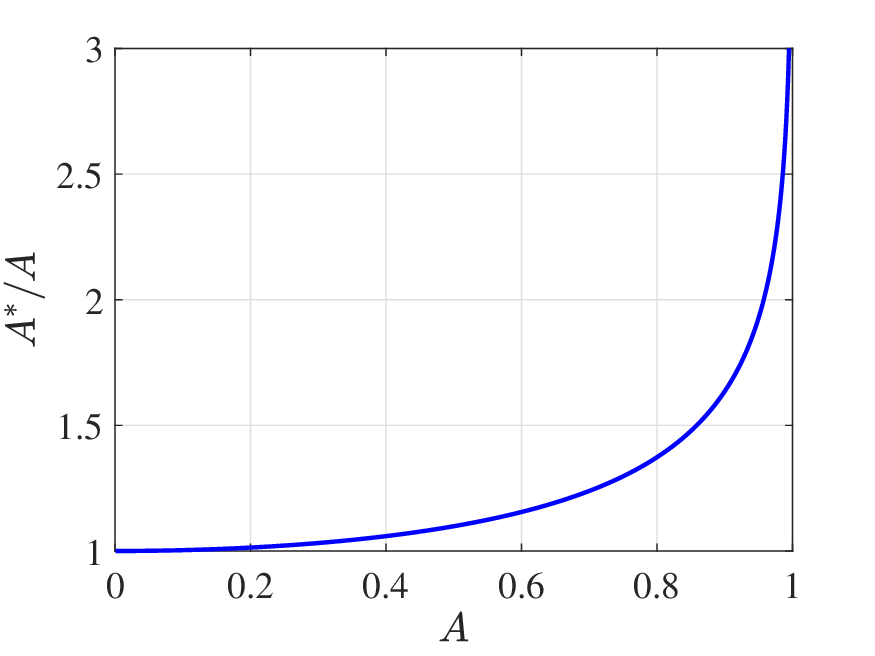}
  \end{minipage}
  \hfill
  \begin{minipage}{0.498\textwidth}
    (\emph{b})\\
    \includegraphics[width=\textwidth]{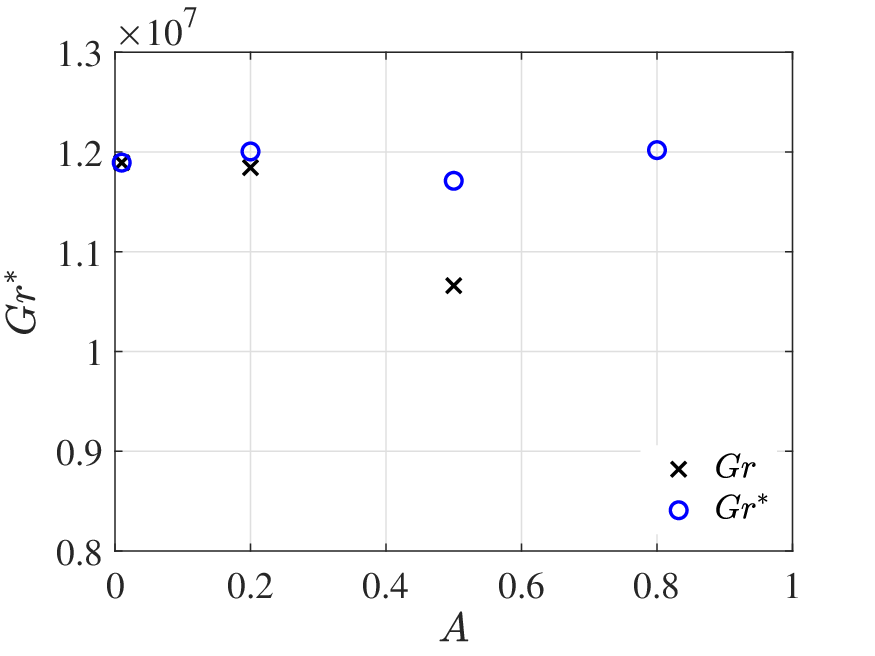}
  \end{minipage}
  \caption{(\emph{a}) Effective Atwood number $A^*$ with respect to $A$; and (\emph{b}) comparison of Grashof number definitions using SRT cases A0--A8. Legend: standard $\Gr$ (black markers), effective $\Gr^*$ (blue circles).
  \label{fig:Astar}}
\end{figure}

The effective Atwood number $A^*$ is equal to $A$ in the Boussinesq limit but differs significantly at higher Atwood numbers, as shown in figure~\ref{fig:Astar}(\emph{a}). Rearranging (\ref{eq:newalphadef}), the effective growth parameter $\alpha_1^*$ is computed using $\alpha^*_1 = h_1/(4 A^*g\tau_0^2)$ and shown in figure~\ref{fig:alpha}(\emph{b}). As intended, $\alpha_1^*=\alpha_1$ for small Atwood numbers. For the SRT cases considered, the variation in $\alpha^*_1$ is 70\% smaller than that of $\alpha_1$, with a near-universal value of $\alpha^*\approx 0.024$. This demonstrates that $A^*$ is a more complete representation of the density effects associated with a variable-density RT flow. This constant-$\alpha_1^*$ model can explain the Atwood number trends of $\alpha_1$ using $\alpha_1 = \alpha_1^*A^*/A \approx 0.024A^*/A$. This estimate is included in figure~\ref{fig:alpha}(\emph{a}), which shows a good agreement with both SRT and TRT data. 

\begin{figure}
  \begin{minipage}{0.498\textwidth}
    (\emph{a})\\
    \includegraphics[width=\textwidth]{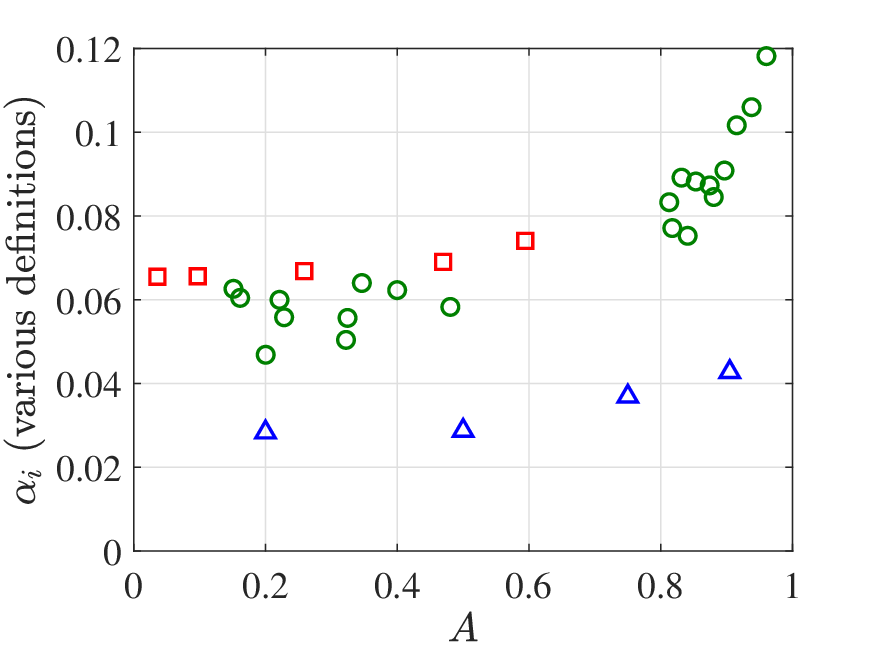}
  \end{minipage}  
  \begin{minipage}{0.498\textwidth}
    (\emph{b})\\
    \includegraphics[width=\textwidth]{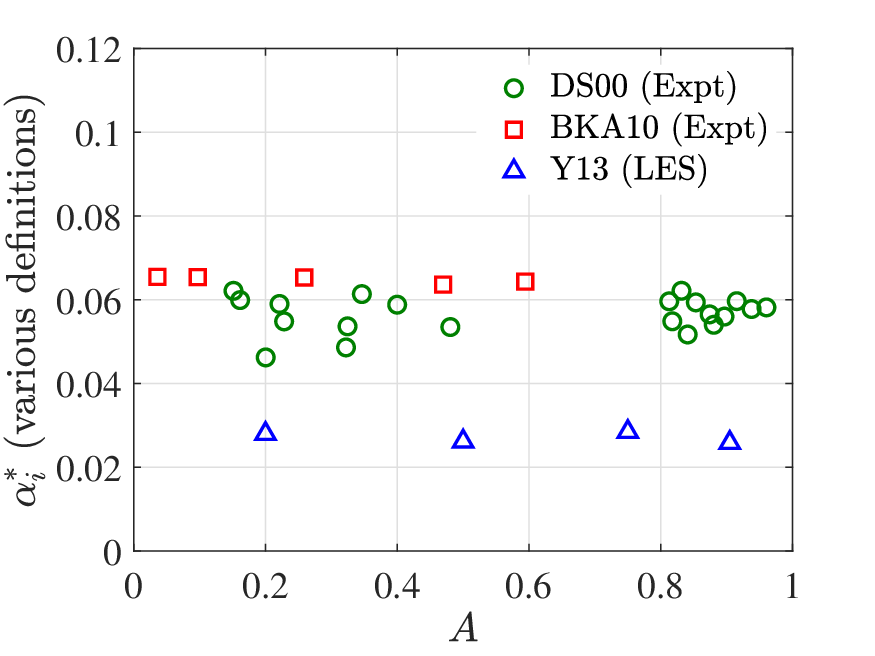}
  \end{minipage}
  \caption{(\emph{a}) Standard growth parameter $\alpha_i$ and (\emph{b}) effective growth parameter $\alpha_i^*$ extracted from other RT studies of Atwood number effects. Legend: \cite{Dimonte2000} (green circles), \cite{banerjee2010detailed} (red squares), and \cite{youngs2013density} (blue triangles). Initial conditions and height definitions vary across studies.
  \label{fig:alpha_compare}}
\end{figure}

In addition to the DNS results, the scaling based on the effective Atwood number is validated against previous experiments \citep{Dimonte2000, banerjee2010detailed} and large-eddy simulations \citep{youngs2013density}. The comparison is limited to studies that explored a range of Atwood numbers within a single experimental or computational setup. In all three studies, a growth parameter representative of the full mixing layer is computed as the mean of the reported bubble ($\alpha_b$) and spike ($\alpha_s$) growth parameters, although each study used different definitions for the bubble and spike heights. These averaged $\alpha_i$ values are presented in figure \ref{fig:alpha_compare}(\emph{a}). It is important to note that, due to differences in initial conditions and height definitions, the absolute values of $\alpha_i$ are not expected to be consistent between studies. Nevertheless, $\alpha_i$ consistently increases with Atwood number for all cases. To explore the validity of $A^*$ scaling, the standard growth parameters are converted to effective growth parameters using $\alpha_i^* = \alpha_iA/A^*$ and shown in figure \ref{fig:alpha_compare}(\emph{b}). Within each study, the values remain approximately constant across the range of Atwood numbers, indicating that the effective Atwood number successfully normalizes the growth behavior. The results also suggest that the scaling relationships developed based on DNS data for $A\le 0.8$ remain valid under conditions beyond this range.

The Atwood number also shows up in the definition of the Grashof number, as shown in (\ref{eq:parameters}). In \S~\ref{sec:simcases}, it was described that the input Grashof number in table~\ref{tab:simcases1} was reduced at higher Atwood numbers in order to maintain a constant Reynolds number (or $h_1/\eta$). This difference in trends between $\Gr$ and $\Rey$ can be explained by substituting (\ref{eq:newalphadef}) into the definition of the Reynolds number,
\begin{equation}
  \Rey^2 = \frac{h_1^4}{\tau_0^2\nu^2} = 4\alpha^* \frac{A^*gh_1^3}{\nu^2} = 4\alpha^*\Gr^*
\end{equation}
where $\Gr^*$ has an identical definition to $\Gr$, except for the replacement of $A$ with $A^*$. It is $\Gr^*$ (and not $\Gr$) that scales with $\Rey^2$. Both definitions of the Grashof number are shown in figure~\ref{fig:Astar}(\emph{b}) for the Atwood number sweep of A0--A8 at constant Reynolds number. Indeed, $\Gr^*$ has an approximately constant value.

\subsection{Summary of scaling relationships and physical insights}
To offer the reader a comprehensive overview of the results, this concluding discussion summarizes the derived scaling relations and highlights the key physical assumptions underlying their derivation.
\label{sec:summary_disc}

\begin{table}
  \begin{center}
  \def~{\hphantom{0}}
    \begin{tabular}{lcc}
      Quantity & Symbol & Normalization\\
      
      Density 
      & $\langle \rho \rangle - \rho_L$ 
      & $\Delta \rho$ \\
    
      Mass flux 
      & $\langle \rho v \rangle$
      & $\Delta \rho h_1/\tau_0 $ \\
      
      Mixed mass
      & $\langle \rho m \rangle$
      & $\rho_m$ \\
    
      Scalar dissipation rate
      & $\langle \rho \chi \rangle$
      & $\rho_m/\tau_0$ \\
    
      Turbulent kinetic energy
      & $\langle \rho k \rangle$
      & $\Delta \rho g h_1$ \\
    
      Buoyancy production
      & $-\langle \rho'vg\rangle$
      & $\Delta \rho g h_1/\tau_0$ \\
    
      Viscous dissipation rate
      & $ \langle \rho \epsilon \rangle$
      & $\Delta \rho g h_1/\tau_0$ \\
    \end{tabular}
  \end{center}
  \caption{Summary of scalings}
  \label{tab:scaling_summary}
\end{table}

The scaling of the budget terms derived in \S~\ref{sec:scaling_budget} is summarized in table \ref{tab:scaling_summary}. The mixed mass budget was normalized by a vertically-averaged mixed mass $M/h_1$, or a turbulent mixing density $\rho_m$, while the TKE budget was normalized by the total potential energy loss of the flow $\Delta \rho g h_1$. Both sets of scalings retained the SRT timescale parameter $\tau_0$. To achieve a true self-similar scaling (where all dynamical properties of the flow depend solely on $h_1$), $\tau_0$ and $\rho_m$ were related to the input variables ($\rho_H$, $\rho_L$, $g$, $h_1$) in \S~\ref{sec:scaling_global} based on the following physical observations on RT turbulence:
\begin{enumerate}
    \item The mean mole fraction profile resembles that of a 1D diffusion problem: this determines the shape of the mean mass flux and the buoyancy production term in the TKE budget. 
    \item The integrated mixing parameter (as defined by $\Theta_m$) is independent of Atwood number. Combined with (i), an (approximate) equivalence is established between between the turbulent mixing density $\rho_m$ and the diffusion mixing density $\rho_m^*$, which has an analytical solution based solely on $\rho_H$ and $\rho_L$.
    \item The growth rate of the mixing layer $h/\tau_0$ is related to the turbulent velocity fluctuations $\tilde{k}^{1/2}$ (and not the mean velocity $\tilde{v}$).
    \item The conserved form of the turbulent kinetic energy $\langle \rho k\rangle$ scales with buoyancy production locally (at each $y$-location) and not just in a volume-integrated sense. As a result, the TKE profile preserves the Gaussian-like shape of the mean mass flux, which is similar to the gradient of the mean density. Combined with an algebraic ``trick'' that writes $1/\langle\rho\rangle$ as ${\rm d}\ln \langle \rho \rangle/{\rm d}\langle \rho \rangle$, the scaling for $\tau_0$ is derived.
\end{enumerate}

Other trends observed in this study can be related back to these four observations. First, the derivation of the effective Atwood number (and effective Grashof number) is simply an algebraic consequence of the scaling of $\tau_0$. Second, observation (i) implies an invariance of the mole fraction $\langle X \rangle$ profile, which dictates (by definition) that Favre-averaged mass fraction $\tilde{Y}$ profiles are shifted towards the light-fluid side. The observed shifts in other mass-fraction-based quantities like the mixed mass and scalar dissipation rate are also consistent with the diffusive description of RT turbulence. Third, the near-symmetric profiles of the TKE budget terms are a combined consequence of observations (i) and (iv)---their Favre-averaged profiles are shifted simply due to division by a monotonically increasing mean density. 

As introduced in \S~\ref{sec:intro} through (\ref{eq:hgrowth_lnR}), the scaling of the mixing layer height with $\ln R$ was first proposed in the theoretical analysis of \citet{belen1965theory}, based on a combination of energy scaling arguments and a spatially-varying turbulent diffusivity model. Although derived independently, the consistency in our results is unsurprising, as the analysis of \citet{belen1965theory} relied on the same physical assumptions (i) and (iv), which had not been verified with physical data until the present study. In \citet{belen1965theory}, these assumptions led to the definition of a self-similar problem for the mean density, which, when constrained to small density ratios ($R\lesssim 4$), has an exact solution that includes $\ln R$. In the present work, the mean mole fraction is approximated by an error function, which leads to the same scaling. To our knowledge, this work provides the first verification of the $\ln R$ scaling using RT turbulence data. Furthermore, we demonstrate that the $\ln R$ scaling remains effective for density ratios well beyond the range initially constrained by \citet{belen1965theory}.

The mean density profile determines the shape of the mean mass flux, buoyancy production, and TKE; it also drives the shift in Favre-averaged profiles towards the light-fluid side. In the present work, approximating the mean mole fraction with an error function proved effective for estimating integrated normalization quantities. However, improving the mean density model could enhance predictions for other profile quantities. \citet{boffetta2010nonlinear} proposed a turbulent diffusivity closure similar to \citet{belen1965theory} that was validated with DNS results of Boussinesq RT turbulence. Extending such models to the general variable-density case could lead to a more complete description of RT turbulence. 

\section{Conclusion}
\label{sec:conclusion}

The present work extends the methodology of \citet{goh2025statistically} to a range of Atwood numbers, $A\le0.8$. By leveraging the low computational cost of the SRT configuration, highly converged statistics were obtained across a broad range of Atwood and Reynolds numbers. Normalizations were developed for all significant non-transport terms in the continuity, mixed mass and turbulent kinetic energy budgets. When applied to the SRT dataset, most quantities exhibit strong collapse in their profiles after normalization. In some cases, the normalized profiles shift toward the light-fluid side with increasing Atwood number, matching trends observed in the one-dimensional variable-density diffusion solution. 

For variable-density RT flows, the mixing layer height is found to scale with $\ln R$, a scaling originally proposed by \citet{belen1965theory}, but which has not been applied in subsequent RT studies. An effective Atwood number is introduced as $A^*=(\ln R)/2$, which reduces to the standard Atwood number in the Boussinesq limit. The corresponding effective growth parameter, $\alpha^*=\dot{h}^2/4A^*gh$, exhibits a significantly reduced dependence on the Atwood number compared to the traditional growth parameter $\alpha$. Further, defining the Grashof number using $A^*$ yields an effective Grashof number $\Gr^*$ that scales directly with  $\Rey^2$, reinforcing the internal consistency of the proposed scaling.

The observed universality of the normalized quantities presents opportunities for broader application. In particular, Reynolds number effects in self-similar RT turbulence appear to be almost entirely captured by the mixing layer growth time scale (or equivalently, $\alpha^*$). The influence of the Atwood number on the integrated quantities is largely reflected in the proposed scalings, while possible shifts in the profiles can be predicted using the 1D analytical solution. These observations highlight the potential for using lower-cost DNS studies of low-Reynolds-number Boussinesq RT flows to estimate flow statistics under more extreme conditions. 

\backsection[Acknowledgements]{
This work used resources provided by Texas Advanced Computing Center (TACC), The University of Texas at Austin through allocation MCH230009 from the Advanced Cyberinfrastructure Coordination Ecosystem: Services \& Support (ACCESS) program \citep{Boerner2023}, which is supported by U.S. National Science Foundation grants \#2138259, \#2138286, \#2138307, \#2137603, and \#2138296. 
This work also used resources of the National Energy Research Scientific Computing Center, a DOE Office of Science User Facility supported by the Office of Science of the U.S. Department of Energy under Contract No. DE-AC02-05CH11231 using NERSC award FES-ERCAP0031321. Additionally, the authors acknowledge the use of Google AI to assist in translating the original Russian document of \citet{belen1965theory} into English.}

\backsection[Funding]{This material is based upon work supported by the National Science Foundation under Grant No. 2422513.}

\backsection[Declaration of interests]{The authors report no conflict of interest.}

\backsection[Author ORCIDs]{C.Y. Goh  https://orcid.org/0009-0007-1426-7581; D. Brito Matehuala https://orcid.org/0009-0004-4301-2375; G. Blanquart  https://orcid.org/0000-0002-5074-9728}

\appendix
\section{Impact of the SRT source terms on energy spectra}
\label{app:src_spectra}
The impact of the SRT source terms (\ref{eq:Sphi}--\ref{eq:Ti}) on the energy spectra is examined. Combining (\ref{eq:mass_SRT}) and (\ref{eq:mom_SRT}), the momentum equation can be written as
\begin{equation}
  \pd{u_i}{s} = -u_j\pd{u_i}{x_j} - \frac{1}{\rho}\left(\pd{p}{x_i} - \pd{\tau_{ij}}{x_j}\right) + \frac{y}{\tau_0}\pd{u_i}{y} - \frac{u_i}{2\tau_0}.
  \label{eq:mom_app}
\end{equation}
Applying a Fourier transform (notated as $\mathcal{F}$) to (\ref{eq:mom_app}) in the homogeneous directions ($x$,$z$) yields
\begin{equation}
  \pd{\hat{u}_i}{s} = \mathcal{F}\left[-u_j\pd{u_i}{x_j} - \frac{1}{\rho}\left(\pd{p}{x_i} - \pd{\tau_{ij}}{x_j}\right)\right] + \frac{y}{\tau_0}\pd{\hat{u}_i}{y} - \frac{\hat{u}_i}{2\tau_0},
  \label{eq:mom_ft}
\end{equation}
where $\hat{u}_i = \mathcal{F}(u_i)$, and the in-plane Fourier transform commutes with the differential operator in $y$. Then, (\ref{eq:mom_ft}) is multiplied by $\hat{u}_i^*$ to derive an equation for the two-dimensional energy spectra $\hat{E}(k_x,y,k_z,s) = (\hat{u}_i\hat{u}_i^*)/2$. The result is
\begin{equation}
  \pd{\hat{E}}{s} = \hat{u}_i^*\cdot\mathcal{F}\left[-u_j\pd{u_i}{x_j} - \frac{1}{\rho}\left(\pd{p}{x_i} - \pd{\tau_{ij}}{x_j}\right)\right] + \frac{1}{\tau_0}\left(y\pd{}{y} - 1\right) \hat{E}.
  \label{eq:tke_fourier}
\end{equation}
Represented by the rightmost term in (\ref{eq:tke_fourier}), the source terms act proportionally to the in-plane TKE spectra, and do not selectively inject or remove energy in any specific wavenumber range. Graphically, the effect of the source terms corresponds to a continuous downward shift of the entire energy spectra (when displayed on a logarithmic scale).

\section{Analytical solution to variable-density one-dimensional diffusion}
\label{app:1dvd}

The one-dimensional variable-density temporal diffusion problem in $(t,y)$ coordinates is entirely governed by continuity, scalar transport, and the equation of state. They are
\begin{gather}
  \pd{\rho}{t} + \pd{\rho v}{y} = 0,
  \label{eq:cont1d}
  \\ 
  \pd{\rho Y}{t} + \pd{\rho v Y}{y} = D\pd{}{y}\left(\rho \pd{Y}{y}\right), 
  \label{eq:scalar1d}
  \\  
  \rho(Y) = \frac{\rho_H \rho_L}{\rho_H- Y\Delta \rho}.
  \label{eq:eos1d}
\end{gather}
Equations (\ref{eq:cont1d})--(\ref{eq:eos1d}) determine $\rho$, $v$, and $Y$ completely. The solution is diffusion driven and does not depend on gravity. With $\rho$ and $v$ determined, the pressure can be obtained from the momentum equation using known values of $g$ and $\nu$.

\begin{figure}
  \begin{minipage}{0.498\textwidth}
    (\emph{a})\\
    \includegraphics[width=\textwidth]{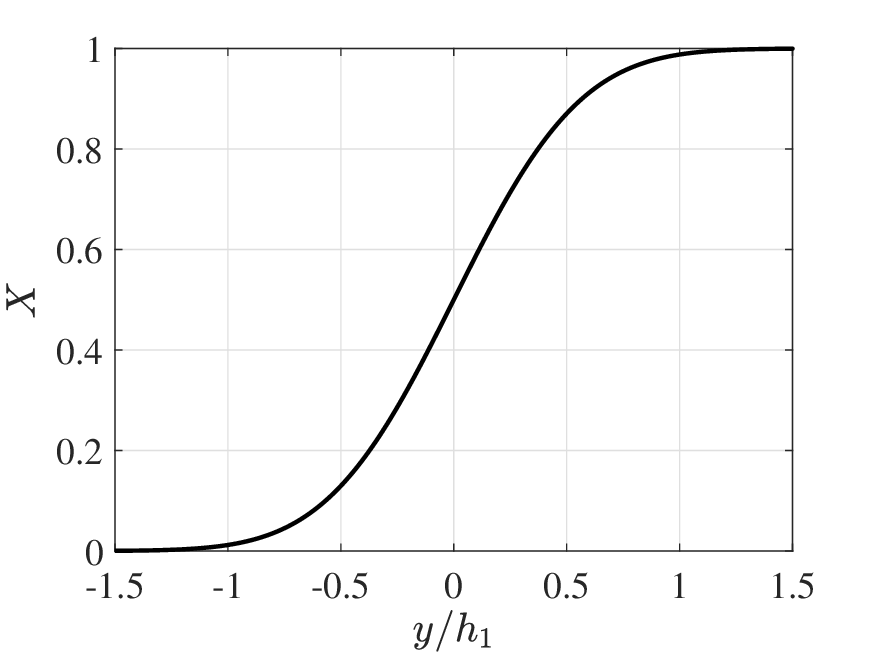}
  \end{minipage}
  \hfill
  \begin{minipage}{0.498\textwidth}
    (\emph{b})\\
    \includegraphics[width=\textwidth]{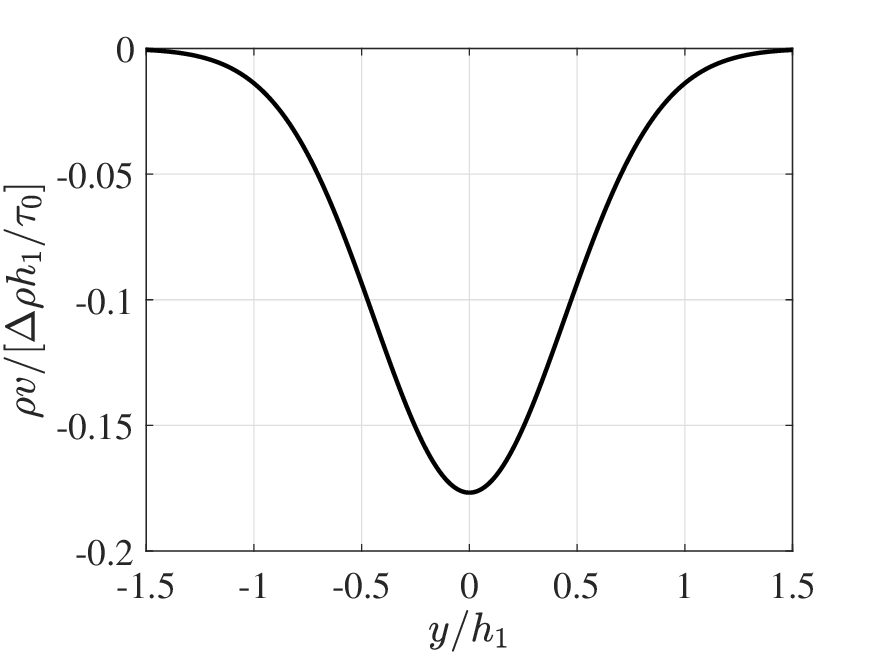}
  \end{minipage}
  \begin{minipage}{0.498\textwidth}
    (\emph{c})\\
    \includegraphics[width=\textwidth]{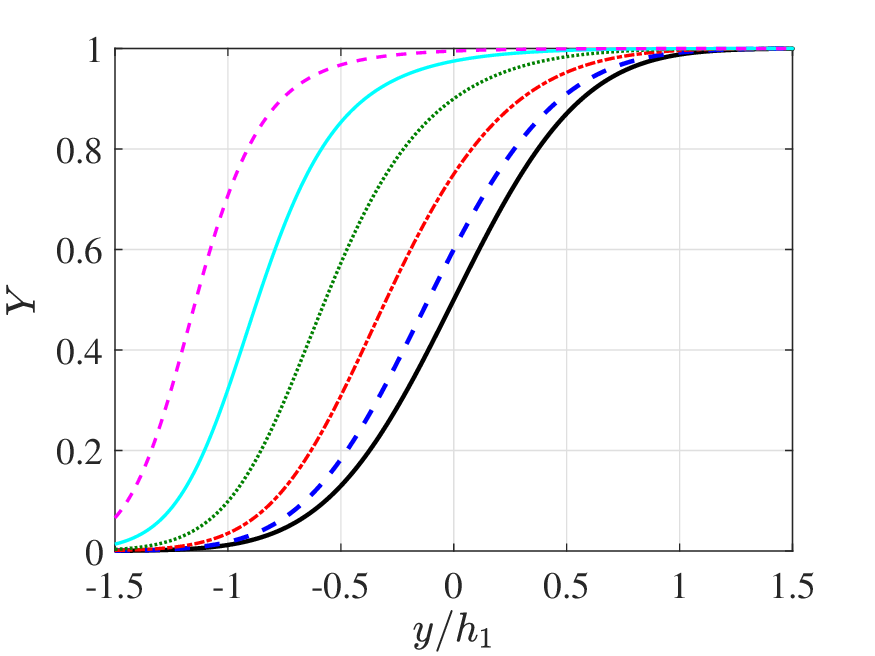}
  \end{minipage}
  \hfill
  \begin{minipage}{0.498\textwidth}
    (\emph{d})\\
    \includegraphics[width=\textwidth]{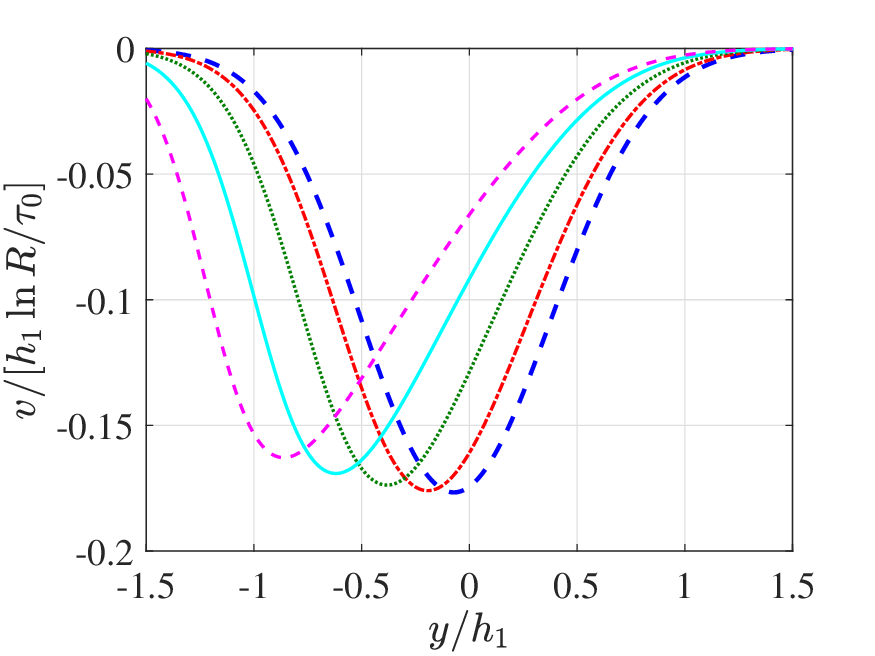}
  \end{minipage}
  \begin{minipage}{0.498\textwidth}
    (\emph{e})\\
    \includegraphics[width=\textwidth]{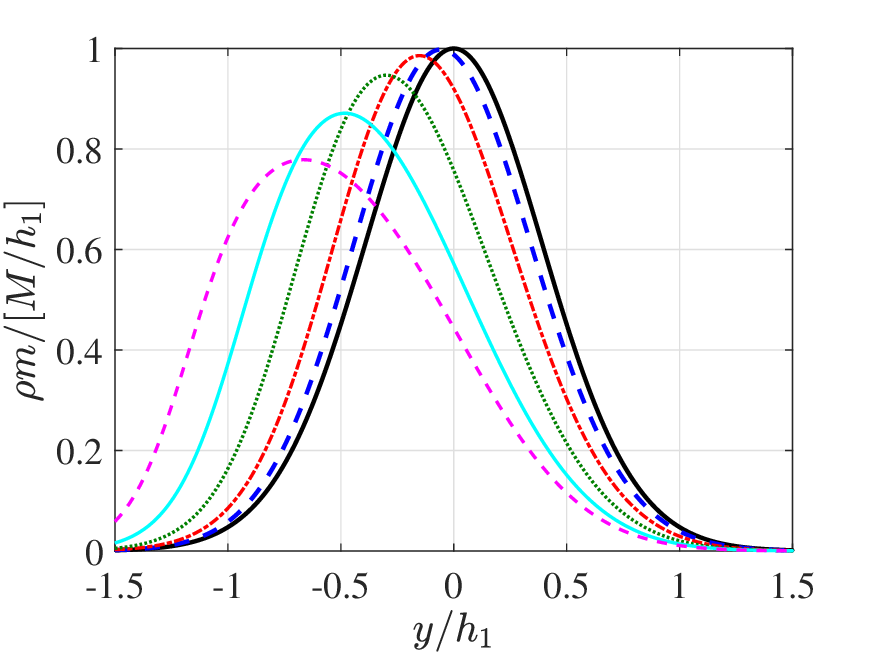}
  \end{minipage}
  \hfill
  \begin{minipage}{0.498\textwidth}
    (\emph{f})\\
    \includegraphics[width=\textwidth]{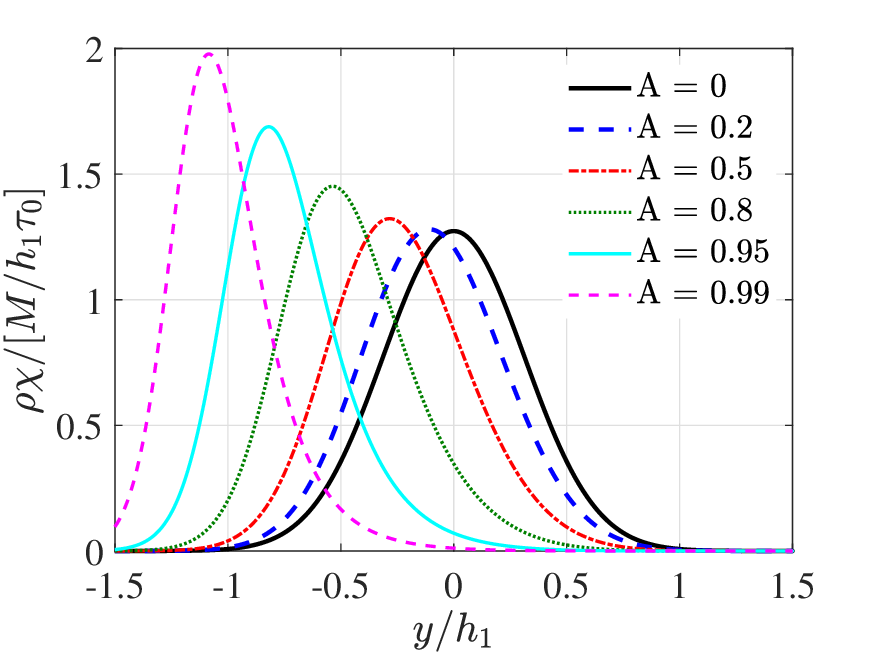}
  \end{minipage}
  \caption{Analytical solution to the one-dimensional variable-density diffusion problem in normalized variables: (\emph{a}) mole fraction (independent of $A$), (\emph{b}) mass flux (independent of $A$), (\emph{c}) mass fraction, (\emph{d}) velocity, (\emph{e}) mixed mass, and (\emph{f}) scalar dissipation rate. Legend: $A=0$ (thick black solid), 0.2 (thick blue dashed); 0.5 (red dash-dotted); 0.8 (green dotted); 0.95 (cyan solid); 0.99 (magenta dashed)
  \label{fig:sol1D}}  
\end{figure}

The analytical self-similar solution to the variable-density diffusion problem, expressed in terms of the similarity variable, $\zeta = y/\sqrt{4Dt} = y/h_D$, (notated with subscript $D$ for ``diffusion''), is
\begin{gather}
  X = \frac{\rho-\rho_L}{\Delta \rho} = \frac{1}{2}(1 + \erf \zeta), 
  \label{eq:X1d}
  \\ 
  \frac{\rho v}{\Delta \rho D/h_D} = -\dd{X}{\zeta} = -\frac{1}{\sqrt{\pi}} e^{-\zeta^2},
  \label{eq:rhou1d}
\end{gather}
where the solution for the mole fraction is identical to that of the constant-density problem, and variable-density effects are only observed in the velocity. The mole fraction $X$ is solely a function of $\zeta$ and independent of the reservoir densities.  
In contrast, the mass fraction $Y$, which is algebraically related to $X$ by the equation of state, $Y(X, R) = RX/[1 + (1-R)X]$, does vary with the density ratio (i.e. Atwood number). By extension, this is generally true for all functions of $Y$, including the one-dimensional mixed mass, $m=\rho Y(1-Y)$, and scalar dissipation rate, $\chi = 2\rho D(\partial Y/\partial y)^2$, as shown in figure~\ref{fig:sol1D}. 
For ease of comparison with the SRT results, quantities in figure~\ref{fig:sol1D} are normalized with the equivalent variables, where $h_1 = \sqrt{8/\pi} h_D$, $M = \int \rho m \,{\rm d}y$, and $\tau_0 = h_D/\dot{h}_D = h_D^2/2D = \pi h_1^2/16D$. As $A$ increases, the mass fraction and velocity shifts leftwards, the mixed mass profile broadens with a reduction in its peak value, and the scalar dissipation narrows while experiencing an increase in its peak value.

\section{Closure equation for $h_1$}
\label{app:closure_h1}
To derive a time evolution equation for the instantaneous mixing layer height $h_1(s)$, density is written in terms of the mole fraction, $\rho = \rho_L + \Delta \rho X$, and substituted into the continuity equation to yield
\begin{equation}
    \pd{\langle X \rangle_{1,3}}{s} + \frac{1}{\Delta \rho}\pd{\langle \rho v \rangle_{1,3}}{y} = \frac{y}{\tau_0} \pd{\langle X \rangle_{1,3}}{y}.
    \label{eq:Xbar}
\end{equation}
An equation for the transport of $\langle X \rangle_{1,3}^2$ is obtained by multiplying (\ref{eq:Xbar}) with $2\langle X \rangle_{1,3}$ to get 
\begin{equation}
    \pd{\langle X \rangle_{1,3} ^2}{s} + \frac{2}{\Delta \rho} \langle X \rangle_{1,3} \pd{\langle\rho v\rangle_{1,3}}{y} = \frac{y}{\tau_0}\pd{\langle X \rangle_{1,3}^2}{y}.
    \label{eq:Xbar2}
\end{equation}
Combining (\ref{eq:Xbar}) and (\ref{eq:Xbar2}) results in 
\begin{equation}
    \dd{h_1}{s} = \dd{}{s} \int 4\left(\langle X \rangle_{1,3} - \langle X \rangle_{1,3}^2\right)\,{\rm d}y= \frac{8}{\Delta \rho}\int \langle X \rangle_{1,3} \pd{\langle \rho v \rangle_{1,3}}{y} \,{\rm d}y- \frac{h_1}{\tau_0}.
    \label{eq:dh1_dsA}
\end{equation}
    
The SRT parameter $\tau_0$ is implemented by assuming a stationary solution, $dh_1/ds=0$, to (\ref{eq:dh1_dsA}) and substituting $h_1 = \hat{h}_1$ so that 
\begin{equation}
    \frac{1}{\tau_0(s)} = \frac{8}{\Delta \rho \hat{h}_1} \int \langle X \rangle_{1,3} \pd{\langle \rho v \rangle_{1,3}}{y} \,{\rm d}y.  
    \label{eq:closureh1_app}
\end{equation}
Substituting (\ref{eq:closureh1_app}) into (\ref{eq:dh1_dsA}) results in a time evolution equation for $h_1$ that is written as 
\begin{equation}
    \dd{h_1}{s} = \dd{}{s} \int 4 \langle X \rangle_{1,3} \left(1- \langle X \rangle_{1,3}\right)\,{\rm d}y= \left[\frac{8}{\Delta \rho}\int \langle X \rangle_{1,3} \pd{\langle \rho v \rangle_{1,3}}{y} \,{\rm d}y\right] \left(1 - \frac{h_1}{\hat{h}_1}\right).
    \label{eq:dh1ds}
\end{equation}
The integral factor in the RHS of (\ref{eq:dh1ds}) is positive for an unstable RT flow configuration, hence the use of (\ref{eq:closureh1_app}) results in a stable relaxation equation for $h_1$. The use of closure equations such as (\ref{eq:closure}) or (\ref{eq:closureh1_app}) (as opposed to a constant $\tau_0$) is simply a practical way of achieving a specified mixing layer height without the need to guess the value of $\tau_0$ \emph{a priori}. It was shown in Appendix A of \citet{goh2025statistically} that the ensemble dynamics of SRT flow is equivalent for different implementations of $\tau_0$ as long as the value of the time average $\bar{\tau}_0$ is the same. 

\section{Local scaling of turbulent kinetic energy with buoyancy production}
\label{app:tke_scaling}

\begin{figure}
  \begin{center}
  \includegraphics[width=0.498\textwidth]{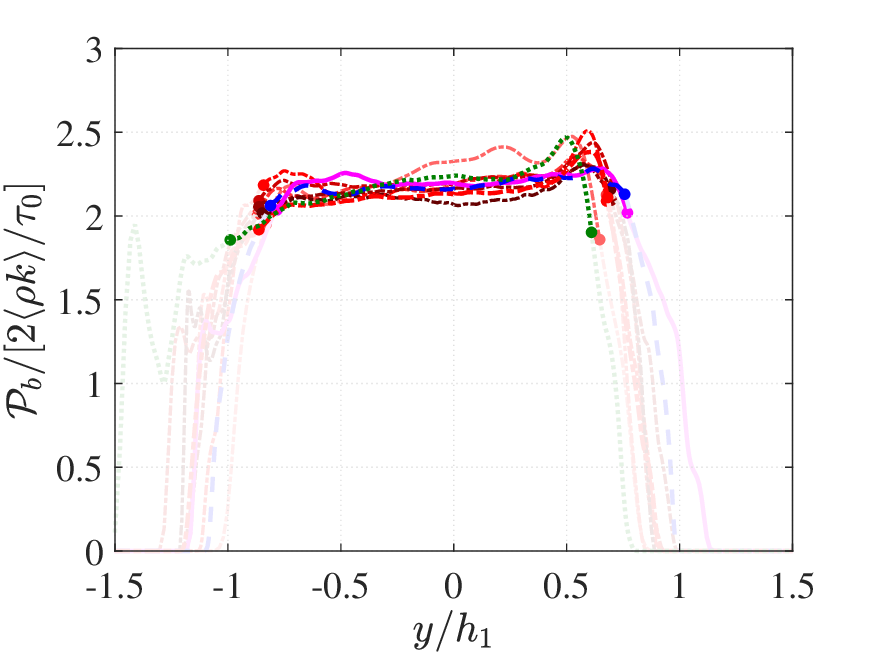}
  \caption{Ratio of buoyancy production to TKE growth. Atwood number effects are represented by $A=$ 0.01 (magenta solid), 0.2 (blue dashed), 0.5 (red dash-dotted), 0.8 (green dotted); and Reynolds number effects are represented by dash-dotted lines of different shades of red. Only regions where buoyancy production is at least 10\% of its maximum value are shown with full opacity.}
  \label{fig:PKratio}
  \end{center}
\end{figure}

A qualitative observation about the TKE budget profiles in figures~\ref{fig:tke} and \ref{fig:tke_Re} is that buoyancy production and TKE profiles have similar shapes that resemble a Gaussian function. To assess their similarities more quantitatively, buoyancy production is normalized by TKE and shown in figure \ref{fig:PKratio}. Towards the edge of the mixing layer, both buoyancy production and TKE approach 0, and the division of small values may lead to significant noise. To focus on the behavior of the bulk of the mixing layer, only regions where buoyancy production is at least 10\% of its maximum value are shown.

The normalized ratio of buoyancy production to TKE growth is remarkably constant across the mixing layer at around 2.2, which suggests that the buoyancy production scaling applies not only globally, but also locally, and that the TKE profile largely preserves the shape of buoyancy production (and by extension, the mean mass flux). This local scaling of TKE with buoyancy production is a key observation that is utilized in the derivation of the scaling for $\tau_0$ in \S~\ref{sec:scaling_global}.

\bibliographystyle{jfm}
\bibliography{references}

\end{document}